\documentclass[a4paper,11pt]{article}
\usepackage{aaskaiid}
\usepackage{pdflscape} 
\usepackage{afterpage} 
\usepackage{orcidlink} 

\graphicspath{{./}{figs/}}

\newcommand\Sec[1]{Sec.~\ref{#1}}

\DeclareRobustCommand{\VAN}[3]{#2}
\let\VANthebibliography\thebibliography
\def\thebibliography{\DeclareRobustCommand{\VAN}[3]{##3}\VANthebibliography}

\newcommand{\lt}{\ell_{\rm turb}} 

\newcommand{\cm}{{\rm cm}}    
\newcommand{\pc}{{\rm pc}}     
\newcommand{\kpc}{{\rm kpc}}  

\newcommand{\MHz}{{\rm MHz}} 
\newcommand{\yr}{{\rm yr}}    

\newcommand{\muG}{\mu{\rm G}} 

\newcommand{\eV}{{\rm eV}}  
\newcommand{\GeV}{{\rm GeV}}  
\newcommand{\nJy}{{\rm nJy}}    

\newcommand{\RM}{\text{RM}}

\newcommand{\Ekin}{E_{\rm kin}}

\newcommand{\Emag}{E_{\rm mag}}

\title{Small-scale Magnetic Fields in the Milky Way and Nearby Galaxies}
\ShortTitle{Small-scale galactic magnetic fields}
\ShortName{Ma, Seta et al.}

\author[1]{Yik~Ki~Ma\,\orcidlink{0000-0003-0742-2006}$^{*}$}
\author[2]{Amit~Seta\,\orcidlink{0000-0001-9708-0286}$^{*}$}
\author[3,1]{Aritra~Basu\,\orcidlink{0000-0003-2030-3394}}
\author[4]{Sebastian~Hutschenreuter\,\orcidlink{0000-0002-6952-9688}}
\author[5]{Marco~Padovani\,\orcidlink{0000-0003-2303-0096}}
\author[6]{Georgia~V.~Panopoulou\,\orcidlink{0000-0001-7482-5759}}
\author[7]{Jeroen~M.~Stil\,\orcidlink{}}
\author[2]{Craig~S.~Anderson\,\orcidlink{0000-0002-6243-7879}}
\author[8]{Lucia~Armillotta\,\orcidlink{0000-0002-5708-1927}}
\author[9,10,11]{Jennifer~Y.~H.~Chan\,\orcidlink{0000-0003-0314-7027}}
\author[12]{Marijke Haverkorn\,\orcidlink{0000-0002-5288-312X}}
\author[2]{Roland~M.~Crocker\,\orcidlink{0000-0002-2036-2426}}
\author[13]{Timea~O.~Kovacs\,\orcidlink{0000-0001-6649-8559}}
\author[14]{Sunil Malik\,\orcidlink{0000-0003-4147-626X}}
\author[1]{S.~A.~Mao\,\orcidlink{0000-0001-8906-7866}}
\author[7]{Kierra~J.~Weatherhead\,\orcidlink{0009-0004-9607-721X}}

\affiliation[*]{These authors jointly co-lead the Chapter and contributed equally to the work.}
\affiliation[1]{Max-Planck-Institut f\"ur Radioastronomie, Auf dem H\"ugel 69, 53121 Bonn, Germany}
\emailAdd{ykma@mpifr-bonn.mpg.de}
\affiliation[2]{Research School of Astronomy \& Astrophysics, Australian National University, Canberra, ACT 2611, Australia}
\emailAdd{amit.seta@anu.edu.au}
\affiliation[3]{Th\"uringer Landessternwarte, Sternwarte 5, 07778 Tautenburg, Germany}
\affiliation[4]{University of Vienna, Department of Astrophysics, T\"urkenschanzstra{\ss}e 17, 1180 Vienna, Austria}
\affiliation[5]{INAF-Osservatorio Astrofisico di Arcetri, Largo E. Fermi 5, 50125 Firenze, Italy}
\affiliation[6]{Department of Space, Earth and Environment, Chalmers University of Technology, 412 93, G\"{o}teborg, Sweden}
\affiliation[7]{Department of Physics and Astronomy, The University of Calgary, 2500 University Drive NW, Calgary, AB T2N 1N4, Canada}
\affiliation[8]{University of Florence, Department of Physics and Astronomy, via G.\ Sansone 1, 50019 Sesto Fiorentino, Firenze, Italy}
\affiliation[9]{Department of Physics and Astronomy, Oberlin College, Oberlin, OH 44074, USA}
\affiliation[10]{Dunlap Institute for Astronomy and Astrophysics, 50 St. George Street, Toronto, Ontario, M5S 3H4, Canada}
\affiliation[11]{Canadian Institute for Theoretical Astrophysics, University of Toronto, 60 St George St, Toronto, ON M5S 3H8, Canada}
\affiliation[12]{Department of Astrophysics/IMAPP, Radboud University, PO Box 9010, 6500 GL Nijmegen, The Netherlands}
\affiliation[13]{Max-Planck-Institut f\"ur Astronomie, K\"onigstuhl 17, 69117 Heidelberg, Germany}
\affiliation[14]{Departamento de Física de la Tierra y Astrofísica \& IPARCOS-UCM, Universidad Complutense de Madrid, 28040 Madrid, Spain}

\abstract{
Magnetic fields in galaxies span decades in physical scale, from the coherent magnetic fields on galactic scales ($>{\rm kpc}$) to the random magnetic fields from $100\,{\rm pc}$ to the resistive scale of the galactic plasma (i.e.\,$\sim 10^{6}\,{\rm cm}$). While many radio studies to date have placed more emphasis on the large-scale galactic magnetic fields than the small-scale counterparts, the emerging SKA will greatly facilitate accurate, detailed studies of the small-scale ($\lesssim 100\,{\rm pc}$) galactic magnetic fields. In this Chapter, we highlight the importance of understanding the small-scale galactic magnetic fields in furthering our understanding of star formation, galaxy evolution, and the fundamental physics of magnetohydrodynamics. Furthermore, we discuss some open questions in the research field and outline several possible large observation programmes with the SKA Array Assembly 4 (AA4).}

\begin{document}
\newcommand{\actaa}{Acta Astron.} 
\newcommand{\araa}{ARA\&A} 
\newcommand{\aar}{A\&ARv} 
\newcommand{\aapr}{A\&ARv} 
\newcommand{\ab}{Astrobiol.} 
\newcommand{\aj}{AJ} 
\newcommand{\apj}{ApJ} 
\newcommand{\apjl}{ApJL} 
\newcommand{\apjs}{ApJSS} 
\newcommand{\ao}{Appl. Opt.} 
\newcommand{\apss}{Astro. \& Space Sci.} 
\newcommand{\aap}{A\&A} 
\newcommand{\aaps}{A\&AS.} 
\newcommand{\baas}{Bull. Am. Astron. Soc.} 
\newcommand{\caa}{Chinese A\&A} 
\newcommand{\cjaa}{Chinese J. A\&A} 
\newcommand{\cqg}{Class. Quantum Gravity} 
\newcommand{\gal}{Galaxies} 
\newcommand{\gca}{Geo. Cosmo. Acta} 
\newcommand{\icarus}{Icarus} 
\newcommand{\jcap}{JCAP} 
\newcommand{\jgr}{J. Geophys. Res.} 
\newcommand{\jgrp}{J. Geophys. Res. Planets} 
\newcommand{\jqsrt}{J. Quant. Spectrosc. Radiat. Transf.} 
\newcommand{\memsai}{Mem. SAIt} 
\newcommand{\mnras}{MNRAS} 
\newcommand{\nat}{Nature} 
\newcommand{\nastro}{Nat. Astron.} 
\newcommand{\ncomms}{Nat. Commun.} 
\newcommand{\nphys}{Nat. Phys.} 
\newcommand{\na}{New Astron.} 
\newcommand{\nar}{New Astron. Rev.} 
\newcommand{\physrep}{Phys. Rep.} 
\newcommand{\pra}{Phys. Rev. A} 
\newcommand{\prb}{Phys. Rev. B} 
\newcommand{\prc}{Phys. Rev. C} 
\newcommand{\prd}{Phys. Rev. D} 
\newcommand{\pre}{Phys. Rev. E} 
\newcommand{\prx}{Phys. Rev. X} 
\newcommand{\prl}{Phys. Rev. Let.} 
\newcommand{\psj}{Planet. Sci. J.} 
\newcommand{\planss}{Planet. Space Sci.} 
\newcommand{\pnas}{Proc. Natl Acad. Sci. USA} 
\newcommand{\procspie}{Proc. SPIE} 
\newcommand{\pasa}{PASA} 
\newcommand{\pasj}{PASJ} 
\newcommand{\pasp}{PASP} 
\newcommand{\rmxaa}{RMXAA} 
\newcommand{\sci}{Science} 
\newcommand{\sciadv}{Sci. Adv.} 
\newcommand{\solphys}{Sol. Phys.} 
\newcommand{\sovast}{Soviet Ast.} 
\newcommand{\ssr}{Space Sci. Rev.} 
\newcommand{\uni}{Universe} 

\maketitle

\section{Introduction} \label{sec:intro}
Magnetic fields are a dynamically important component of the interstellar medium (ISM) of galaxies, as they play a crucial role in star formation \citep{PattleEA2023}, regulate the propagation of cosmic rays \citep{RuszkowskiP2023}, and influence gas morphology, distribution, and flow \citep{PlanckXXXV_2016}. The magnetic energy density in the Milky Way ($\sim 1\,\eV\,\cm^{-3}$) is comparable to the thermal, turbulent kinetic, and cosmic ray energy densities, making it an energetically significant component of the ISM. However, the properties of ISM magnetic fields, in both the Milky Way and external galaxies, are not well known. This is particularly true on small scales ($\lesssim 100\,\pc$), making it difficult to fully understand their role in star formation and galaxy evolution -- two outstanding questions in modern astrophysics. This Chapter discusses the current theoretical and observational understanding of these small-scale magnetic fields in the Milky Way and nearby galaxies and describes the transformative science that will be enabled by the SKA in its Array Assembly 4 (AA4).

Magnetic fields are directly connected to velocity fields, and for disk galaxies, this includes both the more systematic differential rotation and chaotic, random, turbulent motions. Turbulence in the ISM of galaxies is driven by a variety of mechanisms operating across a range of scales. On smaller scales ($\lesssim 100\,\pc$), stellar feedback processes (namely, protostellar jets, stellar outflows, and supernova explosions; each could be at a very different scale within $\sim 100\,\pc$) are the primary drivers of turbulence \citep{ElmegreenS2004}. On larger, $\kpc$-scales, shear from galactic differential rotation and gravitational instabilities also introduce turbulence \citep{KrumholzEA2018}. Energetically, supernova explosions are the most powerful drivers of ISM turbulence; thus, the typical driving scale of turbulence, $\lt$, is often taken to be the average supernova remnant size, approximately $100\,\pc$ \citep[e.g., see][]{GentEA2013I}. Based on $\lt$, galactic magnetic fields are typically divided into two types: large-scale, coherent (or ``regular'') fields with length scales of a few $\kpc$ \citep[$B_{\rm reg}$; usually approximately aligned with the spiral arms or interarm regions; see, e.g.,][]{FletcherEA2011, Beck2015}, and small-scale, turbulent fields on scales of $\lesssim 100\,\pc$ ($b_{\rm iso}$ and $b_{\rm ani}$; see below). We emphasise that magnetic fields in galaxies are inherently multi-scale with power at a range of scales (usually characterised by a power spectrum). When distinguishing between small- and large-scale components, we primarily refer to their respective correlation scales; the range of spatial scales over which power is distributed can still overlap between the two components. This work primarily focuses on the science of the small-scale fields as probed by the SKA and its precursors.

Magnetic fields in the early Universe ($\approx 10^{-10}\,\muG$) and in protogalaxies ($\approx 10^{-4}\,\muG$) are significantly weaker \citep{Subramanian2016} than those observed in the present day galaxies \citep[$1$–$10\,\muG$,][]{Beck2016}. This amplification and subsequent maintenance of magnetic fields are due to dynamos, the process of converting turbulent kinetic energy into magnetic energy \citep{RuzmaikinEA1988, BrandenburgS2005, Rincon2019, ShukurovS2021}. Based on scale, dynamos are classified into two types: large- and small-scale dynamos. The small-scale dynamo (also referred to as the fluctuation dynamo) converts turbulent kinetic energy into magnetic energy on short eddy-turnover timescales ($\approx 10^{7}\,\yr$) at scales smaller than the turbulent driving scale of $\approx 100\,\pc$ \citep{Kazantsev1968, SchekochihinEA2004, HaugenEA2004, SetaEA2020, SetaF2021dyn}. The resultant galactic magnetic field component is commonly referred to as the isotropic turbulent magnetic field ($b_{\rm iso}$). Such magnetic fields can be further enhanced and ordered by the large-scale dynamo (also referred to as the mean-field or $\alpha$–$\Omega$ dynamo), which operates on galactic rotation timescales ($\approx 10^{8}\,\yr$) and over length scales of a few $\kpc$, due to the combined effects of density stratification along the disk height ($\alpha$) and differential rotation ($\Omega$) \citep{RuzmaikinEA1988, BeckEA1996, BrandenburgS2005, Rincon2019, ShukurovS2021}. The large-scale field generated by the large-scale dynamo can also be tangled by ISM turbulence, further producing small-scale random fields (see Appendix~A in \citealt{SetaEA2018} and Sec.~4.1 in \citealt{SetaF2020}). Meanwhile, spiral shocks, differential rotation, and gaseous shearing motion can compress the small-scale turbulent magnetic field \citep{Laing1980}, causing the resulting magnetic fields to follow a single orientation (but can flip in direction by $180^\circ$). This magnetic component of galaxies is referred to as the anisotropic turbulent / ordered random / striated magnetic field \citep[$b_{\rm ani}$; see e.g.][]{Jaffe2010,Jansson2012b,Haverkorn2015,Beck2016,KierdorfEA2020,MaEA2025}. Here, we primarily discuss how radio polarisation observations can help us understand the properties of the small-scale fields, the small-scale dynamo, and their connection with the galactic ecosystem.

This Chapter is structured as follows. In \Sec{sec:obs}, we briefly summarise the observational probes of small-scale magnetic fields that one can employ with the SKA. In \Sec{sec:sci}, we discuss the high-level scientific goals in this research area, and in \Sec{sec:ques}, we explore the specific astrophysics questions that can be addressed with the SKA. \Sec{sec:ska} highlights the preparation works that the community ought to focus on in the coming years in the lead-up to the SKA (\Sec{sec:pre}), describes in general the types of SKA observations that can benefit small-scale galactic magnetic field studies (\Sec{sec:SKA}), presents several prospective large observation programmes with the SKA AA4 (\Sec{sec:largescale}), and highlights multiple anticipated advancements beyond the radio wavelengths that synergise with our proposed science with SKA  (\Sec{sec:synergies}).

\section{Observing Galactic Small-scale Magnetic Fields} \label{sec:obs}

We provide a brief overview of the observational tracers of galactic small-scale magnetic fields that we will be able to utilise with the SKA. In addition, we highlight the key differences in the use of these tracers for the cases of the Milky Way and external galaxies. Furthermore, we point out how recent advancements have enabled direct comparisons between upcoming SKA results with the numerical simulations counterparts.

\subsection{Overview of the Observational Tracers}

\subsubsection{Synchrotron Emission} \label{sec:syn}

Synchrotron emission is radiated by ultra-relativistic charged particles (primarily electrons, in an astrophysical context) subjected to magnetic fields, with the (intrinsic) linear polarisation plane being perpendicular to the magnetic field orientation \citep{RL79}. Below, we separately discuss the two main components of the small-scale magnetic fields in galaxies \citep[see also][]{Jaffe2010,Haverkorn2015,Beck2016,BeckEA2019}.
\begin{itemize}
\item \textit{Anisotropic turbulent magnetic fields ($b_{\rm ani}$):} Despite the $180^\circ$ flips in the magnetic field direction on small scales (believed to be $\lesssim 100$\,pc), $b_{\rm ani}$ follows the same orientation on $\gtrsim$\,kpc scales. Thus, $b_{\rm ani}$ contributes significantly to the linearly polarised synchrotron emission seen in nearby galaxies \cite[e.g.][]{BeckEA2019,KierdorfEA2020}, and possibly also that in the Milky Way \citep[][]{Jansson2012b}\footnote{Note that the \cite{Jansson2012b} formulation assumes that $b_{\rm ani}$ (``striated field'' in their nomenclature) is proportional to $B_{\rm reg}$ in strength, with the two following the same orientation. Recent modelling works suggest that such striated field may not be required \citep{KorochkinEA25}. However, we point out that $b_{\rm ani}$ need not be directly related to $B_{\rm reg}$ in strength or orientation \cite[e.g.][]{BeckEA2019}.}. The plane-of-sky component of $b_{\rm ani}$ is traced by the intrinsic $B$-vector ($= E$-vector + $90^\circ$) of the linearly polarised electromagnetic waves, while the strength ($|b_{\rm ani}|$) can be estimated by considering both the polarisation fraction of the emission and the equipartition magnetic field strength, followed by subtracting the $B_{\rm reg}$ contribution.

\item \textit{Isotropic turbulent magnetic fields ($b_{\rm iso}$):} In typical observations that cannot spatially resolve $b_{\rm iso}$ (i.e.\ spatial resolution poorer than $\approx 100\,{\rm pc}$), the observed synchrotron emission will appear unpolarised due to the (wavelength-independent\footnote{We exclude here the frequency-dependent nature of the beam size of typical radio observations.}) beam depolarisation effect. This results from the intrinsic polarisation position angle of the synchrotron emission within the beam volume being isotropic, hence yielding a net zero observed polarisation \citep[see e.g.][Sec.\ 3.5]{Beck2016}. However, high-spatial-resolution SKA observations of the diffuse synchrotron emission of the Milky Way \citep[see Chapter by][]{Sun01.2026.SKA} can have the prospect of recovering the linearly polarised emission from $b_{\rm iso}$, thereby resolving its structure. Regarding the magnetic field strength ($|b_{\rm iso}|$), the \textit{total} intensity of the synchrotron emission can be used to estimate the \textit{total} magnetic field strength. This requires the assumption of energy equipartition between magnetic fields and cosmic ray particles \citep[with caveats; see][]{BeckK2005,SetaB2019,PonnadaEA2024}. Subsequently, with the subtraction of the contributions by the other magnetic components \citep[namely $b_{\rm ani}$ and $B_{\rm reg}$; e.g.][]{BeckEA2025}, the $|b_{\rm iso}|$ values can be obtained.
\end{itemize}

An intriguing aspect of synchrotron radiation relevant to the SKA is its circular polarisation (CP). Unlike linear polarisation, CP depends on all magnetic field components at the emission site \citep{PandyaEA2016}. It can also be generated through Faraday conversion, where linear polarisation is transformed into circular polarisation as the signal propagates through a magnetised plasma \citep[e.g.][]{MacquartM2000}. Detecting CP, however, is observationally challenging due to its typically very low fractional levels and the strong systematic biases that can obscure such weak signals, challenges that the SKA’s unprecedented sensitivity and polarimetric precision are well suited to address \citep{EnsslinEA2017}.

\subsubsection{Synchrotron Polarisation Statistics}
 The diffuse synchrotron polarisation observations of the ISM, Galactic Centre, and extended objects such as large supernova remnants at high resolution, facilitated by SKA, can also be utilised to infer MHD turbulence characteristics. Many of the established techniques have been driven by previous large-scale radio polarisation surveys, such as the Southern Galactic Plane Survey \citep[SGPS;][]{SGPS} and the Canadian Galactic Plane Survey \citep[CGPS;][]{TaylorEA2003,CGPS}. One approach is to apply the radio polarisation gradient technique to Stokes \textit{Q} and \textit{U} maps \citep{GaenslarEA2011,Burkhart2012}, which can potentially constrain several MHD characteristics (e.g.\ the sonic Mach number, magnetic field strength, and Reynolds number). Alternatively, one can perform power spectrum analysis to the Galactic diffuse polarised emission \citep{stutzEA2014}. Another approach was recently discussed in \citet{Malik2023}, based on the relative anisotropy of the global correlation functions constructed using the Stokes parameters as $I + Q$ and $I - Q$, referred to as the Y-parameter. To develop this approach, synthetic synchrotron maps from MHD turbulence data cubes having various plasma properties were used. This method is supported by the spatial separation of the Y-parameter values for decomposed Alfv\'enic (A-mode) and compressible (C-mode) MHD modes, which respectively decrease and increase with the inclination angle between the mean magnetic field and the line of sight, $\theta_\lambda$. A statistical demarcation at ${\rm Y_{turb}} \sim 1.5$ (with ${\rm Y_{turb}} > 1.5$ for A-mode and ${\rm Y_{turb}} < 1.5$ for C-mode) is used to identify the dominant MHD turbulence mode in a given region. Moreover, the unique sensitivity of the $\rm Y_{turb}$ parameter to $\theta_\lambda$ in different MHD mode–dominated regimes (see Fig.~4 in \citealt{Malik2023}) enables estimation of $\theta_\lambda$ for certain morphologies. The method has been further applied and validated in subsequent studies using the Effelsberg polarisation observations of the extended Monogem pulsar wind nebula and the Cygnus-loop SNR  \citep{Malik2024monogem, malik2026}, demonstrating its robustness. However, the original technique \citep{Malik2023} was developed under conditions of very weak Faraday rotation.

The arcsecond angular resolution of SKA AA4 at a Galactic Centre distance of $\sim 8.5$ kpc can resolve scales of about $\sim 0.04$ pc, well suited to the required high physical resolution (sub-pc to a few tens of pc) in this approach of the Y-parameter method. Furthermore, the compressible MHD turbulence mode has been identified as a major contributor to particle acceleration and scattering, leading to high-energy emissions in the GeV and TeV range, particularly the extended emission from pulsar wind nebulae and supernova remnants up to 50 pc in extent \citep{AbeysekaraEA2017, LiuEA2019}. With SKA's resolution and sensitivity, MHD turbulence within the Galactic plane, including the Galactic Centre, can be mapped extensively, which is crucial for understanding cosmic-ray scattering and acceleration in detail. 

\subsubsection{Faraday Rotation -- Rotation Measure} \label{sec:rm}

As linearly polarised emission propagates through a medium of ionised plasma with magnetic fields along the propagation direction, the emission experiences the Faraday rotation effect that leads to a rotation of the polarisation plane. Through multi-wavelength polarisation measurements, the Faraday rotation effect can be quantified by a parameter called the rotation measure (RM):
\begin{equation}
\Psi(\lambda^2) = \Psi(0) + 0.81 \int_L^0 n_e(s) B_\parallel (s)\,{\rm d}s \cdot \lambda^2 \equiv \Psi(0) + {\rm RM} \cdot \lambda^2{\rm ,}
\end{equation}
where $\Psi$ [rad] is the linear polarisation position angle at the electromagnetic-wavelength-squared $\lambda^2$ [m$^2$], $L$ [pc] is the physical distance from the observer to the synchrotron-emitting source along the line-of-sight $s$, $n_e$ [cm$^{-3}$] is the thermal electron number density, and $B_\parallel$ [$\mu$G] is the magnetic field component along $s$ \citep[e.g.~see][]{FerriereEA2021}. In galactic magnetism studies, RM is often obtained through radio observations of synchrotron-emitting sources such as galactic diffuse emission \citep[e.g.][]{VanEck2017}, pulsars \citep[e.g.][]{Sobey2021}, active galactic nuclei (AGN) and radio galaxies \citep[e.g.][]{MaEA2020}, and fast radio bursts \citep[FRBs; e.g.][]{Pandhi2024}. Subsequently, the magnetic field strength, scale, and structure in the intervening medium can be retrieved from the RM values, given ancillary information about the spatial distribution of the thermal electrons.

\begin{figure}[t]
\includegraphics[width=0.99\textwidth]{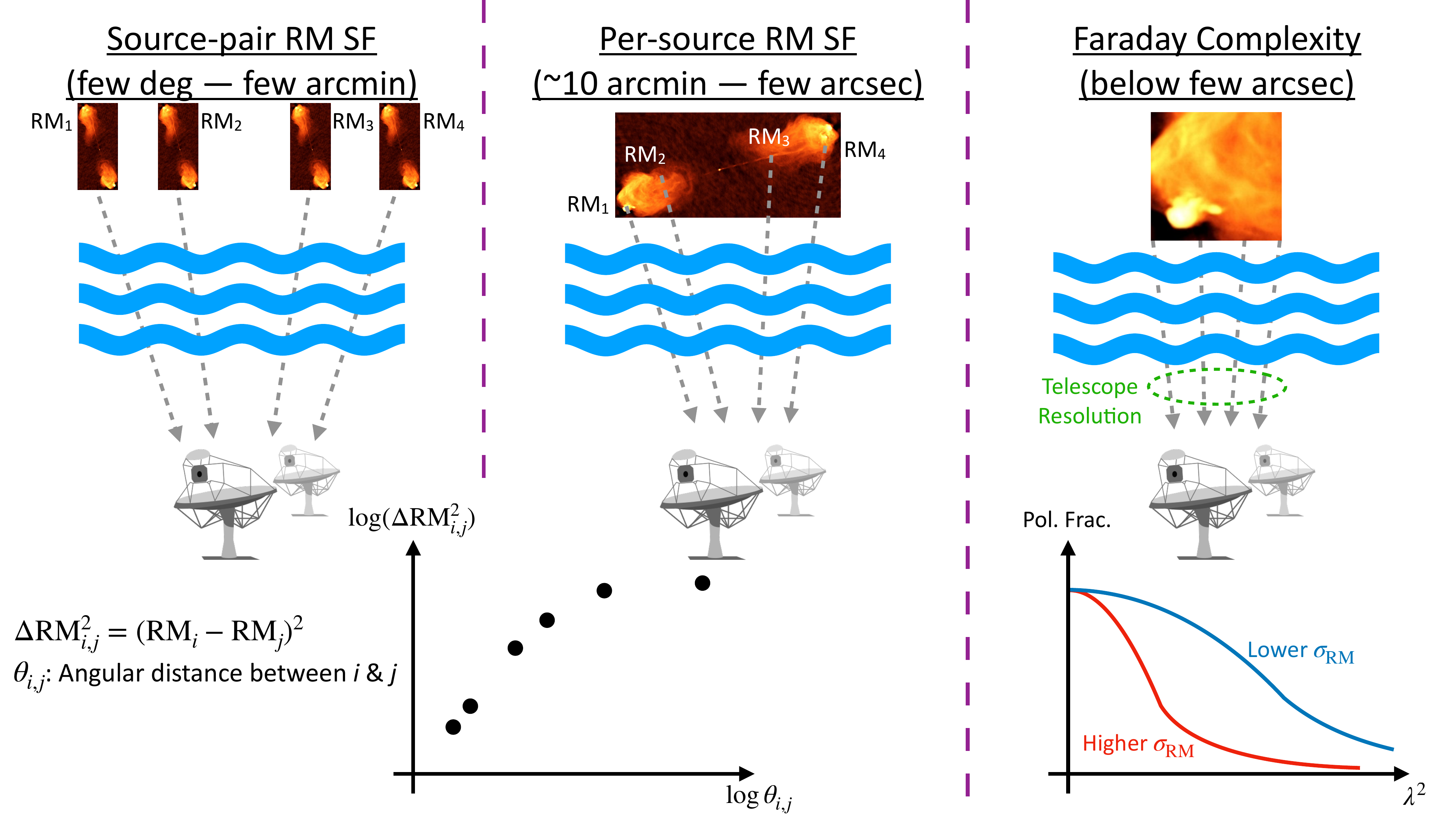}
\centering
\caption{Schematics illustrating the Faraday rotation analysis methods, namely rotation measure structure function (RM SF; Sec.~\ref{sec:rm}) and Faraday complexity (Sec.~\ref{sec:complex}), that can be deployed with the SKA for small-scale galactic magnetic field studies. The corresponding angular scale range that each method is sensitive to is included in the title. The blue wave patterns represent the intervening galactic ISM. The formula in the figure presents RM SF computation with two points and this could also be generalised to $n$-points, which enables testing convergence in statistical properties of the small-scale magnetic fields from observations \citep{SetaEA2023, Seta2024}. Image credits: the Australian Academy of Science (SKA-Mid dishes clip-art); NRAO/AUI (Cygnus A radio image).} \label{fig:methods}
\end{figure}

With a dense sampling of background source RM across the sky, one can form a so-called RM-grid \citep{Gaensler2004} that reveals the average magnetic fields in the intervening volume across spatial scales. The minimum scales probed are limited by the source density of the RM-grid. Given the presence of small-scale magnetic structures in the intervening galactic volume ($b_{\rm iso}$ and $b_{\rm ani}$), the RM values can be found to correspondingly exhibit spatial variations. Thus, the scale of magnetic fields in galaxies can be quantitatively studied by careful consideration of the spatial RM fluctuations. Specifically, the source-pair RM structure function (SF) analysis, as depicted in Figure~\ref{fig:methods} left column, is frequently utilised to statistically measure the spatial RM variations \citep[e.g.,][]{MinterS1996,StilEA2011,Seta2024}. With a polarised source density of $\approx 100\,{\rm deg}^{-2}$ that we can expect from SKA AA4 surveys, source-pair RM SF studies will be sensitive to magnetic structures from $\gg\,{\rm deg}$ to $\approx {\rm arcmin}$ scales (the corresponding physical scale in the system will depend on the distance).

Alternatively, RM fluctuations can be revealed by the RM maps of spatially resolved polarised emission \citep[e.g.][]{Haverkorn2004,MaoEA2015}. In principle, the (per-source) RM SF analysis (see Figure~\ref{fig:methods} middle column) can be applied to RM maps of background AGN and radio galaxies to reveal galactic magnetic structures at scales between those of the radio sources (up to $\approx 10$\,arcmin) and the angular resolution of the observations ($\approx\,{\rm arcsec}$ with the SKA AA4). However, this has not yet been attempted, as such studies require the combination of (1) high angular resolution ($\ll\,{\rm arcmin}$), (2) high sensitivity (translating to per-beam RM uncertainty of $\ll 10\,{\rm rad\,m}^{-2}$), (3) moderate sky-area coverage ($\gg\,{\rm deg}^2$; to gather a statistically meaningful sample), and (4) a sufficient control sample that do not experience spatial RM variations due to foreground galaxies (see below). This has just been enabled recently by the exquisite polarimetric data from the many on-going surveys conducted with SKA precursor telescopes, such as MMGPS \citep{MMGPS} with MeerKAT and SPICE-RACS \citep{ThomsonEA2023,ThomsonEA2026} and POSSUM \citep{GaenslerEA2025} with ASKAP, and will certainly warrant focus over the coming years in preparation for the SKA (see Sec.~\ref{sec:pre}). Specifically, the exact details of the method will need to be fine-tuned using the SKA precursor survey data. Firstly, it is known that the magneto-ionic medium within/surrounding the background sources can contribute to the observed spatial RM variations \citep[e.g.][]{MaEA2019a,BaidooEA2023}. This component of RM fluctuations will therefore need to be carefully removed in a statistical manner (see Sec.~\ref{sec:pre}), using information from a representative control sample (as mentioned above), to yield the galactic RM fluctuations that we are after. In addition, the RM SF from a single source is expected to be subjected to the stochastic nature of small-scale magnetic structures and therefore \textit{not} representative to the overall characteristics of $b_{\rm iso}$ or $b_{\rm ani}$. Hence, the per-source RM SF from a moderate number of sources that are closely separated in the sky will need to be combined in some ways to unveil the underlying statistical properties of the galactic small-scale magnetic fields.

Another challenge in the analysis of RM SF arises from the numerical methodology. Standard two-point calculation may not provide accurate results, particularly when small-scale fluctuations have a larger power spectral slope and higher-order stencils involving more than two points can be required \citep[see Sec. 2 in][]{SetaEA2023}. Such multi-point calculations also improve the separation of large- and small-scale magnetic field components \citep{Seta2024} and could also potentially help remove the contribution from the large-scale field structures and gradients. In addition, they enable convergence tests by verifying whether the inferred statistical properties of magnetic fields remain stable as the number of points in the computation increases \citep{SetaEA2023}.

\subsubsection{Faraday Rotation -- Wavelength-dependent Depolarisation (Faraday Complexity)} \label{sec:complex}

As discussed above in Sec.~\ref{sec:rm}, the galactic small-scale magnetic fields can lead to spatial variations of the observed RM. For cases where there are significant RM variations at angular scales smaller than that of the background sources or the angular resolution (whichever is smaller), the polarised emission from the background source through different lines-of-sight within the telescope beam can individually experience a random walk in Faraday rotation \citep{Burn1966,SokoloffEA1998}. Because of this, even if the polarisation planes of the emission are originally aligned among the different lines-of-sight, the eventual emission reaching the telescope can have misaligned polarisation planes due to the differing amount of Faraday rotation\footnote{This can also be regarded as a mix of linearly polarised emission experiencing different RM within the beam volume.}. This leads to the observed wavelength-dependent depolarisation effect, also commonly called Faraday complexity or Faraday depolarisation. By considering the observed linear polarisation signal across a broad frequency range using techniques such as RM-Synthesis \citep{BrentjensB2005} and Stokes \textit{QU}-fitting \citep{FarnsworthEA2011,OSullivanEA2012}, one can infer the magnitude of RM variations at spatial scales smaller than the telescope beam (see Figure~\ref{fig:methods} right column), enabling the statistical study of the intricate small-scale magnetic fields in the foreground volume.

Thus far, the technique of wavelength-dependent depolarisation has been applied to the study of the small-scale magnetic fields in the Milky Way \citep[via background sources:][and via the Galactic diffuse emission: \citealt{VanEckEA2019,ErcegEA2022}]{LivingstonEA2021,RanchodEA2024,MaEA2025}, as well as external galaxies near \citep[via diffuse emission of the target galaxies;][]{GaenslerEA2005,MaoEA2015,BasuEA2017} and far \citep[via background sources;][]{MaoEA2017,KovacsEA2025}. The key advancements that the SKA AA4 will bring are the broad instantaneous frequency coverage\footnote{With a broad coverage in $\lambda^2$, one can characterise Faraday complexity with higher accuracy with RM-Synthesis and Stokes \textit{QU}-fitting \citep[e.g.][]{BrentjensB2005,AndersonEA2016,KovacsEA2025,VanEckEA2026}.}, surface brightness sensitivity, and angular resolution. All these combined will enable accurate studies of magnetic structures that are at least an order of magnitude smaller than is feasible with current-generation instruments. While the $\approx\,{\rm arcsec}$ angular resolution of SKA AA4 will already set useful upper limits in the physical scale of the magnetic structures probed through Faraday complexity, we point out the advantages of stringent angular scale measurements (in the $\ll\,{\rm arcsec}$ regime) that can be made available by other radio studies. These include (but are not limited to) direct measurements from (wide-field) Very Long Baseline Interferometry (VLBI), as well as indirect inferences from interplanetary scintillation (IPS) studies \citep{MorganEA2018}. Similar to the case of per-source RM SF studies (above in Sec.~\ref{sec:rm}), a precise probe of the galactic small-scale magnetic fields through Faraday complexity analysis will first require adequate knowledge of the intrinsic properties of the Faraday complexity exhibited by the background sources themselves \citep{MaEA2025}.

Finally, we highlight the prospects of the extreme case of wavelength-dependent depolarisation, where many of the intrinsically polarised sources are completely depolarised. This has already been illustrated with the ASKAP POSSUM survey \citep{GaenslerEA2025} for the case of the Galactic mid-plane \citep{VanderwoudeEA2024} and for the Small Magellanic Cloud (SMC; Price et al., \textit{submitted}). In particular, the effect of Faraday depolarisation can be deduced from the decrease in the polarised source density. Such analysis method will require (1) sensitive observations (yielding a polarised source density of $\gg 10\,{\rm deg}^{-2}$ without depolarisation), (2) moderate sky-area coverage ($\gg 10\,{\rm deg}^2$), and (3) excellent knowledge of the polarised source population. All three are expected to improve with SKA.

\subsubsection{Zeeman Effect}

As summarised in a concurrent SKA Science Chapter \citep[see][]{Robishaw01.2026.SKA}, spectral line measurements of the Zeeman effect (in particular for H\textsc{i} and OH; both covered by the SKA-Mid Band 2) will be a powerful, direct \textit{in-situ} tracer of the line-of-sight component of magnetic fields in the Milky Way and nearby galaxies. The H\textsc{i} Zeeman effect can be realistically detected in both the diffuse emission and absorption towards bright background sources, while that in OH can be performed towards (mega-)masers. We reiterate the comment from both the previous \citep{RobishawEA2015} and the contemporary \citep{Robishaw01.2026.SKA} SKA Science Chapters that high precision wide-field correction in the instrumental circular polarisation will be pivotal to the success of future Zeeman experiments with the SKA-Mid.

\subsection{Special Considerations}

Observational probes of the small-scale magnetic field are different for the Milky Way and for other nearby galaxies. One of the most obvious differences is the physical scales concerned when using the diffuse synchrotron emission -- greater than a few 100s of pc for most nearby galaxy studies \citep[e.g.][]{KierdorfEA2020}, while in the Milky Way, we can be sensitive to $\lesssim {\rm pc}$ scales. Another key difference between the Milky Way and other galaxies is the line-of-sight depth of the field compared with the distance, applicable to both diffuse emission and RM-grid studies. Adjacent lines of sight are located on a cone with an opening angle equal to the angular distance between sky positions. For nearby galaxies, even in the Local Group, the intersection of that cone with the object is, to a good approximation, cylindrical, while for the Milky Way, the intersection is almost always a cone that probes smaller scales in the local ISM and larger scales in the more distant ISM.

\subsubsection{Milky Way}
The most significant probe of small-scale structure in the magnetic fields of the Milky Way by the SKA will be through Faraday rotation of a very large number of extragalactic sources (Sec.~\ref{sec:rm}), supplemented by that from a large number of pulsars with reliable distance estimates \citep[e.g.][]{SobeyEA2019}. The SKA will provide a giant step forward through a combination of factors. The smallest scale structures that can be probed by an RM-grid depend critically on the polarised source density. This is because of both the direct need for a fine spatial sample, and also because averaging across multiple sources is required to remove the stochastic nature of ISM turbulence (and the intrinsic RM contributions). With the large number of pulsars expected to be discovered with the SKA \citep[e.g.\ $\approx$ 9,400 with the SKA-Low;][]{XueEA2017}, their RM and dispersion measure (DM) combined, in addition to parallax distances, will allow for a detailed 3D probe of magnetic fields and thermal electron density in spiral arms and interarm regions \citep[see][and references therein]{Han2017,CurtinEA2024, DhakalS2025}.

\subsubsection{Nearby galaxies}
Thus far, the study of the magnetic fields in external galaxies via RM-grids is only feasible for the Magellanic Clouds \citep[][Price et al.\ \textit{submitted}, and references therein]{GaenslerEA2005,LivingstonEA2022,LivingstonEA2025}, M~31 \citep[][as well as the Local Group L-Band Survey: Kovacs et al.\ \textit{in prep.}]{HanEA1998}, and a handful of lensing galaxy systems \citep{MaoEA2017,KovacsEA2025}. The SKA will improve significantly on this, with the potential to reveal the small-scale magnetic field in the outer halo of many Local Group galaxies.

The energy injection scale in the interstellar medium, roughly 100 pc, corresponds to $4^{\prime\prime}$ at a distance of 5 Mpc. For galaxies outside the Local Group, they can impose significant wavelength-dependent (Faraday) depolarisation to the background polarised sources (Sec.~\ref{sec:complex}). High angular resolution is key to improving the contrast between the background source and the emission of the foreground galaxy. Probes of halo and/or circumgalactic medium magnetic fields in nearby galaxies are still possible, as demonstrated by the statistical analysis of \citet{HeesenEA2023}.

\subsection{Connecting SKA Observation with Numerical Simulations} \label{sec:radtrans}

To fully harness the diagnostic power of broadband polarimetry, parallel theoretical efforts are advancing. Recent developments \citep[in covariant polarised radiative transfer formalism;][see also Chapter of \citealt{Chan01.2026.SKA}]{Chan2019MNRAS, On2019MNRAS} have unlocked the possibility of direct interfacing with galactic or cosmological magneto-hydrodynamics (MHD) simulations to produce accurate predictions of the polarised sky. This forward-modelling framework can treat point sources and diffuse magnetised media either together or separately, and computes the full Stokes spectra and high-fidelity polarisation and RM maps, enabling realistic synthesis of the complex Faraday behaviour observed in cosmic media. Naturally, this can allow production of mock SKA observations from numerical simulation results. This integration of forward modelling, measurement modelling, and inverse inference establishes a self-consistent framework for decoding Faraday complexity, allowing emission and propagation effects to be disentangled within the magneto-ionic interstellar medium. By closely integrating both observations and simulations, this approach provides a reliable platform for discovery, and predictive exploration of magnetic field evolution and Faraday complexity across galactic environments and beyond, fully leveraging the capabilities of SKA.

\section{Scientific Goals} \label{sec:sci}
To understand the physics of the small-scale turbulent dynamo and the role of small-scale magnetic fields on a range of scales, especially in star formation and galaxy evolution (Sec.~\ref{sec:ques}), it is essential to examine their strength, scale, and structure within galaxies. In this Section, we summarise the current state of knowledge and highlight the open questions in each of these areas.

\subsection{Strength} \label{sec:stren}
From depolarisation observations and Faraday rotation measurements, it is estimated that the small-scale magnetic fields in the Milky Way and nearby galaxies are of the order of $5$ -- $20\,\muG$ \citep{GaenslerEA2005, MaoEA2008, Haverkorn2015, Beck2016, BeckEA2019, LivingstonEA2022, SetaEA2023, LivingstonEA2025}. The small-scale field strength is often comparable to, or even greater than, that of the large-scale field. Such a dynamically strong field can influence local gas flows \citep{PlanckXXXV_2016} and provide additional pressure that inhibits the collapse of gas into stars, thereby reducing the star formation efficiency. A key open question is what fraction of this field strength arises from amplification by the small-scale dynamo and what fraction results from the tangling of the large-scale field. Addressing this will constrain the physics of the small-scale dynamo and provide insights into the strength and properties of the turbulence responsible for tangling the large-scale field.

Another important prediction from small-scale turbulent dynamo theory and numerical simulations concerns the ratio of magnetic ($\Emag$) to turbulent kinetic ($\Ekin$) energy densities and its dependence on the ISM phase \citep[$\Emag/\Ekin \approx 0.1$ in the warm/hot phase and $\approx 0.01$ in the cold phase is inferred from numerical simulations, see][]{SetaF2022, GentEA2023}. These numerical experiments, and the associated estimates of the $\Emag/\Ekin$ ratio, primarily refer to small-scale magnetic fields amplified by the small-scale turbulent dynamo and may change when additional processes, such as the large-scale dynamo and the tangling of large-scale magnetic fields, are included. Observationally, based on pulsar and Zeeman measurements, it has recently been shown that the magnetic energy density scales with the turbulent kinetic energy \citep{SetaMCG2025}, although the precise fraction, $\Emag/\Ekin$, as a function of ISM phase \citep[and possibly also on the galactocentric radius, as shown for the total field in][]{Beck2007, Beck2015}, remains difficult to determine. Constraining this ratio would provide a direct measure of the efficiency of the small- and large-scale dynamo in different phases and strengthen our understanding of the ISM–magnetic field connection.

\subsection{Scale} \label{sec:scale}
The correlation scale of the small-scale ISM magnetic fields in the synchrotron-emitting and Faraday-rotating plasma is primarily estimated using two methods: $\RM$ structure functions and depolarisation models. For the Milky Way, this scale lies in the broad range $1$--$100\,\pc$ depending on location in the Galactic disk \citep{HaverkornEA2008}, although recently estimated to be within $20$--$30\,\pc$ using pulsars for the volume-filling, ionised ISM \citep{DhakalS2025}. For nearby irregular galaxies, it is larger, with values of $\approx 160\,\pc$ for the Small Magellanic Cloud and $\approx 90\,\pc$ for the Large Magellanic Cloud \citep{GaenslerEA2005, SetaEA2023}. For grand-design spiral galaxies, this has been estimated to be about 50\,pc \citep{FletcherEA2011}. In contrast to field strength, where the physics of converting turbulent kinetic energy into magnetic energy and the associated local magnetic field compression is relatively well understood, the description of magnetic field scales in terms of ISM properties remains missing. Establishing a predictive relationship between magnetic field scale and ISM properties would clarify the role of the ISM in shaping these scales and provide stronger constraints on cosmic-ray propagation (e.g.~see \Sec{sec:mol}).

The smallest magnetic field scales observable in the ISM must naturally be positioned close to the Sun. One of the nearby highly magnetised structures predominantly lies along the walls of the Local Bubble \citep{AlvesEA2018,ONeillEA2025}. These regions are shaped by recent star formation activity and the cumulative effects of stellar winds and supernova explosions, which have carved out and compressed the surrounding medium. Prominent examples include H \textsc{ii} regions such as Sh2-27 \citep{RaychevaEA2022}, star-forming molecular clouds like the Orion complex, and various supernova remnants that trace small-scale, turbulent magnetic fields on sub-parsec scales.

All theories of MHD turbulence and the small-scale dynamo predict a power-law magnetic field power spectrum \citep{Schekochihin2022}, with a slope that depends on the specific theory. Reliably determining this slope from radio observations, potentially using multi-point $\RM$ structure functions with higher-order stencils \citep{SetaEA2023, Seta2024}, would provide further insight into the plasma physics of the ISM.

\subsection{Structure} \label{sec:struc}
Based on MHD turbulence and small-scale dynamo theories, the small-scale magnetic fields in galaxies are expected to be random but non-Gaussian, with strong fields concentrated in filamentary or sheet-like structures \citep{ZeldovichEA1987, Subramanian1998, SchekochihinEA02}. This behaviour is seen in a variety of simulations \citep{SchekochihinEA2004, HaugenEA2004, SetaEA2020, SetaF2021dyn, Basu2021, SurS2024}, and some ISM observations also reveal filamentary structures \citep[e.g.,~][]{ZaroubiEA2015, ErcegEA2022, HeywoodEA2022, Yusef-ZadehEA2022}. However, the lack of finely spaced data has so far prevented a detailed observational study of magnetic field structure. Characterising the morphology of these structures using topological methods \citep[e.g.,~Minkowski functions, Betti numbers, and persistence diagrams, see][]{WilkinBS2007, MakarenkoEA2018II, MakarenkoEA2018I, SetaEA2020, DwivediEA2024, DuttaEA2024} would not only constrain theoretical parameters but also allow us to better quantify the local impact of magnetic fields on gas flows and star formation.

Moreover, a precise understanding of the structure of the magneto-ionic medium of the Milky Way, even in a statistical sense, is crucial for effective foreground determination, particularly in the context of Faraday rotation. Across angular scales ranging from degrees (comparable to the apparent size of nearby galaxy clusters) down to the effective point-source level, Galactic structures can easily be mistaken for larger-scale RMs \citep[such as from galaxy groups or clusters, see][for a discussion]{AndersonEA2024}, and vice versa. Furthermore, the anisotropy of the magnetic field structure, which may be linked to gaseous motions in the ISM, ought to be further understood and its origins clarified.

For studies of extragalactic structures, this issue is commonly mitigated by focusing on high Galactic latitudes and employing large-scale models to subtract the Milky Way contribution. Conversely, the inference of the Galactic RM has traditionally relied on the assumption that extragalactic RMs are uncorrelated and point-like in nature \citep{HutschenreuterEA2022}. However, at the angular resolutions achievable with SKA (both Low and Mid), and given our current understanding of Milky Way structure, neither assumption is likely to remain valid to the required precision. Thus, robust methods for characterising magnetic structures both in the source and foreground and their standardisation with MHD simulations are necessary to fully exploit the upcoming SKA data.

The quality of foreground removal depends sensitively on both the source density and analysis methods employed \citep[e.g.~the interpolation kernel, as discussed in][]{KhadirEA2024}. Notably, the structure-function analysis, discussed in \Sec{sec:rm}, is also directly connected to the underlying correlation function of foreground structures, which in turn determines the interpolation kernel for this purpose.

\section{Highlights of Specific Scientific Questions} \label{sec:ques}

In this Section, we highlight a few key outstanding scientific questions in the field of small-scale magnetic fields in galaxies. Then in \Sec{sec:largescale}, we will describe how future SKA observations will significantly advance our understanding in these areas.

\subsection{Magnetic Fields in Molecular Clouds and \textit{in-situ} Synchrotron Emission} \label{sec:mol}
One of the most powerful probes of magnetic fields in the ISM is synchrotron emission (see \Sec{sec:syn}). However, no detectable synchrotron emission has been observed from the molecular cloud cores \citep[see recent work regarding possible synchrotron emission from the boundary of molecular gas in][]{BraccoEA2023}, even though Zeeman effect observations show strong magnetic fields \citep[$\gtrsim 10\,\muG$, e.g.,][]{CrutcherEA2010}. This raises further questions about the strength and geometry of magnetic fields, the sites of star formation, and the propagation of low-energy cosmic ray electrons in molecular clouds.

The most recent data from the \textit{Voyager} probes \citep{Cummings+2016,Stone+2019}, together with detections provided by the \textit{Fermi Large Area Telescope} (\textit{Fermi-LAT}; \citealt{Ackermann+2010}), the balloon-borne Pamela experiment \citep{Adriani+2011}, and the Alpha Magnetic Spectrometer (AMS-02) on board the International Space Station \citep{Aguilar+2014}, constrain the local cosmic-ray electron flux down to energies relevant for synchrotron emission.

The average magnetic field strength in molecular clouds could be weaker than that shown by Zeeman measurements, which primarily trace the dense regions. Ambipolar diffusion might be the reason for the average weaker fields \citep{Tritsis2025}. A more significant issue, however, is the penetration of low-energy (primarily $\GeV$) cosmic ray electrons \citep{DickinsonEA2015}. The ISM is filled with low-energy cosmic ray electrons, and magnetic field lines connecting the relatively warm ISM around the molecular clouds to the cold gas within the molecular clouds should allow them to enter. There are, however, two related issues. First, cosmic ray electrons lose energy rapidly, with the loss rate proportional to the square of the magnetic field strength, which remains poorly constrained in and around molecular clouds. Second, the magnetic field structure at cloud boundaries may create particle traps \citep{CesarskyV1978, Chandran2000, SetaEA2018, SilsbeeEA2018}, further enhancing energy losses through increased confinement times. Both effects limit cosmic ray electron penetration. In that case, synchrotron emission would originate mainly from the boundary or outer regions of molecular clouds, but probably at levels too low to be detected with current facilities \citep{WollebenR2004, BraccoEA2025}. With the superior sensitivity of the SKA-Low, \textit{in-situ} synchrotron emission from molecular clouds, likely from their envelopes, will likely become detectable (Sec.~\ref{sec:large_mc}). Such observations will constrain the strength and geometry of magnetic fields in and around molecular clouds and shed light on the propagation of cosmic ray electrons.

Understanding how cosmic rays propagate through the ISM and dense clouds offers valuable insight into their distribution, and therefore, their coupling with the gas and potential feedback effects. If free-streaming penetration of electrons in molecular clouds holds \citep{Takayanagi1973,Padovani+2009}, the relativistic part of the cosmic-ray electron spectrum turns out to be unaffected during the propagation up to H$_2$ column densities of approximately $10^{23}$~cm$^{-2}$, which are characteristic of the densest regions of molecular clouds \citep{Padovani+2018a}. In this case, \cite{Padovani+2018c} showed that for typical expected magnetic field strength in starless cores ($<10^3\,\muG$), the detection of synchrotron radiation depends on the integrated line-of-sight value of the magnetic field strength, which, in turn, is determined by the density profile. As a consequence, synchrotron flux should be larger for a dense core with a shallow density profile compared to a core with a peaked density profile. For example, SKA should detect synchrotron emission at frequencies of $\sim200$ MHz reaching a signal-to-noise up to $\sim7$ and $\sim2$ in one hour of integration for starless cores such as FeSt 1-457 in the Pipe Nebula and Barnard 68, respectively. Low-frequency SKA observations, combined with other innovative methods for estimating magnetic field strength, will therefore provide useful information on the propagation of cosmic rays in molecular clouds, clarifying whether free-streaming is a good approximation for cosmic-ray transport. If not, it would imply that cosmic-ray motion is slowed down by scattering off magnetic field fluctuations on scales comparable to their gyroradius.

One final important aspect to consider is that the description of synchrotron emission is usually made assuming a single energy slope for the CR electron spectrum. However, \textit{Voyager} data indicate that the local Galactic cosmic-ray electron spectrum bends at low energies. If this is the case, the standard frequency-magnetic field-energy relation \citep[see e.g.][]{Longair2011} does not hold anymore and there is no unique correspondence between energy slope and  spectral index slope. Thanks to SKA's high frequency resolution, it will be possible to compare independent estimates of the spectral index obtained over a series of narrow frequency intervals, and then trace back the variation in the energy slope of the cosmic-ray electron spectrum
\citep{Padovani+2021b, Linzer+2025}.

\subsection{The Interplay between Small-scale Magnetic Fields and Star Formation Processes}

It has been well established that turbulence and magnetic fields can regulate the star formation rate and influence the stellar initial mass function \citep{Federrath2012,Krumholz2019,MathewEA2023,PattleEA2023}. While the effects of the molecular cloud magnetic fields have been explored observationally \citep[via dust and starlight polarisation studies;][]{PlanckXXXV_2016,LiEA2017}, the roles of the magnetic fields throughout the baryon cycle, especially the cooling and condensation of the warm/hot ISM phases to the cold gas phases, remain poorly understood. Although measurements of the magnetic field strength across ISM phases exist \citep[e.g.][]{CrutcherEA2010,Harvey-SmithEA2011,ThomsonEA2018,SetaMCG2025}, it is crucial to obtain observationally derived information on the magnetic field strength, scale, and structure \emph{combined} across all ISM phases (Sec.~\ref{sec:sci}). This will be the key to understanding how magnetic fields co-evolve with galaxies, addressing one of the central questions in modern astrophysics. This is particularly the case for the small-scale ($b_{\rm iso}$ and $b_{\rm ani}$) components of the magnetic fields (Sec.~\ref{sec:intro}), given the relevance in spatial domain (at $\lesssim 100\,{\rm pc}$), as well as their dominance in strength over the large-scale coherent magnetic field \citep{Beck2016}.

Further, the exact astrophysical origin of the radio-infrared correlation in star-forming galaxies remain obscured \citep[see][]{BeckWielebinski2013,Beck2016}. While it is understood that both emission are linked to star-formation activities, the details for the radio (synchrotron) emission are unclear. Specifically, the total (dominated by $b_{\rm iso}$ and $b_{\rm ani}$) magnetic field strength, which is responsible for the observed synchrotron intensity, can either be amplified by shocks in star-formation feedback processes, or by compression (due to gravity) during the stage of cloud collapse. This can be answered by future high angular resolution, high sensitivity observations of Galactic and extragalactic star-forming regions.

While we anticipate that the on-going SKA precursor polarisation surveys \citep{MMGPS,GaenslerEA2025} can begin to address the above in the coming few years, it will almost certainly lead to discoveries that warrant follow-up studies. Compared to the SKA precursor telescopes, the SKA will offer greatly enhanced sensitivity and angular resolution, allow us to explore the galactic magnetic fields at much smaller spatial scales and at much higher accuracy than is currently feasible (Sec.~\ref{sec:SKA}).

\subsection{Magnetic Fields in Elliptical Galaxies} \label{sec:ell}

Most studies of magnetic fields focus on star-forming galaxies \citep{Beck2016}, but it is also important to examine magnetic fields in ellipticals that, at most, are forming stars near their centres at low rates. Since ellipticals lack significant differential rotation, the conventional $\alpha$-$\Omega$ large-scale dynamo does not operate, and the small-scale field produced by the tangling of a large-scale field is negligible. This makes ellipticals a more pristine environment for probing the properties of small-scale dynamo \citep{MossS1996, SetaEA2021}.

Using the Laing-Garrington effect \citep{Laing1988, Garrington1988} for ellipticals with an active host and a statistical study of intervening elliptical galaxies, it is inferred that the typical magnetic fields in ellipticals ($\lesssim 1\,\muG$) are about an order of magnitude weaker than those in star-forming disk galaxies \citep{SetaEA2021, ShahS2021}. Alternative methods are hindered by various limitations, such as the detection of the diffuse synchrotron emission from the elliptical galaxies can be contaminated by the central sources (namely, AGN) and can be challenging given the sparse cosmic ray electrons, while the use of physically unrelated background source RM is limited by the insufficient RM-grid density. The expected synchrotron emission from the ISM of ellipticals at $\approx 100\,\MHz$ is only $\approx 25\,\nJy$ \citep[see Sec.~5.3 in][]{SetaEA2021}, which will remain challenging for a direct detection even with SKA. Thus, the most effective probe of magnetic fields in ellipticals will be the high $\RM$ source density expected with SKA (Sec.~\ref{sec:ell2}).

\subsection{Magnetic Fields in Intermediate-redshift Galaxies} \label{sec:red}

To constrain the parameters of the dynamo theories (both small-scale, fluctuation dynamo and large-scale, $\alpha$-$\Omega$ dynamo), it is important to observe magnetic fields in galaxies across cosmic time. This can be done by measuring the magnetic field growth rate and scale as a function of redshift. Recent efforts studying gravitationally lensing galaxies \citep{MaoEA2017,KovacsEA2025} have captured both the large- and small-scale magnetic fields in the foreground lensing galaxies, highlighting the prospects of the innovative technique. We refer the readers to the \cite{Mao01.2026.SKA} Chapter for discussions on the large-scale magnetic fields and focus on the small-scale magnetic fields here (see also Sec.~\ref{sec:large_intermediatez}). With the SKA, we shall greatly expand the number of sampled lensing galaxy systems, and thereby explore the properties of the small-scale galactic magnetic fields across redshift and various galactic parameters (e.g.\ star formation rate and stellar mass).

\section{Preparation for and Plans with the SKA} \label{sec:ska}

\subsection{Preparation with SKA Precursors} \label{sec:pre}

From early-mid 2020s, high quality polarimetric data from SKA precursor surveys, such as MMGPS with MeerKAT \citep{MMGPS}, POSSUM with ASKAP \citep{GaenslerEA2025}, and LoTSS with LOFAR \citep{OsullivanEA2023}, have been made available in abundance. These data will be crucial in the lead up to the SKA AA* and AA4 for the study of small-scale magnetic fields in galaxies, as well as cosmic magnetism in general, as we describe in more details below.

The anticipated advent of the SKA will bring a multitude of advancements in the context of the study of the galactic small-scale magnetic fields with respect to the current state-of-the-art (Sec.~\ref{sec:obs} and \ref{sec:SKA}). However, the SKA's full potential can only be realised given adequate preparations in both analysis algorithm development and our astrophysical knowledge in galactic magnetism. While the source-pair RM SF analysis is already sufficiently mature \citep[e.g.][]{SetaEA2023}, the per-source RM SF method (Sec.~\ref{sec:rm}) that will allow us to explore the sub-arcmin regime will still require significant progress. This includes both the removal of the RM SF contributions by the emitting AGN and radio galaxies, and the specifics of averaging across multiple sources to uncover the underlying statistical signal. Similarly, the Faraday complexity (wavelength-dependent depolarisation) technique (Sec.~\ref{sec:complex}) will require further development in the coming years. While the general analysis framework has already been set by recent works \citep{LivingstonEA2021,RanchodEA2024,MaEA2025}, the need for an accurate removal of the intrinsic Faraday complexity has been recognised.

Finally, as the study of small-scale galactic magnetic fields will likely see the most significant advancements from deep, small-sky-area SKA surveys rather than shallow, wide-area studies (Sec.~\ref{sec:SKA}), we ought to make use of the on-going wide-area SKA precursor surveys to identify (1) interesting Galactic regions, and (2) specific galaxy types \citep[by, e.g., stellar mass, star formation rate, and redshift;  also, see Chapter by][]{Mao01.2026.SKA}, for future SKA AA* and AA4 studies.

\afterpage{
\begin{landscape}
\begin{center}
\begin{table}[t]
\caption{Expected linear polarisation performance of SKA-Low and SKA-Mid AA4}
\begin{tabular}{c|c|cccccc}
& SKA-Low & \multicolumn{6}{c}{SKA-Mid}\\
& & Band 1 & Band 2 & Band 3 & Band 4 & Band 5a & Band 5b \\
\cline{1-8}
Frequency (MHz) & 50--350 & 350--1050 & 950--1760 & 1650--3050 & 2800--5180 & 4600--8500 & 8300--15400 \\
RMTF FWHM (${\rm rad\,m}^{-2}$)$^a$ & 0.1 & 4.8 & 51 & 140 & 420 & 1200 & 3700 \\
RM uncertainty (${\rm rad\,m}^{-2}$)$^{a,b}$ & 0.008 & 0.4 & 4 & 12 & 35 & 95 & 311 \\
RM max-scale (${\rm rad\,m}^{-2}$)$^a$ & 4 & 38 & 108 & 325 & 937 & 2522 & 8278 \\
Angular resolution (arcsec)$^c$ & 9.1 & 1.3 & 0.8 & 0.4 & 0.2 & 0.1 & 0.08 \\
Image rms noise (1\,hr; $\mu{\rm Jy/beam}$)$^c$ & 7.0 & 2.3 & 1.1 & ? & ? & 0.7 & 0.8 \\
Pol src density (1 hr; ${\rm deg}^{-2}$)$^{d,e}$ & 1--10? & 80--130 & 100--130 & ? & ? & 70--90 & 50--60 \\
\cline{1-8}
\multicolumn{8}{l}{$^a$: Following Eq.~61 and 62 of \cite{BrentjensB2005}.}\\
\multicolumn{8}{l}{$^b$: Evaluated at a linear polarisation signal-to-noise ratio of 6.}\\
\multicolumn{8}{l}{$^c$: Using Briggs weighting with \texttt{robust} $=0$, without \textit{uv}-tapering.}\\
\multicolumn{8}{l}{$^d$: Estimated using \cite{RudnickOwen2014}, extrapolated across frequency assuming a spectral index of $-0.7$.}\\
\multicolumn{8}{l}{$^e$: See \href{https://github.com/jackieykma/SKA-pol-density}{https://github.com/jackieykma/SKA-pol-density}, for a script estimating the polarised source density.}
\end{tabular} \label{table:pol_performance}
\end{table}
\end{center}
\end{landscape}
}

\subsection{Considerations for SKA Observations} \label{sec:SKA}

Here, we discuss the specific types of SKA observations that can benefit the study of small-scale galactic magnetic fields, in order to facilitate future commensal observations. The approximate linear polarisation performance of the SKA AA4 across different frequency bands is summarised in Table~\ref{table:pol_performance}. Below in Sec.~\ref{sec:largescale}, we will outline potential large-scale SKA survey projects primarily driven by our particular science topic. 

\subsubsection{SKA-Low (50--350\,MHz)}

The most exciting aspect of SKA-Low will be the study of the local Galactic ISM through its diffuse polarised synchrotron emission \citep{VanEckEA2019, ErcegEA2022}. The $\approx 9\,{\rm arcsec}$ ($\approx 4 \times 10^{-3}\,{\rm pc}$ at $100\,{\rm pc}$) resolution will reveal the highly intricate magnetic structures in the Solar neighbourhood and enable joint analysis with other tracers of the (magnetised) ISM such as dust polarisation \citep[e.g.][]{Planck2016fil}, starlight polarisation \citep[e.g.][]{AngaritaEA2024,PanopoulouEA2025}, and H\textsc{i} structures \citep[e.g.][]{BraccoEA2020,MaEA2023}.

Meanwhile, we expect that RM SF studies will see limited progress with the SKA-Low, given the scarce polarised source density \citep[about $1$--$10\,{\rm deg}^{-2}$; see Chapter by][]{OSullivan01.2026.SKA}\footnote{This number is highly uncertain given our current limited knowledge in the polarised source population at such low radio frequencies. With LOFAR, the LoTSS RM catalogue DR2 \citep{OsullivanEA2023} has achieved a polarised source density of $0.43\,{\rm deg}^{-2}$ at 144\,MHz with a sensitivity of $83\,\mu{\rm Jy/beam}$; while a stacking study with the LoTSS data has provided $1.24\,{\rm deg}^{-2}$ with a sensitivity of $19\,\mu{\rm Jy/beam}$ \citep{PirasEA2024}.}. Although the SKA-Low will provide extremely precise RM values (measurement uncertainty $\lesssim 8 \times 10^{-3}\,{\rm rad\,m}^{-2}$), this key advantage will be less relevant for our case here given the RM uncertainty contributed by the intrinsic RM scatter of the extragalactic sources \citep[$\approx 6\,{\rm rad\,m}^{-2}$;][]{Schnitzeler2010}. Though, we recognise that pulsars are believed to have minimal intrinsic RM \citep{WangEA2011} and Faraday complexity \citep{SobeyEA2019}. This makes the superior RM precision afforded by the SKA-Low to be highly advantageous for pulsar RM studies, which in turn can provide significant insight into the small-scale Galactic magnetic fields \citep[using and advancing the methods such as in][]{DhakalS2025}. For example, year/decade-long monitoring of pulsars with high proper motion can be used to trace the $\ll 0.01\,{\rm pc}$-scale magneto-ionic medium, given that the ionospheric RM is properly accounted for \citep{PoraykoEA2019}. 

Finally, we identify the prospects of SKA-Low Faraday depolarisation studies. With the high RM precision and low maximum Faraday depth extent ($\approx 4\,{\rm rad\,m}^{-2}$), SKA-Low observations of polarised extragalactic sources will be incredibly sensitive to the presence of weak small-scale magnetic fields that would otherwise be undetectable with higher frequency instruments (e.g.\ ASKAP, MeerKAT, \& SKA-Mid). This is demonstrated in Figure~\ref{fig:depol}, from which one can see that $\sigma_{\rm RM} \lesssim 1\,{\rm rad\,m}^{-2}$ can only be detected with high confidence the SKA-Low. 

Overall, we consider a wide-area SKA-Low polarisation survey with a uniform sensitivity to be highly relevant to our science goal, provided that various undesirable effects such as ionospheric RM contributions and wide-field instrumental polarisation are carefully removed from the science-ready data.

\subsubsection{SKA-Mid Bands 1 (350--1050\,MHz) \& 2 (950--1760\,MHz)}

These two frequency bands are likely the most relevant for the study of the small-scale galactic magnetic fields. We focus on the prospects of RM SF and Faraday complexity studies here, and refer the readers to Chapters by \cite{Sun01.2026.SKA} and \cite{Mao01.2026.SKA} for potential polarised diffuse emission works for the Milky Way and external galaxies, respectively. We further refer readers to Chapter by \cite{Robishaw01.2026.SKA} for details on direct magnetic field strength measurements with the Zeeman effect (noting that SKA-Mid Band 2 will cover both the H\textsc{i} and OH spectral lines).

\begin{figure}[t]
\includegraphics[width=0.99\textwidth]{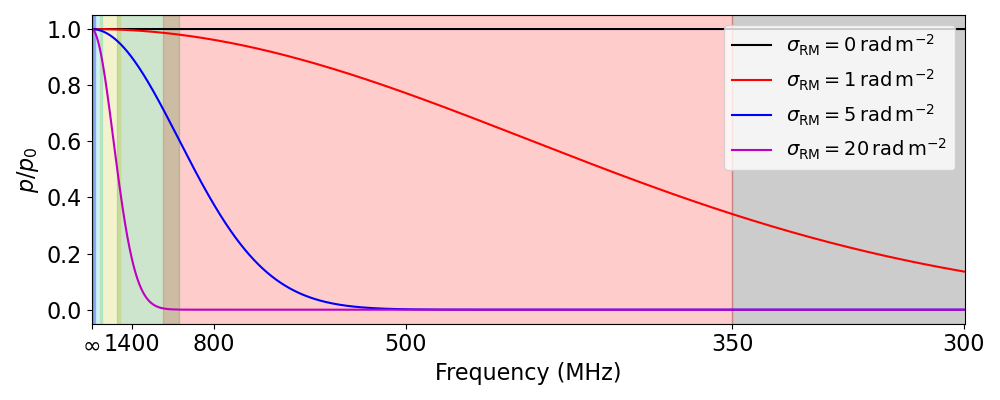}
\centering
\caption{The effect of external Faraday dispersion \citep[which exhibits $p \propto \exp(-2\sigma_{\rm RM}^2 \lambda^4)$;][]{SokoloffEA1998} across the SKA frequency bands, with the solid coloured lines showing the results of different $\sigma_{\rm RM}$ values. The $y$-axis represents the ratio of observed-to-intrinsic polarisation fraction (1 for no depolarisation; 0 for complete depolarisation), and the $x$-axis is linear in $\lambda^2$ domain. The shaded areas represent (from right to left) SKA-Low (black; 50--350\,MHz), SKA-Mid Band 1 (red; 350--1050\,MHz), SKA-Mid Band 2 (green; 950--1760\,MHz), SKA-Mid Band 3 (yellow; 1650--3050\,MHz), SKA-Mid Band 4 (cyan; 2800--5180\,MHz), and SKA-Mid Band 5a (blue; 4600--8500\,MHz). SKA-Low is truncated at 300\,MHz, and SKA-Mid 5b is omitted in this illustration.} \label{fig:depol}
\end{figure}

\begin{figure}[t]
\includegraphics[width=0.99\textwidth]{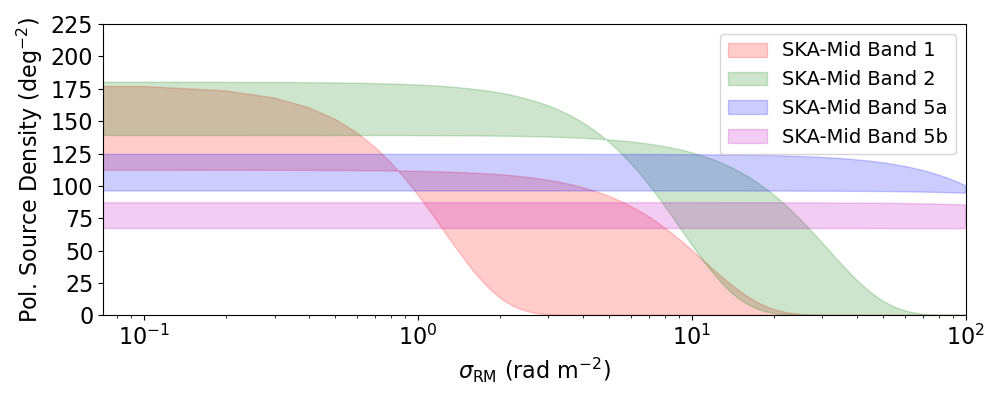}
\centering
\caption{The number density of linearly polarised source that we expect to detect with a fiducial 1\,hr SKA-Mid AA4 observation, as a function of the spatial RM fluctuation ($\sigma_{\rm RM}$). The depolarisation due to $\sigma_{\rm RM}$ follows $\exp(-2\sigma_{\rm RM}^2\lambda^4)$ \citep[external Faraday dispersion;][]{SokoloffEA1998}. For any given frequency band and $\sigma_{\rm RM}$ value, the span in $y$-value captures two frequency-dependent effects in the band -- radio spectral index and Faraday depolarisation.} \label{fig:srcdensity}
\end{figure}

Both of these frequency bands will provide excellent RM precisions ($\lesssim 0.4$ and $\lesssim 4\,{\rm rad\,m}^{-2}$ for Bands 1 and 2, respectively), sensitivity to Faraday complexity (with maximum Faraday depth extents of $38$ and $108\,{\rm rad\,m}^{-2}$, respectively), and similar expected polarised source density (for a 1\,hr observation, Bands 1 and 2 will likely yield $80$--$130\,{\rm deg}^{-2}$ and $100$--$130\,{\rm deg}^{-2}$, respectively). Given that Band 1 will be more sensitive to small RM fluctuations ($\sim 1\,{\rm rad\,m}^{-2}$; see Figure~\ref{fig:depol}), it will be more suited for studies of sky regions with minimal RM fluctuations from the small-scale galactic magnetic fields (intermediate Galactic latitudes of beyond $\pm 10^\circ$ for the Milky Way; halos or circumgalactic medium for external galaxies). Meanwhile, Band 2 will be more robust against complete Faraday depolarisation (see Figure~\ref{fig:srcdensity}) and therefore will be the better choice for studies of sky areas with modest RM fluctuations ($\sim 10\,{\rm rad\,m}^{-2}$) such as the Galactic plane (within $\pm 10^\circ$ in latitude) as well as the galactic disks of nearby galaxies.

We further highlight the prospects of polarimetric studies of scintillating pulsars. Specifically, time-resolved measurements of Faraday rotation (both RM and Faraday complexity) can provide information on the small-scale magnetic fields on $\ll 1000\,{\rm AU}$ scale. This will require the development of analysis methods analogous to the power spectrum of the pulsar dynamic spectrum \citep[the ``secondary spectrum''; e.g.][]{SprengerEA2022}. We expect that observations with the SKA-Mid Bands 1 or 2 (or Band 3; for pulsars in extreme environments such as the Galactic Centre or along spiral arm tangents) will be the most useful.

\subsubsection{SKA-Mid Bands 3 (1650--3050\,MHz) \& 4 (2800--5180\,MHz)}

These two SKA-Mid bands, which will be made available contingent on funding, are crucial for studying galactic areas with high levels of RM fluctuations ($\sim 50\,{\rm rad\,m}^{-2}$). This conclusion comes from their linear polarisation performance, with RM uncertainties of $\lesssim 12\,{\rm rad\,m}^{-2}$ (Band 3) and $\lesssim 35\,{\rm rad\,m}^{-2}$ (Band 4), as well as the maximum Faraday depth extents of $325\,{\rm rad\,m}^{-2}$ (Band 3) and $937\,{\rm rad\,m}^{-2}$ (Band 4). We note that, as at the time of writing, the exact sensitivities of the receivers remain unknown, but we assume that (similar to all other SKA-Mid frequency bands) a polarised source density of $\approx 100\,{\rm deg}^{-2}$ can be reasonably achieved with a 1\,hr observation. As mentioned above, Bands 3 and 4 observations will be especially relevant for in-depth studies of extreme Galactic regions, such as the Galactic mid-plane (within $\pm 1^\circ$ in latitude), the Galactic Centre region, and ionised Galactic structures resulting from stellar feedback processes.

\subsubsection{SKA-Mid Bands 5a (4600--8500\,MHz) \& 5b (8300--15400\,MHz)}

Finally, these two bands at the high end of the SKA spectrum can lead us to discover the unknowns~--- regions with extreme small-scale galactic magnetic fields that exhibit as small-scale RM fluctuations of $\gg 100\,{\rm rad\,m}^{-2}$ on $\lesssim$\,arcmin scale. Such potential breakthroughs can be enabled by, for example, a Band 5a/5b Galactic plane survey in full polarisation.
\subsubsection{SKA-Mid Survey for Small-scale Galactic Magnetic Fields --- Wide or Deep?}

Summarising the above information, and considering the performance of the on-going wide-area polarisation surveys with SKA-Mid precursors that will yield polarised source density of $\approx 30\,{\rm deg}^{-2}$ \citep{MMGPS,GaenslerEA2025}, we believe that shallow ($\approx 1\,{\rm hr}$ per pointing), wide-area SKA-Mid surveys may not represent the generational leap that will be required to enable transformational discoveries in cosmic magnetism studies\footnote{One exception to this being a wide-area SKA-Mid Band 1 survey, as this frequency range can be highly potent for Faraday complexity studies (see above), yet poorly explored beyond the MMGPS-UHF survey. While the Band 5a/5b frequency range is similarly under-utilised, the low survey speed can make a wide-area survey prohibitive.}. Therefore, we consider that deep, small-area SKA-Mid surveys will be the most promising for advancing our knowledge of the small-scale magnetic fields in galaxies. We point out that to bring an order-of-magnitude increase in the polarised source density with respect to the current-generation surveys (i.e.\ to $\gtrsim 300\,{\rm deg}^{-2}$), we will require SKA-Mid AA4 observations with per-pointing integration time (target sensitivity) of 20--80\,hr (0.26--0.52\,$\mu{\rm Jy/beam}$) for Band 1, 20--40\,hr (0.18--0.26\,$\mu{\rm Jy/beam}$) for Band 2, 60--140\,hr (0.06--0.09\,$\mu{\rm Jy/beam}$) for Band 5a, and 200--450\,hr (0.04--0.06\,$\mu{\rm Jy/beam}$) for Band 5b.

\subsection{Potential Large-scale SKA Programmes} \label{sec:largescale}

\subsubsection{Magnetic Fields in Molecular Clouds} \label{sec:large_mc}

The detection of magnetic fields in diffuse molecular clouds across multiple radio tracers (namely, Zeeman effect, synchrotron emission, and Faraday rotation) \emph{combined} will certainly shed light on magnetism's roles in ISM phase transitions and star formation (see also \citealt{Tahani01.2026.SKA} Chapter). In particular, studies of nearby (distance $\approx 100\,{\rm pc}$), off-plane ($|b| > 10^\circ$) systems in the southern sky such as the Chamaeleon and the Corona Australis will likely be the most fruitful. This is because of the combined advantages from (i) high resolution in physical scales, (ii) high background polarised source count, (iii) minimal confusion with signals from the Galactic mid-plane, and (iv) availability of 3D density maps in the Solar neighbourhood \citep{EdenhoferEA2024, SodingEA2025}. The typical angular extent of such nearby molecular complexes is $\sim 10\,{\rm deg}^2$, meaning that deep observations ($\gtrsim 10\,{\rm hr}$ per pointing) surveying their entirety are feasible with the SKA AA4. We note that synchrotron emission and Faraday rotation may not be sensitive to the magnetic fields in the densest regions of molecular clouds\footnote{This is because of the expected low number density of cosmic rays (for synchrotron; because of shielding effects) and thermal electrons (for Faraday rotation; because of the low ionisation fraction).} but nonetheless may be able to probe the intermediate/outer layers of such systems.

The aforementioned pursuit of the detailed characterisation of the magnetic fields in molecular clouds can be achieved with observations with the SKA-Low and the SKA-Mid in Bands 1 and 2. In particular, for the detection of synchrotron emission from molecular clouds, a 50\,hr per-pointing SKA-Low survey will yield a sensitivity of $6.6\,\mu{\rm Jy}$ per $4^{\prime\prime}$ beam, which is a significant improvement over the current state-of-the-art \citep[e.g.\ $83\,\mu{\rm Jy}$ per $6^{\prime\prime}$ beam with LoTSS;][]{ShimwellEA2022}. RM and Faraday complexity studies can be best performed with SKA-Mid Band 1, thanks to its high RM precision and sensitivity to Faraday depolarisation (Sec.~\ref{sec:SKA}). A 50\,hr per-pointing Band 1 survey will deliver a sensitivity of $0.3\,\mu{\rm Jy}$ per $1.3^{\prime\prime}$ beam, corresponding to a polarised source density of 380--610\,${\rm deg}^{-2}$, and thus leading to 3,800--6,100 RM and Faraday complexity measurements across the extent of each $10\,{\rm deg}^2$ nearby molecular cloud. Finally, the SKA-Mid Band 2 covers both the H\textsc{i} and OH spectral lines, enabling direct measurements of the line-of-sight magnetic field strength and direction via the Zeeman effect. A concurrent recording of the Band~2 full-Stokes continuum data will further complement the Band~1 survey mentioned just above.

\subsubsection{Interesting Galactic Regions; Magellanic Clouds; M~31}

As described above in Sec.~\ref{sec:pre}, the many on-going wide-area polarisation surveys with SKA precursor telescopes \citep[e.g.][]{MMGPS,GaenslerEA2025} will almost certainly lead to discoveries regarding the small-scale magnetic fields in the Milky Way ISM, similar to the extreme |RM| towards the Sagittarius spiral arm tangent \citep{ShanahanEA2019} recently unravelled by the Very Large Array (VLA) The H\textsc{i}/OH/Recombination line survey of the Milky Way \citep[THOR;][]{BeutherEA2016}. The mysteries surrounding the new discoveries can be resolved with deep observations of the Milky Way and our cosmic neighbours (in particular, the Magellanic Clouds and M~31) with the SKA-Mid, leveraging on its exceptional sensitivity, high angular resolution, and broad instantaneous bandwidth combined.

The exact frequency band choice, as pointed out in Sec.~\ref{sec:SKA}, will depend on the properties of the sky area of interest. Specifically, regions subjected to severe Faraday depolarisation will require observations in the higher frequency bands (Band~2 or even Bands~3 and 4) at the cost of lower RM precisions; while regions exhibiting low levels of Faraday depolarisation will require observations in the low frequency range (Band~1) to detect the associated depolarisation effects, with the added benefit of higher RM accuracies. Meanwhile, the required per-pointing integration time will depend on the science requirements of the polarised source density (likely in the 100--1000$\,{\rm deg}^{-2}$ range).

While we cannot predict the exact science directions that the SKA precursor surveys will steer us towards in the coming years, we list below a few potential Galactic areas that we may pursue with the SKA-Mid AA4:

\begin{itemize}
\item \textit{Galactic Centre \& Central Molecular Zone} --- These are amongst the most extreme environments that can be found within the Milky Way \citep{EatoughEA2013,HeywoodEA2022,PareEA2024}, allowing us to explore the interplays between the ISM and Sgr A*, Galactic outflows, and intense star formation. Moreover, the observed synchrotron filaments in the MeerKAT observations \citep{HeywoodEA2022, Yusef-ZadehEA2022} suggest this region to be promising for studying the magnetic field structure (discussed in \Sec{sec:struc}).

\item \textit{Galactic Anti-centre} --- By looking away from the Galactic Centre region, one observes through a relatively calm sky area that is not saturated by stellar feedback processes (due to the shorter path-length and the lower star-formation rate), as well as less affected by the regular / anisotropic turbulent magnetic fields (since both are approximately perpendicular to the line-of-sight). Therefore, the Galactic Anti-centre can be an excellent target area for studying how individual star-forming regions can produce/sustain the isotropic turbulent magnetic fields.

\item \textit{Off-plane control regions} --- We stress the importance of observing a control region away from the Galactic mid-plane that has minimal contributions by the Galactic small-scale magnetic fields. This is required to isolate the intrinsic Faraday complexity of the background extragalactic sources (Sec.~\ref{sec:complex} and \ref{sec:pre}). Given that the observed Faraday complexity can depend on the frequency coverage \citep{AndersonEA2016} and signal-to-noise \citep{OSullivanEA2017}, information gathered from similar control experiments with the SKA precursors ought not be applied to future SKA studies. Instead, the control regions should be observed in an almost identical manner (i.e.\ frequency coverage and integration time) to the science observations.

\item \textit{Our closest neighbours} --- Studies of the ISM of external galaxies are often more straight-forward than those of the Milky Way, thanks to our external perspectives of the nearby galaxies. Amongst our cosmic neighbours, the Magellanic Clouds and M~31\footnote{While M~31 is situated in the northern sky, its declination of $\approx +41^\circ$ is still accessible from the SKA.} can be SKA's prime targets for detailed studies of their small-scale magnetic fields, given their proximities to us ($\approx 50$--$65\,{\rm kpc}$ for the Magellanic Clouds; $\approx 770\,{\rm kpc}$ for M~31). With diffuse polarised emission, an assumed angular resolution of $5^{\prime\prime}$ translates to $1.2$--$1.6\,{\rm pc}$ for the Magellanic Clouds, and $22\,{\rm pc}$ for M~31 -- sufficient for spatially resolving the $\ll 100\,{\rm pc}$ small-scale galactic magnetic structures. With background polarised sources, assuming a fiducial source density of $\approx 100\,{\rm deg}^{-2}$ (which will allow us to probe down to arcmin scales with source-pair RM SF analysis), we can explore down to physical scales of $15$--$19\,{\rm pc}$ for the Magellanic Clouds, and about $220\,{\rm pc}$ for the M~31.

\end{itemize}

\subsubsection{Elliptical Galaxies} \label{sec:ell2}

With the SKA AA4, we expect to be able to attain an in-depth knowledge of the magnetic fields in nearby elliptical galaxies and thus, how small-scale magnetic fields can be amplified and maintained in the absence of large-scale magnetic fields in galaxies (Sec.~\ref{sec:ell}). With the background polarised sources as probes via Faraday rotation effects (by the enhanced RM scatter, and possibly by RM SF), the prime target elliptical galaxies are those that have large angular sizes and are accessible from the southern hemisphere. Examples include NGC~1316, NGC~3115, NGC~4406, and NGC~5102, all with angular diameters $\approx 10^{\prime}$ (areas of $\approx 0.02\,{\rm deg}^2$). Given the amount of RM contribution is expected to be low \citep{SetaEA2021}, the prospective SKA-Mid observations should be performed in Bands 1 and/or 2, with the former being favoured if the Faraday depolarisation effect contributed by the elliptical galaxies is minimal.

With a fiducial integration time of 200\,hr per band and per galaxy, the sensitivity that can be attained with the Band 1/2 observations will be $0.17$/$0.08\,\mu{\rm Jy/beam}$, respectively. These will lead to polarised source densities of 540--860\,deg$^{-2}$ for Band 1 and 680--890\,deg$^{-2}$ for Band 2. With our nominated target elliptical galaxies above (each with $\approx 0.02\,{\rm deg}^2$ area), we can expect 10--20 background polarised sources behind each of those galaxies. These measurements will place strong constraints on magnetic field properties in ellipticals and on the parameters of the small-scale dynamo theory.

\subsubsection{Small-scale Magnetic Fields in Intermediate-redshift Galaxies} \label{sec:large_intermediatez}

The turbulent magnetic field strength on small scales (smaller than the projected size of the background source – usually 10s of pc) can be derived from the $\sigma_{\rm RM}$ difference between the lensed images (with assumptions on the electron density/typical turbulent cell size). We note the importance of higher frequency data (i.e.\ Bands 3 and 4) in detecting the $\sigma_{\rm RM}$ caused by lensing galaxies. It was shown in \cite{KovacsEA2025} that using only L-band data (i.e.\ 1--2\,GHz) can lead to the inability to measure $\sigma_{\rm RM}$ (for the case of the lensing system B1600$+$434).

The SKA is expected to discover $\approx 10^5$ new lensing systems \citep[see e.g.][]{McKeanEA2015}. With the large sample size, it can become feasible to characterise the outer scale of turbulence \citep[see][for detailed discussions]{KovacsEA2025}. However, we note that the exact technique will still need to be developed and verified with numerical simulation data. Finally, we highlight the significant improvements in the sensitivity using the SKA when compared to, e.g., the VLA. From the work of \cite{KovacsEA2025}, the on-source integration time was 13\,min in L-band (1--2\,GHz), 5\,min in S-band (2--4\,GHz), 9\,min in C1-band (4--6\,GHz), and 7\,min in C2-band (6--8\,GHz). Similar exposure time using the SKA-Mid will yield $\approx 5$ times better sensitivity.

\subsubsection{Circumgalactic Medium (CGM) of Low-redshift Galaxies}

The statistical studies of the CGM magnetic fields in nearby galaxies have just been enabled with LOFAR \citep{HeesenEA2023} and MeerKAT \citep{BockmannEA23} surveys. While CGM magnetism studies on a per-galaxy basis can be possible for systems within $\approx 10\,{\rm Mpc}$ using on-going surveys such as POSSUM \citep{GaenslerEA2025}, the analysis of such nearby systems can be challenging. This is because the great angular extents ($\gtrsim$\,deg) can lead to difficulties in decoupling between the CGM and the Milky Way ISM contributions to the spatially fluctuating RM. By targeting more distant galaxies that have smaller angular sizes ($\ll$\,deg), the Galactic RM fluctuations can be reduced, leading to more accurate characterisation of the magnetic fields in the CGM.

With the SKA-Mid AA4, the direct study of the magnetic fields in the CGM of a large sample of low-redshift ($z \lesssim 0.01$) galaxies can be realised (see also \citealt{Mao01.2026.SKA} Chapter for further discussions). In particular, assuming the magnetic halo of the CGM to be about $200\,\kpc$ in diameter \citep{ShahS2021, HeesenEA2023, ShahEA2025} and with a target sample size of 20 polarised sources background to the CGM, we will require a polarised source density of $360\,{\rm deg}^{-2}$ to map the CGM magnetic fields for a galaxy at $z = 0.01$. This translates to a per-galaxy integration time (sensitivity) of 12--60\,hr (0.3--0.7\,$\mu{\rm Jy/beam}$) for Band~1 and 8--30\,hr (0.2--0.4\,$\mu{\rm Jy/beam}$) for Band~2. These figures are reasonable and a Key Science Program sampling a few dozen of such galaxies can be feasible. Alternatively, if one were to dedicate, say, 200\,hr of SKA-Mid time on one single galaxy of interest, the resulting source density (sensitivity) will be $560$--$860\,{\rm deg}^{-2}$ ($0.17\,\mu{\rm Jy/beam}$) for Band 1, and $680$--$890\,{\rm deg}^{-2}$ ($0.08\,\mu{\rm Jy/beam}$) for Band 2. Maintaining the constraint of 20 polarised sources within the $200\,\kpc$ of the central galaxy as assumed above, the redshift limit will be $z \leq 0.013$--$0.016$ with Band 1 and $z \leq 0.014$--$0.016$ with Band 2. 

\subsection{Synergies beyond Radio} \label{sec:synergies}

We highlight here a few select advancements beyond the radio regime in recent years / near future that will be highly relevant to the study of small-scale galactic magnetic fields. First of all, the limitations of the many existing thermal electron number density ($n_e$) models of the Milky Way have been pointed out \citep[e.g.][]{OckerEA2024}. Upcoming optical spectroscopic surveys, such as the Local Volume Mapper \citep[LVM;][]{LVM} with the Sloan Digital Sky Survey V (SDSS-V) that will cover sky areas such as the Galactic mid-plane, Orion, and the Magellanic Clouds, will provide high-quality data that will enable improved constraints of $n_e$. Such information will be highly valuable for galactic magnetism studies via Faraday rotation \citep[both RM and Faraday complexity; e.g.][]{LivingstonEA2022,LivingstonEA2025,MaEA2025}, including the study of the small-scale magnetic fields. Specifically, high-spatial-resolution $n_e$ information will allow one to isolate the small-scale magnetic fields' contribution to spatial RM fluctuations \citep[see e.g.][]{SetaEA2023}.

Next, the combined analysis between the SKA and data from starlight polarisation surveys \citep[e.g.][]{Magalhaes2012AIPC.1429..244M,Tassis2018arXiv181005652T,Clemens2020ApJS..249...23C,Versteeg2023AJ....165...87V}, with the latter supplemented by accurate distance measurements from e.g. \textit{Gaia} parallax \citep{LuriEA2018}, holds the prospect of considerable advancements in our knowledge in the small-scale magnetic fields in the Solar neighbourhood (within a few kpc). 
Recent efforts in conjoining Galactic radio polarised diffuse emission and starlight polarisation data \citep[e.g.][]{PanopoulouEA2021,TuricEA2021} have highlighted the ability in placing distance constraints to the former. Generalising this, and taking into account that both tracers can offer a tomographic view of the magnetic fields in the local ISM \citep[e.g.][]{VanEck2017,PanopoulouEA2019,ThomsonEA2019,PelgrimsEA2024}, future careful comparisons and combinations of these two powerful magnetic tracers will almost certainly provide us with a clear view of the ISM magnetic fields in our vicinity.

Finally, we point out the recent advents of various 3D maps of the local ISM (within a few kpc), including those of dust extinction \citep{EdenhoferEA2024}, H$\alpha$ emission \citep{McCallumEA2025}, and hydrogen (atomic and molecular) number density \citep{SodingEA2025}. These are the results of combining the high quality data from various large-scale surveys of the Milky Way ISM with sophisticated analysis and inference algorithms. Apart from the usefulness of these (and the future) 3D maps in aiding our understanding of the small-scale magnetic fields in the Milky Way, we anticipate that future SKA surveys may in turn contribute to next generations of 3D maps of the Milky Way ISM.

\section{Summary} \label{sec:sum}

In this Chapter, we provide an overview of the prospects of the SKA in advancing our understanding of the small-scale magnetic fields in galaxies, thereby contributing to a better understanding of the fundamental physics of dynamos, as well as the roles that magnetic fields play in star formation and galaxy evolution (Sec.~\ref{sec:intro}~\&~\ref{sec:ques}). The most promising tracers of the small-scale galactic magnetic fields that the SKA AA4 will be sensitive to are (i) diffuse synchrotron emission, (ii) rotation measure (RM), (iii) Faraday complexity, and (iv) the Zeeman effect (as summarised in Sec.~\ref{sec:obs}). Combined with upcoming and innovative analysis methods that will be developed using the ongoing polarisation surveys of SKA precursor telescopes (Sec.~\ref{sec:pre}), we will characterise and better understand the strength, scale, and structure of the galactic magnetic fields with greater accuracy (Sec.~\ref{sec:sci}).

The general SKA observational requirements for our research field are summarised in Sec.~\ref{sec:SKA}. We argue that with the SKA-Low, a uniform wide-area polarisation survey, in addition to deep targeted observations of nearby molecular clouds, will be the most fruitful for our science direction. Meanwhile, with the SKA-Mid, wide-area surveys (except for Band 1) appear unlikely to transform our understanding of the subject, given the expected yields of ongoing polarisation surveys such as POSSUM and MMGPS. Instead, we advocate for deep integration ($10$--$1000\,{\rm hr}$ per pointing) observations of specific sky areas of interest to bring forth a generational leap in our knowledge of astrophysical magnetic fields. We highlight the prospects of SKA-Mid Bands 3 and 4 in studying the extreme Galactic environments (e.g.,\ the Galactic Centre and along the Galactic mid-plane; Sec.~\ref{sec:SKA}). With this thought, we propose and discuss a few possible large-scale science programs with SKA for deciphering the physics and impact of the small-scale magnetic fields in galaxies (Sec.~\ref{sec:largescale}).

\section{Acknowledgements} \label{sec:ack}
We thank Rainer Beck and Tom Landecker for their very useful comments and suggestions on the text. Y.~K.~M.\ is partially supported by the Federal Ministry of Education and Research (BMBF) in Germany via the German Academic Exchange Service (DAAD) Project-based Personnel Exchange Programme (PPP). A. S.\ is supported by the Australian Research Council through the Discovery Early Career Researcher Award (DECRA) Fellowship (project~DE250100003) funded by the Australian Government and the Australia-Germany Joint Research Cooperation Scheme of Universities Australia (UA--DAAD, 2025--2026). S.~H. acknowledges funding by the European Union (ERC, ISM-FLOW, 101055318). Views and opinions expressed are, however, those of the author(s) only and do not necessarily reflect those of the European Union or the European Research Council. Neither the European Union nor the granting authority can be held responsible for them. M.~P.\ acknowledges the INAF grant 2023 MERCATOR (``MultiwavelEngth signatuRes of Cosmic rAys in sTar-fOrming Regions'') and the INAF grant 2024 ENERGIA (``ExploriNg low-Energy cosmic Rays throuGh theoretical InvestigAtions at INAF''). G.~P.\ acknowledges support from the Swedish Research Council (VR) under grant number 2023-04038 and the Knut and Alice Wallenberg Foundation Fellowship program under grant number 2023.0080. L.~A.\ acknowledges support through the Program ``Rita Levi Montalcini'' of the Italian MIUR. R.~M.~C.\ acknowledges support from the Australian Research Council via grant DP230101055 shared with Prof.\ Mark Krumholz. S.~M. acknowledges support from grant PID2023-146372OB-I00, funded by MICIU/AEI/10.13039/501100011033 and by ERDF, EU.

\bibliographystyle{abbrvnat-maxbibnames4}
\bibliography{ms}

@ARTICLE{SilsbeeEA2018,
       author = {{Silsbee}, Kedron and {Ivlev}, Alexei V. and {Padovani}, Marco and {Caselli}, Paola},
        title = "{Magnetic Mirroring and Focusing of Cosmic Rays}",
      journal = {\apj},
         year = 2018,
        month = aug,
       volume = {863},
       number = {2},
          eid = {188},
        pages = {188},
          doi = {10.3847/1538-4357/aad3cf},
archivePrefix = {arXiv},
       eprint = {1807.05025},
 primaryClass = {astro-ph.HE},
       adsurl = {https://ui.adsabs.harvard.edu/abs/2018ApJ...863..188S}
}

@ARTICLE{CesarskyV1978,
       author = {{Cesarsky}, C.~J. and {Volk}, H.~J.},
        title = "{Cosmic Ray Penetration into Molecular Clouds}",
      journal = {\aap},
         year = 1978,
        month = nov,
       volume = {70},
        pages = {367},
       adsurl = {https://ui.adsabs.harvard.edu/abs/1978A&A....70..367C}
}

@ARTICLE{WollebenR2004,
       author = {{Wolleben}, M. and {Reich}, W.},
        title = "{Faraday screens associated with local molecular clouds}",
      journal = {\aap},
         year = 2004,
        month = nov,
       volume = {427},
        pages = {537-548},
          doi = {10.1051/0004-6361:20040561},
archivePrefix = {arXiv},
       eprint = {astro-ph/0409339},
 primaryClass = {astro-ph},
       adsurl = {https://ui.adsabs.harvard.edu/abs/2004A&A...427..537W}
}

@ARTICLE{Harvey-SmithEA2011,
       author = {{Harvey-Smith}, L. and {Madsen}, G.~J. and {Gaensler}, B.~M.},
        title = "{Magnetic Fields in Large-diameter H II Regions Revealed by the Faraday Rotation of Compact Extragalactic Radio Sources}",
      journal = {\apj},
         year = 2011,
        month = aug,
       volume = {736},
       number = {2},
          eid = {83},
        pages = {83},
          doi = {10.1088/0004-637X/736/2/83},
archivePrefix = {arXiv},
       eprint = {1106.0931},
 primaryClass = {astro-ph.GA},
       adsurl = {https://ui.adsabs.harvard.edu/abs/2011ApJ...736...83H}
}

@ARTICLE{RaychevaEA2022,
       author = {{Raycheva}, N.~C. and {Haverkorn}, M. and {Ideguchi}, S. and {Stil}, J.~M. and {Gaensler}, B.~M. and {Sun}, X. and {Han}, J.~L. and {Carretti}, E. and {Gao}, X.~Y. and {Wijte}, T.},
        title = "{Turbulent magnetic field in the H II region Sh 2-27}",
      journal = {\aap},
         year = 2022,
        month = jul,
       volume = {663},
          eid = {A170},
        pages = {A170},
          doi = {10.1051/0004-6361/202039474},
archivePrefix = {arXiv},
       eprint = {2206.01787},
 primaryClass = {astro-ph.GA},
       adsurl = {https://ui.adsabs.harvard.edu/abs/2022A&A...663A.170R}
}

@ARTICLE{Clemens2020ApJS..249...23C,
       author = {{Clemens}, Dan P. and {Cashman}, L.~R. and {Cerny}, C. and {El-Batal}, A.~M. and {Jameson}, K.~E. and {Marchwinski}, R. and {Montgomery}, J. and {Pavel}, M. and {Pinnick}, A. and {Taylor}, B.~W.},
        title = "{The Galactic Plane Infrared Polarization Survey (GPIPS): Data Release 4}",
      journal = {\apjs},
         year = 2020,
        month = aug,
       volume = {249},
       number = {2},
          eid = {23},
        pages = {23},
          doi = {10.3847/1538-4365/ab9f30},
archivePrefix = {arXiv},
       eprint = {2006.15203},
 primaryClass = {astro-ph.GA},
       adsurl = {https://ui.adsabs.harvard.edu/abs/2020ApJS..249...23C}
}

@ARTICLE{Versteeg2023AJ....165...87V,
       author = {{Versteeg}, M.~J.~F. and {Magalh{\~a}es}, A.~M. and {Haverkorn}, M. and {Angarita}, Y. and {Rodrigues}, C.~V. and {Santos-Lima}, R. and {Kawabata}, Koji S.},
        title = "{Interstellar Polarization Survey. II. General Interstellar Medium}",
      journal = {\aj},
         year = 2023,
        month = mar,
       volume = {165},
       number = {3},
          eid = {87},
        pages = {87},
          doi = {10.3847/1538-3881/aca8fd},
archivePrefix = {arXiv},
       eprint = {2212.05985},
 primaryClass = {astro-ph.GA},
       adsurl = {https://ui.adsabs.harvard.edu/abs/2023AJ....165...87V}
}

@ARTICLE{Tassis2018arXiv181005652T,
       author = {{Tassis}, Konstantinos and {Ramaprakash}, Anamparambu N. and {Readhead}, Anthony C.~S. and {Potter}, Stephen B. and {Wehus}, Ingunn K. and {Panopoulou}, Georgia V. and {Blinov}, Dmitry and {Eriksen}, Hans Kristian and {Hensley}, Brandon and {Karakci}, Ata and {Kypriotakis}, John A. and {Maharana}, Siddharth and {Ntormousi}, Evangelia and {Pavlidou}, Vasiliki and {Pearson}, Timothy J. and {Skalidis}, Raphael},
        title = "{PASIPHAE: A high-Galactic-latitude, high-accuracy optopolarimetric survey}",
      journal = {arXiv e-prints},
         year = 2018,
        month = oct,
          eid = {arXiv:1810.05652},
        pages = {arXiv:1810.05652},
          doi = {10.48550/arXiv.1810.05652},
archivePrefix = {arXiv},
       eprint = {1810.05652},
 primaryClass = {astro-ph.IM},
       adsurl = {https://ui.adsabs.harvard.edu/abs/2018arXiv181005652T}
}

@ARTICLE{Malik2024monogem,
       author = {{Malik}, Sunil and {Yuen}, Ka Ho and {Yan}, Huirong},
        title = "{Study of Magnetic Field and Turbulence in the TeV Halo around the Monogem Pulsar}",
      journal = {\apj},
         year = 2024,
        month = apr,
       volume = {965},
       number = {1},
          eid = {65},
        pages = {65},
          doi = {10.3847/1538-4357/ad34d7},
archivePrefix = {arXiv},
       eprint = {2307.13342},
 primaryClass = {astro-ph.HE},
       adsurl = {https://ui.adsabs.harvard.edu/abs/2024ApJ...965...65M}
}

@ARTICLE{LiuEA2019,
       author = {{Liu}, Ruo-Yu and {Yan}, Huirong and {Zhang}, Heshou},
        title = "{Understanding the Multiwavelength Observation of Geminga's Tev Halo: The Role of Anisotropic Diffusion of Particles}",
      journal = {\prl},
         year = 2019,
        month = nov,
       volume = {123},
       number = {22},
          eid = {221103},
        pages = {221103},
          doi = {10.1103/PhysRevLett.123.221103},
archivePrefix = {arXiv},
       eprint = {1904.11536},
 primaryClass = {astro-ph.HE},
       adsurl = {https://ui.adsabs.harvard.edu/abs/2019PhRvL.123v1103L}
}

@ARTICLE{AbeysekaraEA2017,
       author = {{Abeysekara}, A.~U. and {Albert}, A. and {Alfaro}, R. and {Alvarez}, C. and {{\'A}lvarez}, J.~D. and {Arceo}, R. and {Arteaga-Vel{\'a}zquez}, J.~C. and {Avila Rojas}, D. and {Ayala Solares}, H.~A. and {Barber}, A.~S. and {Bautista-Elivar}, N. and {Becerril}, A. and {Belmont-Moreno}, E. and {BenZvi}, S.~Y. and {Berley}, D. and {Bernal}, A. and {Braun}, J. and {Brisbois}, C. and {Caballero-Mora}, K.~S. and {Capistr{\'a}n}, T. and {Carrami{\~n}ana}, A. and {Casanova}, S. and {Castillo}, M. and {Cotti}, U. and {Cotzomi}, J. and {Couti{\~n}o de Le{\'o}n}, S. and {De Le{\'o}n}, C. and {De la Fuente}, E. and {Dingus}, B.~L. and {DuVernois}, M.~A. and {D{\'\i}az-V{\'e}lez}, J.~C. and {Ellsworth}, R.~W. and {Engel}, K. and {Enr{\'\i}quez-Rivera}, O. and {Fiorino}, D.~W. and {Fraija}, N. and {Garc{\'\i}a-Gonz{\'a}lez}, J.~A. and {Garfias}, F. and {Gerhardt}, M. and {Gonz{\'a}lez Mu{\~n}oz}, A. and {Gonz{\'a}lez}, M.~M. and {Goodman}, J.~A. and {Hampel-Arias}, Z. and {Harding}, J.~P. and {Hern{\'a}ndez}, S. and {Hern{\'a}ndez-Almada}, A. and {Hinton}, J. and {Hona}, B. and {Hui}, C.~M. and {H{\"u}ntemeyer}, P. and {Iriarte}, A. and {Jardin-Blicq}, A. and {Joshi}, V. and {Kaufmann}, S. and {Kieda}, D. and {Lara}, A. and {Lauer}, R.~J. and {Lee}, W.~H. and {Lennarz}, D. and {Vargas}, H. Le{\'o}n and {Linnemann}, J.~T. and {Longinotti}, A.~L. and {Luis Raya}, G. and {Luna-Garc{\'\i}a}, R. and {L{\'o}pez-Coto}, R. and {Malone}, K. and {Marinelli}, S.~S. and {Martinez}, O. and {Martinez-Castellanos}, I. and {Mart{\'\i}nez-Castro}, J. and {Mart{\'\i}nez-Huerta}, H. and {Matthews}, J.~A. and {Miranda-Romagnoli}, P. and {Moreno}, E. and {Mostaf{\'a}}, M. and {Nellen}, L. and {Newbold}, M. and {Nisa}, M.~U. and {Noriega-Papaqui}, R. and {Pelayo}, R. and {Pretz}, J. and {P{\'e}rez-P{\'e}rez}, E.~G. and {Ren}, Z. and {Rho}, C.~D. and {Rivi{\`e}re}, C. and {Rosa-Gonz{\'a}lez}, D. and {Rosenberg}, M. and {Ruiz-Velasco}, E. and {Salazar}, H. and {Salesa Greus}, F. and {Sandoval}, A. and {Schneider}, M. and {Schoorlemmer}, H. and {Sinnis}, G. and {Smith}, A.~J. and {Springer}, R.~W. and {Surajbali}, P. and {Taboada}, I. and {Tibolla}, O. and {Tollefson}, K. and {Torres}, I. and {Ukwatta}, T.~N. and {Vianello}, G. and {Weisgarber}, T. and {Westerhoff}, S. and {Wisher}, I.~G. and {Wood}, J. and {Yapici}, T. and {Yodh}, G. and {Younk}, P.~W. and {Zepeda}, A. and {Zhou}, H. and {Guo}, F. and {Hahn}, J. and {Li}, H. and {Zhang}, H.},
        title = "{Extended gamma-ray sources around pulsars constrain the origin of the positron flux at Earth}",
      journal = {Science},
         year = 2017,
        month = nov,
       volume = {358},
       number = {6365},
        pages = {911-914},
          doi = {10.1126/science.aan4880},
archivePrefix = {arXiv},
       eprint = {1711.06223},
 primaryClass = {astro-ph.HE},
       adsurl = {https://ui.adsabs.harvard.edu/abs/2017Sci...358..911A}
}

@ARTICLE{Malik2023,
       author = {{Malik}, Sunil and {Yuen}, Ka Ho and {Yan}, Huirong},
        title = "{Diagnosis of 3D magnetic field and mode composition in MHD turbulence with Y-parameter}",
      journal = {\mnras},
         year = 2023,
        month = oct,
       volume = {524},
       number = {4},
        pages = {6102-6113},
          doi = {10.1093/mnras/stad2225},
archivePrefix = {arXiv},
       eprint = {2303.17282},
 primaryClass = {astro-ph.GA},
       adsurl = {https://ui.adsabs.harvard.edu/abs/2023MNRAS.524.6102M}
}

@INPROCEEDINGS{Magalhaes2012AIPC.1429..244M,
       author = {{Magalh{\~a}es}, A.~M. and {de Oliveira}, C.~M. and {Carciofi}, A. and {Costa}, R. and {Dal Pino}, E.~M.~G. and {Diaz}, M. and {Ferrari}, T. and {Fernandez}, C. and {Gomes}, A.~L. and {Marrara}, L. and {Pereyrac}, A. and {Ribeiro}, N.~L. and {Rodrigues}, C.~V. and {Rubinho}, M.~S. and {Seriacopi}, D.~B. and {Taylor}, K.},
        title = "{South Pol: Revealing the polarized southern sky}",
    booktitle = {Stellar Polarimetry: from Birth to Death},
         year = 2012,
       editor = {{Hoffman}, Jennifer L. and {Bjorkman}, Jon and {Whitney}, Barbara},
       series = {American Institute of Physics Conference Series},
       volume = {1429},
        month = may,
    publisher = {AIP},
        pages = {244-247},
          doi = {10.1063/1.3701933},
archivePrefix = {arXiv},
       eprint = {1201.3639},
 primaryClass = {astro-ph.IM},
       adsurl = {https://ui.adsabs.harvard.edu/abs/2012AIPC.1429..244M}
}

@ARTICLE{ShahEA2025,
       author = {{Shah}, Hilay and {van de Voort}, Freeke and {Seta}, Amit and {Federrath}, Christoph},
        title = "{Understanding gas mixing in the circumgalactic medium}",
      journal = {\mnras},
         year = 2025,
        month = aug,
       volume = {541},
       number = {3},
        pages = {2471-2492},
          doi = {10.1093/mnras/staf1066},
archivePrefix = {arXiv},
       eprint = {2505.21980},
 primaryClass = {astro-ph.GA},
       adsurl = {https://ui.adsabs.harvard.edu/abs/2025MNRAS.541.2471S}
}

@ARTICLE{ONeillEA2025,
       author = {{O'Neill}, Theo J. and {Goodman}, Alyssa A. and {Soler}, Juan D. and {Zucker}, Catherine and {Han}, Jiwon Jesse},
        title = "{A 3D Model of the Local Bubble's Magnetic Field: Insights from Dust and Starlight Polarization}",
      journal = {\apj},
         year = 2025,
        month = aug,
       volume = {988},
       number = {2},
          eid = {191},
        pages = {191},
          doi = {10.3847/1538-4357/ade306},
archivePrefix = {arXiv},
       eprint = {2410.17341},
 primaryClass = {astro-ph.GA},
       adsurl = {https://ui.adsabs.harvard.edu/abs/2025ApJ...988..191O}
}

@ARTICLE{HutschenreuterEA2022,
       author = {{Hutschenreuter}, S. and {Anderson}, C.~S. and {Betti}, S. and {Bower}, G.~C. and {Brown}, J. -A. and {Br{\"u}ggen}, M. and {Carretti}, E. and {Clarke}, T. and {Clegg}, A. and {Costa}, A. and {Croft}, S. and {Van Eck}, C. and {Gaensler}, B.~M. and {de Gasperin}, F. and {Haverkorn}, M. and {Heald}, G. and {Hull}, C.~L.~H. and {Inoue}, M. and {Johnston-Hollitt}, M. and {Kaczmarek}, J. and {Law}, C. and {Ma}, Y.~K. and {MacMahon}, D. and {Mao}, S.~A. and {Riseley}, C. and {Roy}, S. and {Shanahan}, R. and {Shimwell}, T. and {Stil}, J. and {Sobey}, C. and {O'Sullivan}, S.~P. and {Tasse}, C. and {Vacca}, V. and {Vernstrom}, T. and {Williams}, P.~K.~G. and {Wright}, M. and {En{\ss}lin}, T.~A.},
        title = "{The Galactic Faraday rotation sky 2020}",
      journal = {\aap},
         year = 2022,
        month = jan,
       volume = {657},
          eid = {A43},
        pages = {A43},
          doi = {10.1051/0004-6361/202140486},
archivePrefix = {arXiv},
       eprint = {2102.01709},
 primaryClass = {astro-ph.GA},
       adsurl = {https://ui.adsabs.harvard.edu/abs/2022A&A...657A..43H}
}

@ARTICLE{KhadirEA2024,
       author = {{Khadir}, Affan and {Pandhi}, Ayush and {Hutschenreuter}, Sebastian and {Gaensler}, B.~M. and {Vanderwoude}, Shannon and {West}, Jennifer L. and {O'Sullivan}, Shane P.},
        title = "{Interpolation Techniques for Reconstructing Galactic Faraday Rotation}",
      journal = {\apj},
         year = 2024,
        month = dec,
       volume = {977},
       number = {2},
          eid = {276},
        pages = {276},
          doi = {10.3847/1538-4357/ad8ddf},
archivePrefix = {arXiv},
       eprint = {2410.15265},
 primaryClass = {astro-ph.IM},
       adsurl = {https://ui.adsabs.harvard.edu/abs/2024ApJ...977..276K}
}

@article{EnsslinEA2017,
  title = {The Galaxy in circular polarization: All-sky radio prediction, detection strategy, and the charge of the leptonic cosmic rays},
  author = {En\ss{}lin, Torsten A. and Hutschenreuter, Sebastian and Vacca, Valentina and Oppermann, Niels},
  journal = {Phys. Rev. D},
  volume = {96},
  issue = {4},
  pages = {043021},
  numpages = {11},
  year = {2017},
  month = {Aug},
  publisher = {American Physical Society},
  doi = {10.1103/PhysRevD.96.043021},
  url = {https://link.aps.org/doi/10.1103/PhysRevD.96.043021}
}

@ARTICLE{PandyaEA2016,
       author = {{Pandya}, Alex and {Zhang}, Zhaowei and {Chandra}, Mani and {Gammie}, Charles F.},
        title = "{Polarized Synchrotron Emissivities and Absorptivities for Relativistic Thermal, Power-law, and Kappa Distribution Functions}",
      journal = {\apj},
         year = 2016,
        month = may,
       volume = {822},
       number = {1},
          eid = {34},
        pages = {34},
          doi = {10.3847/0004-637X/822/1/34},
archivePrefix = {arXiv},
       eprint = {1602.08749},
 primaryClass = {astro-ph.HE},
       adsurl = {https://ui.adsabs.harvard.edu/abs/2016ApJ...822...34P}
}

@ARTICLE{MacquartM2000,
       author = {{Macquart}, J. -P. and {Melrose}, D.~B.},
        title = "{Scintillation-induced Circular Polarization in Pulsars and Quasars}",
      journal = {\apj},
         year = 2000,
        month = dec,
       volume = {545},
       number = {2},
        pages = {798-806},
          doi = {10.1086/317852},
archivePrefix = {arXiv},
       eprint = {astro-ph/0007429},
 primaryClass = {astro-ph},
       adsurl = {https://ui.adsabs.harvard.edu/abs/2000ApJ...545..798M}
}

@ARTICLE{Chan2019MNRAS,
       author = {{Chan}, Jennifer Y.~H. and {Wu}, Kinwah and {On}, Alvina Y.~L. and {Barnes}, David J. and {McEwen}, Jason D. and {Kitching}, Thomas D.},
        title = "{Covariant polarized radiative transfer on cosmological scales for investigating large-scale magnetic field structures}",
      journal = {\mnras},
         year = 2019,
        month = apr,
       volume = {484},
       number = {2},
        pages = {1427-1455},
          doi = {10.1093/mnras/sty3498},
archivePrefix = {arXiv},
       eprint = {1901.04581},
 primaryClass = {astro-ph.CO},
       adsurl = {https://ui.adsabs.harvard.edu/abs/2019MNRAS.484.1427C}
}

@ARTICLE{On2019MNRAS,
       author = {{On}, Alvina Y.~L. and {Chan}, Jennifer Y.~H. and {Wu}, Kinwah and {Saxton}, Curtis J. and {van Driel-Gesztelyi}, Lidia},
        title = "{Polarized radiative transfer, rotation measure fluctuations, and large-scale magnetic fields}",
      journal = {\mnras},
         year = 2019,
        month = dec,
       volume = {490},
       number = {2},
        pages = {1697-1713},
          doi = {10.1093/mnras/stz2683},
archivePrefix = {arXiv},
       eprint = {1909.06703},
 primaryClass = {astro-ph.HE},
       adsurl = {https://ui.adsabs.harvard.edu/abs/2019MNRAS.490.1697O}
}

@ARTICLE{BraccoEA2023,
       author = {{Bracco}, Andrea and {Padovani}, Marco and {Soler}, Juan D.},
        title = "{The Orion-Taurus ridge: A synchrotron radio loop at the edge of the Orion-Eridanus superbubble}",
      journal = {\aap},
         year = 2023,
        month = sep,
       volume = {677},
          eid = {L11},
        pages = {L11},
          doi = {10.1051/0004-6361/202347283},
archivePrefix = {arXiv},
       eprint = {2308.16663},
 primaryClass = {astro-ph.GA},
       adsurl = {https://ui.adsabs.harvard.edu/abs/2023A&A...677L..11B}
}

@ARTICLE{AndersonEA2024,
       author = {{Anderson}, Craig S. and {McClure-Griffiths}, N.~M. and {Rudnick}, L. and {Gaensler}, B.~M. and {O'Sullivan}, S.~P. and {Bradbury}, S. and {Akahori}, T. and {Baidoo}, L. and {Bruggen}, M. and {Carretti}, E. and {Duchesne}, S. and {Heald}, G. and {Jung}, S.~L. and {Kaczmarek}, J. and {Leahy}, D. and {Loi}, F. and {Ma}, Y.~K. and {Osinga}, E. and {Seta}, A. and {Stuardi}, C. and {Thomson}, A.~J.~M. and {Van Eck}, C. and {Vernstrom}, T. and {West}, J.},
        title = "{Probing the magnetized gas distribution in galaxy groups and the cosmic web with POSSUM Faraday rotation measures}",
      journal = {\mnras},
         year = 2024,
        month = oct,
       volume = {533},
       number = {4},
        pages = {4068-4080},
          doi = {10.1093/mnras/stae1954},
archivePrefix = {arXiv},
       eprint = {2407.20325},
 primaryClass = {astro-ph.GA},
       adsurl = {https://ui.adsabs.harvard.edu/abs/2024MNRAS.533.4068A}
}

@ARTICLE{MakarenkoEA2018II,
       author = {{Makarenko}, I. and {Bushby}, P. and {Fletcher}, A. and {Henderson}, R. and {Makarenko}, N. and {Shukurov}, A.},
        title = "{Topological data analysis and diagnostics of compressible magnetohydrodynamic turbulence}",
      journal = {Journal of Plasma Physics},
         year = 2018,
        month = aug,
       volume = {84},
       number = {4},
          eid = {735840403},
        pages = {735840403},
          doi = {10.1017/S0022377818000752},
archivePrefix = {arXiv},
       eprint = {1804.04688},
 primaryClass = {physics.data-an},
       adsurl = {https://ui.adsabs.harvard.edu/abs/2018JPlPh..84d7303M}
}

@ARTICLE{DwivediEA2024,
       author = {{Dwivedi}, Shreya and {Anandavijayan}, Chandranathan and {Bhat}, Pallavi},
        title = "{Quasi-two-dimensionality of three-dimensional, magnetically dominated, decaying turbulence}",
      journal = {The Open Journal of Astrophysics},
         year = 2024,
        month = sep,
       volume = {7},
          eid = {75},
        pages = {75},
          doi = {10.33232/001c.122855},
archivePrefix = {arXiv},
       eprint = {2401.01965},
 primaryClass = {astro-ph.HE},
       adsurl = {https://ui.adsabs.harvard.edu/abs/2024OJAp....7E..75D}
}

@ARTICLE{MakarenkoEA2018I,
       author = {{Makarenko}, Irina and {Shukurov}, Anvar and {Henderson}, Robin and {Rodrigues}, Luiz F.~S. and {Bushby}, Paul and {Fletcher}, Andrew},
        title = "{Topological signatures of interstellar magnetic fields - I. Betti numbers and persistence diagrams}",
      journal = {\mnras},
         year = 2018,
        month = apr,
       volume = {475},
       number = {2},
        pages = {1843-1858},
          doi = {10.1093/mnras/stx3337},
archivePrefix = {arXiv},
       eprint = {1708.04558},
 primaryClass = {astro-ph.GA},
       adsurl = {https://ui.adsabs.harvard.edu/abs/2018MNRAS.475.1843M}
}

@ARTICLE{Yusef-ZadehEA2022,
       author = {{Yusef-Zadeh}, F. and {Arendt}, R.~G. and {Wardle}, M. and {Heywood}, I. and {Cotton}, W. and {Camilo}, F.},
        title = "{Statistical Properties of the Population of the Galactic Center Filaments: the Spectral Index and Equipartition Magnetic Field}",
      journal = {\apjl},
         year = 2022,
        month = feb,
       volume = {925},
       number = {2},
          eid = {L18},
        pages = {L18},
          doi = {10.3847/2041-8213/ac4802},
archivePrefix = {arXiv},
       eprint = {2201.10552},
 primaryClass = {astro-ph.GA},
       adsurl = {https://ui.adsabs.harvard.edu/abs/2022ApJ...925L..18Y}
}

@ARTICLE{DhakalS2025,
       author = {{Dhakal}, Saakshi and {Seta}, Amit},
        title = "{Probing the magneto-ionic medium of the Milky Way using pulsars}",
      journal = {\mnras},
         year = 2025,
        month = oct,
          doi = {10.1093/mnras/staf1816},
archivePrefix = {arXiv},
       eprint = {2510.12991},
 primaryClass = {astro-ph.GA},
       adsurl = {https://ui.adsabs.harvard.edu/abs/2025MNRAS.tmp.1707D}
}

@ARTICLE{KovacsEA2025,
       author = {{Kovacs}, Timea Orsolya and {Mao}, Sui Ann and {Basu}, Aritra and {Ma}, Yik Ki and {Gaensler}, Bryan M.},
        title = "{The halo magnetic field of a spiral galaxy at z = 0.414}",
      journal = {\aap},
         year = 2026,
        month = jan,
       volume = {705},
          eid = {A28},
        pages = {A28},
          doi = {10.1051/0004-6361/202453502},
archivePrefix = {arXiv},
       eprint = {2507.12542},
 primaryClass = {astro-ph.GA},
       adsurl = {https://ui.adsabs.harvard.edu/abs/2026A&A...705A..28K}
}

@ARTICLE{BraccoEA2025,
       author = {{Bracco}, Andrea and {Padovani}, Marco and {Galli}, Daniele and {Pezzuto}, Stefania and {Cipriani}, Alexandre and {Drabent}, Alexander},
        title = "{Are stellar embryos in Perseus radio-synchrotron emitters?: Statistical data analysis with Herschel and LOFAR paving the way for the SKA}",
      journal = {\aap},
         year = 2025,
        month = feb,
       volume = {694},
          eid = {A148},
        pages = {A148},
          doi = {10.1051/0004-6361/202452010},
archivePrefix = {arXiv},
       eprint = {2411.19573},
 primaryClass = {astro-ph.GA},
       adsurl = {https://ui.adsabs.harvard.edu/abs/2025A&A...694A.148B}
}

@INPROCEEDINGS{DickinsonEA2015,
       author = {{Dickinson}, C. and {Beck}, R. and {Crocker}, R. and {Crutcher}, R.~M. and {Davies}, R.~D. and {Ferri{\`e}re}, K. and {Fuller}, G. and {Jaffe}, T.~R. and {Jones}, D. and {Leahy}, P. and {Murphy}, E. and {Peel}, M.~W. and {Orlando}, E. and {Porter}, T. and {Protheroe}, R.~J. and {Strong}, A. and {Robishaw}, T. and {Watson}, R.~A. and {Yusef-Zadeh}, F.},
        title = "{SKA studies of in situ synchrotron radiation from molecular clouds}",
    booktitle = {Advancing Astrophysics with the Square Kilometre Array (AASKA14)},
         year = 2015,
        month = apr,
          eid = {102},
        pages = {102},
          doi = {10.22323/1.215.0102},
archivePrefix = {arXiv},
       eprint = {1501.00804},
 primaryClass = {astro-ph.GA},
       adsurl = {https://ui.adsabs.harvard.edu/abs/2015aska.confE.102D}
}

@ARTICLE{Tritsis2025,
       author = {{Tritsis}, Aris},
        title = "{Ambipolar diffusion and the mass-to-flux ratio in a turbulent collapsing cloud}",
      journal = {\aap},
         year = 2025,
        month = aug,
       volume = {700},
          eid = {A152},
        pages = {A152},
          doi = {10.1051/0004-6361/202554422},
archivePrefix = {arXiv},
       eprint = {2505.20391},
 primaryClass = {astro-ph.GA},
       adsurl = {https://ui.adsabs.harvard.edu/abs/2025A&A...700A.152T}
}

@ARTICLE{ErcegEA2022,
       author = {{Erceg}, Ana and {Jeli{\'c}}, Vibor and {Haverkorn}, Marijke and {Bracco}, Andrea and {Shimwell}, Timothy W. and {Tasse}, Cyril and {Dickey}, John M. and {Ceraj}, Lana and {Drabent}, Alexander and {Hardcastle}, Martin J. and {Turi{\'c}}, Luka},
        title = "{Faraday tomography of LoTSS-DR2 data. I. Faraday moments in the high-latitude outer Galaxy and revealing Loop III in polarisation}",
      journal = {\aap},
         year = 2022,
        month = jul,
       volume = {663},
          eid = {A7},
        pages = {A7},
          doi = {10.1051/0004-6361/202142244},
archivePrefix = {arXiv},
       eprint = {2203.01351},
 primaryClass = {astro-ph.GA},
       adsurl = {https://ui.adsabs.harvard.edu/abs/2022A&A...663A...7E}
}

@ARTICLE{SetaMCG2025,
       author = {{Seta}, Amit and {McClure-Griffiths}, N.~M.},
        title = "{Magnetic fields in the multiphase interstellar medium of the Milky Way: turbulent kinetic and magnetic energy density relation}",
      journal = {\mnras},
         year = 2025,
        month = may,
       volume = {539},
       number = {2},
        pages = {1024-1039},
          doi = {10.1093/mnras/staf520},
archivePrefix = {arXiv},
       eprint = {2503.23634},
 primaryClass = {astro-ph.GA},
       adsurl = {https://ui.adsabs.harvard.edu/abs/2025MNRAS.539.1024S}
}

@ARTICLE{LivingstonEA2025,
       author = {{Livingston}, J.~D. and {McClure-Griffiths}, N.~M. and {Ma}, Y.~K. and {Bustard}, C. and {Mao}, S.~A. and {Gaensler}, B.~M. and {Kaczmarek}, J.},
        title = "{Magnetic fields in the Large Magellanic Cloud and their connection to the Magellanic System}",
      journal = {\mnras},
         year = 2024,
        month = dec,
       volume = {535},
       number = {2},
        pages = {1944-1963},
          doi = {10.1093/mnras/stae2416},
       adsurl = {https://ui.adsabs.harvard.edu/abs/2024MNRAS.535.1944L}
}

@ARTICLE{ThomsonEA2023,
       author = {{Thomson}, Alec J.~M. and {McConnell}, David and {Lenc}, Emil and {Galvin}, Timothy J. and {Rudnick}, Lawrence and {Heald}, George and {Hale}, Catherine L. and {Duchesne}, Stefan W. and {Anderson}, Craig S. and {Carretti}, Ettore and {Federrath}, Christoph and {Gaensler}, B.~M. and {Harvey-Smith}, Lisa and {Haverkorn}, Marijke and {Hotan}, Aidan W. and {Ma}, Yik Ki and {Murphy}, Tara and {McClure-Griffiths}, N.~M. and {Moss}, Vanessa A. and {O'Sullivan}, Shane P. and {Raja}, Wasim and {Seta}, Amit and {Van Eck}, Cameron L. and {West}, Jennifer L. and {Whiting}, Matthew T. and {Wieringa}, Mark H.},
        title = "{The Rapid ASKAP Continuum Survey III: Spectra and Polarisation In Cutouts of Extragalactic Sources (SPICE-RACS) first data release}",
      journal = {\pasa},
         year = 2023,
        month = aug,
       volume = {40},
          eid = {e040},
        pages = {e040},
          doi = {10.1017/pasa.2023.38},
archivePrefix = {arXiv},
       eprint = {2307.07207},
 primaryClass = {astro-ph.GA},
       adsurl = {https://ui.adsabs.harvard.edu/abs/2023PASA...40...40T}
}

@ARTICLE{MathewEA2023,
       author = {{Mathew}, Sajay Sunny and {Federrath}, Christoph and {Seta}, Amit},
        title = "{The role of the turbulence driving mode for the initial mass function}",
      journal = {\mnras},
         year = 2023,
        month = feb,
       volume = {518},
       number = {4},
        pages = {5190-5214},
          doi = {10.1093/mnras/stac3415},
archivePrefix = {arXiv},
       eprint = {2208.08802},
 primaryClass = {astro-ph.GA},
       adsurl = {https://ui.adsabs.harvard.edu/abs/2023MNRAS.518.5190M}
}

@ARTICLE{SurS2024,
       author = {{Sur}, Sharanya and {Subramanian}, Kandaswamy},
        title = "{Role of magnetic pressure forces in fluctuation dynamo saturation}",
      journal = {\mnras},
         year = 2024,
        month = jan,
       volume = {527},
       number = {2},
        pages = {3968-3981},
          doi = {10.1093/mnras/stad3535},
archivePrefix = {arXiv},
       eprint = {2305.09969},
 primaryClass = {astro-ph.GA},
       adsurl = {https://ui.adsabs.harvard.edu/abs/2024MNRAS.527.3968S}
}

@ARTICLE{TuricEA2021,
       author = {{Turi{\'c}}, Luka and {Jeli{\'c}}, Vibor and {Jaspers}, Rutger and {Haverkorn}, Marijke and {Bracco}, Andrea and {Erceg}, Ana and {Ceraj}, Lana and {van Eck}, Cameron and {Zaroubi}, Saleem},
        title = "{Multi-tracer analysis of straight depolarisation canals in the surroundings of the 3C 196 field}",
      journal = {\aap},
         year = 2021,
        month = oct,
       volume = {654},
          eid = {A5},
        pages = {A5},
          doi = {10.1051/0004-6361/202141071},
archivePrefix = {arXiv},
       eprint = {2108.10679},
 primaryClass = {astro-ph.GA},
       adsurl = {https://ui.adsabs.harvard.edu/abs/2021A&A...654A...5T}
}

@ARTICLE{BeckEA2019,
       author = {{Beck}, Rainer and {Chamandy}, Luke and {Elson}, Ed and {Blackman}, Eric G.},
        title = "{Synthesizing Observations and Theory to Understand Galactic Magnetic Fields: Progress and Challenges}",
      journal = {Galaxies},
         year = 2019,
        month = dec,
       volume = {8},
       number = {1},
          eid = {4},
        pages = {4},
          doi = {10.3390/galaxies8010004},
archivePrefix = {arXiv},
       eprint = {1912.08962},
 primaryClass = {astro-ph.GA},
       adsurl = {https://ui.adsabs.harvard.edu/abs/2019Galax...8....4B}
}

@ARTICLE{VanderwoudeEA2024,
       author = {{Vanderwoude}, S. and {West}, J.~L. and {Gaensler}, B.~M. and {Rudnick}, L. and {Van Eck}, C.~L. and {Thomson}, A.~J.~M. and {Andernach}, H. and {Anderson}, C.~S. and {Carretti}, E. and {Heald}, G.~H. and {Leahy}, J.~P. and {McClure-Griffiths}, N.~M. and {O'Sullivan}, S.~P. and {Tahani}, M. and {Willis}, A.~G.},
        title = "{Prototype Faraday Rotation Measure Catalogs from the Polarisation Sky Survey of the Universe's Magnetism (POSSUM) Pilot Observations}",
      journal = {\aj},
         year = 2024,
        month = may,
       volume = {167},
       number = {5},
          eid = {226},
        pages = {226},
          doi = {10.3847/1538-3881/ad2fc8},
archivePrefix = {arXiv},
       eprint = {2403.15668},
 primaryClass = {astro-ph.GA},
       adsurl = {https://ui.adsabs.harvard.edu/abs/2024AJ....167..226V}
}

@ARTICLE{Beck2007,
       author = {{Beck}, R.},
        title = "{Magnetism in the spiral galaxy NGC 6946: magnetic arms, depolarization rings, dynamo modes, and helical fields}",
      journal = {\aap},
         year = 2007,
        month = aug,
       volume = {470},
       number = {2},
        pages = {539-556},
          doi = {10.1051/0004-6361:20066988},
archivePrefix = {arXiv},
       eprint = {0705.4163},
 primaryClass = {astro-ph},
       adsurl = {https://ui.adsabs.harvard.edu/abs/2007A&A...470..539B}
}

@ARTICLE{Schekochihin2022,
       author = {{Schekochihin}, Alexander A.},
        title = "{MHD turbulence: a biased review}",
      journal = {Journal of Plasma Physics},
         year = 2022,
        month = oct,
       volume = {88},
       number = {5},
          eid = {155880501},
        pages = {155880501},
          doi = {10.1017/S0022377822000721},
archivePrefix = {arXiv},
       eprint = {2010.00699},
 primaryClass = {physics.plasm-ph},
       adsurl = {https://ui.adsabs.harvard.edu/abs/2022JPlPh..88e1501S}
}

@ARTICLE{RuszkowskiP2023,
       author = {{Ruszkowski}, Mateusz and {Pfrommer}, Christoph},
        title = "{Cosmic ray feedback in galaxies and galaxy clusters}",
      journal = {\aapr},
         year = 2023,
        month = dec,
       volume = {31},
       number = {1},
          eid = {4},
        pages = {4},
          doi = {10.1007/s00159-023-00149-2},
archivePrefix = {arXiv},
       eprint = {2306.03141},
 primaryClass = {astro-ph.HE},
       adsurl = {https://ui.adsabs.harvard.edu/abs/2023A&ARv..31....4R}
}

@ARTICLE{BraccoEA2020,
       author = {{Bracco}, A. and {Jeli{\'c}}, V. and {Marchal}, A. and {Turi{\'c}}, L. and {Erceg}, A. and {Miville-Desch{\^e}nes}, M. -A. and {Bellomi}, E.},
        title = "{The multiphase and magnetized neutral hydrogen seen by LOFAR}",
      journal = {\aap},
         year = 2020,
        month = dec,
       volume = {644},
          eid = {L3},
        pages = {L3},
          doi = {10.1051/0004-6361/202039283},
archivePrefix = {arXiv},
       eprint = {2011.05647},
 primaryClass = {astro-ph.GA},
       adsurl = {https://ui.adsabs.harvard.edu/abs/2020A&A...644L...3B}
}

@ARTICLE{GentEA2023,
       author = {{Gent}, Frederick A. and {Mac Low}, Mordecai-Mark and {Korpi-Lagg}, Maarit J. and {Singh}, Nishant K.},
        title = "{The Small-scale Dynamo in a Multiphase Supernova-driven Medium}",
      journal = {\apj},
         year = 2023,
        month = feb,
       volume = {943},
       number = {2},
          eid = {176},
        pages = {176},
          doi = {10.3847/1538-4357/acac20},
archivePrefix = {arXiv},
       eprint = {2210.04460},
 primaryClass = {astro-ph.GA},
       adsurl = {https://ui.adsabs.harvard.edu/abs/2023ApJ...943..176G}
}

@INPROCEEDINGS{PattleEA2023,
       author = {{Pattle}, K. and {Fissel}, L. and {Tahani}, M. and {Liu}, T. and {Ntormousi}, E.},
        title = "{Magnetic Fields in Star Formation: from Clouds to Cores}",
    booktitle = {Protostars and Planets VII},
         year = 2023,
       editor = {{Inutsuka}, S. and {Aikawa}, Y. and {Muto}, T. and {Tomida}, K. and {Tamura}, M.},
       series = {Astronomical Society of the Pacific Conference Series},
       volume = {534},
        month = jul,
        pages = {193},
          doi = {10.48550/arXiv.2203.11179},
archivePrefix = {arXiv},
       eprint = {2203.11179},
 primaryClass = {astro-ph.GA},
       adsurl = {https://ui.adsabs.harvard.edu/abs/2023ASPC..534..193P}
}

@ARTICLE{PlanckXXXV_2016,
       author = {{Planck Collaboration} and {Ade}, P.~A.~R. and {Aghanim}, N. and {Alves}, M.~I.~R. and {Arnaud}, M. and {Arzoumanian}, D. and {Ashdown}, M. and {Aumont}, J. and {Baccigalupi}, C. and {Banday}, A.~J. and {Barreiro}, R.~B. and {Bartolo}, N. and {Battaner}, E. and {Benabed}, K. and {Beno{\^\i}t}, A. and {Benoit-L{\'e}vy}, A. and {Bernard}, J. -P. and {Bersanelli}, M. and {Bielewicz}, P. and {Bock}, J.~J. and {Bonavera}, L. and {Bond}, J.~R. and {Borrill}, J. and {Bouchet}, F.~R. and {Boulanger}, F. and {Bracco}, A. and {Burigana}, C. and {Calabrese}, E. and {Cardoso}, J. -F. and {Catalano}, A. and {Chiang}, H.~C. and {Christensen}, P.~R. and {Colombo}, L.~P.~L. and {Combet}, C. and {Couchot}, F. and {Crill}, B.~P. and {Curto}, A. and {Cuttaia}, F. and {Danese}, L. and {Davies}, R.~D. and {Davis}, R.~J. and {de Bernardis}, P. and {de Rosa}, A. and {de Zotti}, G. and {Delabrouille}, J. and {Dickinson}, C. and {Diego}, J.~M. and {Dole}, H. and {Donzelli}, S. and {Dor{\'e}}, O. and {Douspis}, M. and {Ducout}, A. and {Dupac}, X. and {Efstathiou}, G. and {Elsner}, F. and {En{\ss}lin}, T.~A. and {Eriksen}, H.~K. and {Falceta-Gon{\c{c}}alves}, D. and {Falgarone}, E. and {Ferri{\`e}re}, K. and {Finelli}, F. and {Forni}, O. and {Frailis}, M. and {Fraisse}, A.~A. and {Franceschi}, E. and {Frejsel}, A. and {Galeotta}, S. and {Galli}, S. and {Ganga}, K. and {Ghosh}, T. and {Giard}, M. and {Gjerl{\o}w}, E. and {Gonz{\'a}lez-Nuevo}, J. and {G{\'o}rski}, K.~M. and {Gregorio}, A. and {Gruppuso}, A. and {Gudmundsson}, J.~E. and {Guillet}, V. and {Harrison}, D.~L. and {Helou}, G. and {Hennebelle}, P. and {Henrot-Versill{\'e}}, S. and {Hern{\'a}ndez-Monteagudo}, C. and {Herranz}, D. and {Hildebrandt}, S.~R. and {Hivon}, E. and {Holmes}, W.~A. and {Hornstrup}, A. and {Huffenberger}, K.~M. and {Hurier}, G. and {Jaffe}, A.~H. and {Jaffe}, T.~R. and {Jones}, W.~C. and {Juvela}, M. and {Keih{\"a}nen}, E. and {Keskitalo}, R. and {Kisner}, T.~S. and {Knoche}, J. and {Kunz}, M. and {Kurki-Suonio}, H. and {Lagache}, G. and {Lamarre}, J. -M. and {Lasenby}, A. and {Lattanzi}, M. and {Lawrence}, C.~R. and {Leonardi}, R. and {Levrier}, F. and {Liguori}, M. and {Lilje}, P.~B. and {Linden-V{\o}rnle}, M. and {L{\'o}pez-Caniego}, M. and {Lubin}, P.~M. and {Mac{\'\i}as-P{\'e}rez}, J.~F. and {Maino}, D. and {Mandolesi}, N. and {Mangilli}, A. and {Maris}, M. and {Martin}, P.~G. and {Mart{\'\i}nez-Gonz{\'a}lez}, E. and {Masi}, S. and {Matarrese}, S. and {Melchiorri}, A. and {Mendes}, L. and {Mennella}, A. and {Migliaccio}, M. and {Miville-Desch{\^e}nes}, M. -A. and {Moneti}, A. and {Montier}, L. and {Morgante}, G. and {Mortlock}, D. and {Munshi}, D. and {Murphy}, J.~A. and {Naselsky}, P. and {Nati}, F. and {Netterfield}, C.~B. and {Noviello}, F. and {Novikov}, D. and {Novikov}, I. and {Oppermann}, N. and {Oxborrow}, C.~A. and {Pagano}, L. and {Pajot}, F. and {Paladini}, R. and {Paoletti}, D. and {Pasian}, F. and {Perotto}, L. and {Pettorino}, V. and {Piacentini}, F. and {Piat}, M. and {Pierpaoli}, E. and {Pietrobon}, D. and {Plaszczynski}, S. and {Pointecouteau}, E. and {Polenta}, G. and {Ponthieu}, N. and {Pratt}, G.~W. and {Prunet}, S. and {Puget}, J. -L. and {Rachen}, J.~P. and {Reinecke}, M. and {Remazeilles}, M. and {Renault}, C. and {Renzi}, A. and {Ristorcelli}, I. and {Rocha}, G. and {Rossetti}, M. and {Roudier}, G. and {Rubi{\~n}o-Mart{\'\i}n}, J.~A. and {Rusholme}, B. and {Sandri}, M. and {Santos}, D. and {Savelainen}, M. and {Savini}, G. and {Scott}, D. and {Soler}, J.~D. and {Stolyarov}, V. and {Sudiwala}, R. and {Sutton}, D. and {Suur-Uski}, A. -S. and {Sygnet}, J. -F. and {Tauber}, J.~A. and {Terenzi}, L. and {Toffolatti}, L. and {Tomasi}, M. and {Tristram}, M. and {Tucci}, M. and {Umana}, G. and {Valenziano}, L. and {Valiviita}, J. and {Van Tent}, B. and {Vielva}, P. and {Villa}, F. and {Wade}, L.~A. and {Wandelt}, B.~D. and {Wehus}, I.~K. and {Ysard}, N. and {Yvon}, D. and {Zonca}, A.},
        title = "{Planck intermediate results. XXXV. Probing the role of the magnetic field in the formation of structure in molecular clouds}",
      journal = {\aap},
         year = 2016,
        month = feb,
       volume = {586},
          eid = {A138},
        pages = {A138},
          doi = {10.1051/0004-6361/201525896},
archivePrefix = {arXiv},
       eprint = {1502.04123},
 primaryClass = {astro-ph.GA},
       adsurl = {https://ui.adsabs.harvard.edu/abs/2016A&A...586A.138P}
}

@ARTICLE{HeesenEA2023,
       author = {{Heesen}, V. and {O'Sullivan}, S.~P. and {Br{\"u}ggen}, M. and {Basu}, A. and {Beck}, R. and {Seta}, A. and {Carretti}, E. and {Krause}, M.~G.~H. and {Haverkorn}, M. and {Hutschenreuter}, S. and {Bracco}, A. and {Stein}, M. and {Bomans}, D.~J. and {Dettmar}, R. -J. and {Chy{\.z}y}, K.~T. and {Heald}, G.~H. and {Paladino}, R. and {Horellou}, C.},
        title = "{Detection of magnetic fields in the circumgalactic medium of nearby galaxies using Faraday rotation}",
      journal = {\aap},
         year = 2023,
        month = feb,
       volume = {670},
          eid = {L23},
        pages = {L23},
          doi = {10.1051/0004-6361/202346008},
archivePrefix = {arXiv},
       eprint = {2302.06617},
 primaryClass = {astro-ph.GA},
       adsurl = {https://ui.adsabs.harvard.edu/abs/2023A&A...670L..23H}
}

@ARTICLE{MaEA2020,
       author = {{Ma}, Y.~K. and {Mao}, S.~A. and {Ordog}, A. and {Brown}, J.~C.},
        title = "{The complex large-scale magnetic fields in the first Galactic quadrant as revealed by the Faraday depth profile disparity}",
      journal = {\mnras},
         year = 2020,
        month = sep,
       volume = {497},
       number = {3},
        pages = {3097-3117},
          doi = {10.1093/mnras/staa2105},
archivePrefix = {arXiv},
       eprint = {2007.07893},
 primaryClass = {astro-ph.GA},
       adsurl = {https://ui.adsabs.harvard.edu/abs/2020MNRAS.497.3097M}
}

@ARTICLE{MaEA2023,
       author = {{Ma}, Y.~K. and {McClure-Griffiths}, N.~M. and {Clark}, S.~E. and {Gibson}, S.~J. and {van Loon}, J. Th and {Soler}, J.~D. and {Putman}, M.~E. and {Dickey}, J.~M. and {Lee}, M. -Y. and {Jameson}, K.~E. and {Uscanga}, L. and {Dempsey}, J. and {D{\'e}nes}, H. and {Lynn}, C. and {Pingel}, N.~M.},
        title = "{H I filaments as potential compass needles? Comparing the magnetic field structure of the Small Magellanic Cloud to the orientation of GASKAP-H I filaments}",
      journal = {\mnras},
         year = 2023,
        month = may,
       volume = {521},
       number = {1},
        pages = {60-83},
          doi = {10.1093/mnras/stad462},
archivePrefix = {arXiv},
       eprint = {2302.04880},
 primaryClass = {astro-ph.GA},
       adsurl = {https://ui.adsabs.harvard.edu/abs/2023MNRAS.521...60M}
}

@ARTICLE{FerriereEA2021,
       author = {{Ferri{\`e}re}, K. and {West}, J.~L. and {Jaffe}, T.~R.},
        title = "{The correct sense of Faraday rotation}",
      journal = {\mnras},
         year = 2021,
        month = nov,
       volume = {507},
       number = {4},
        pages = {4968-4982},
          doi = {10.1093/mnras/stab1641},
archivePrefix = {arXiv},
       eprint = {2106.03074},
 primaryClass = {astro-ph.GA},
       adsurl = {https://ui.adsabs.harvard.edu/abs/2021MNRAS.507.4968F}
}

@ARTICLE{GentEA2013I,
       author = {{Gent}, F.~A. and {Shukurov}, A. and {Fletcher}, A. and {Sarson}, G.~R. and {Mantere}, M.~J.},
        title = "{The supernova-regulated ISM - I. The multiphase structure}",
      journal = {\mnras},
         year = 2013,
        month = jun,
       volume = {432},
       number = {2},
        pages = {1396-1423},
          doi = {10.1093/mnras/stt560},
archivePrefix = {arXiv},
       eprint = {1204.3567},
 primaryClass = {astro-ph.GA},
       adsurl = {https://ui.adsabs.harvard.edu/abs/2013MNRAS.432.1396G}
}

@ARTICLE{SetaEA2023,
       author = {{Seta}, Amit and {Federrath}, Christoph and {Livingston}, Jack D. and {McClure-Griffiths}, N.~M.},
        title = "{Rotation measure structure functions with higher-order stencils as a probe of small-scale magnetic fluctuations and its application to the Small and Large Magellanic Clouds}",
      journal = {\mnras},
         year = 2023,
        month = jan,
       volume = {518},
       number = {1},
        pages = {919-944},
          doi = {10.1093/mnras/stac2972},
archivePrefix = {arXiv},
       eprint = {2206.13798},
 primaryClass = {astro-ph.GA},
       adsurl = {https://ui.adsabs.harvard.edu/abs/2023MNRAS.518..919S}
}

@BOOK{ShukurovS2021, 
  place={Cambridge}, 
  series={Cambridge Astrophysics}, 
  title={Astrophysical Magnetic Fields: From Galaxies to the Early Universe}, 
  publisher={Cambridge University Press}, 
  author={Shukurov, Anvar and Subramanian, Kandaswamy}, 
  year={2021}, 
  collection={Cambridge Astrophysics}}

@ARTICLE{KrumholzEA2018,
       author = {{Krumholz}, Mark R. and {Burkhart}, Blakesley and {Forbes}, John C. and {Crocker}, Roland M.},
        title = "{A unified model for galactic discs: star formation, turbulence driving, and mass transport}",
      journal = {\mnras},
         year = 2018,
        month = jun,
       volume = {477},
       number = {2},
        pages = {2716-2740},
          doi = {10.1093/mnras/sty852},
archivePrefix = {arXiv},
       eprint = {1706.00106},
 primaryClass = {astro-ph.GA},
       adsurl = {https://ui.adsabs.harvard.edu/abs/2018MNRAS.477.2716K}
}

@ARTICLE{AlvesEA2018,
       author = {{Alves}, M.~I.~R. and {Boulanger}, F. and {Ferri{\`e}re}, K. and {Montier}, L.},
        title = "{The Local Bubble: a magnetic veil to our Galaxy}",
      journal = {\aap},
         year = 2018,
        month = mar,
       volume = {611},
          eid = {L5},
        pages = {L5},
          doi = {10.1051/0004-6361/201832637},
archivePrefix = {arXiv},
       eprint = {1803.05251},
 primaryClass = {astro-ph.GA},
       adsurl = {https://ui.adsabs.harvard.edu/abs/2018A&A...611L...5A}
}

@ARTICLE{BeckK2005,
   author = {{Beck}, R. and {Krause}, M.},
    title = "{Revised equipartition and minimum energy formula for magnetic field strength estimates from radio synchrotron observations}",
  journal = {\an},
   eprint = {astro-ph/0507367},
     year = 2005,
    month = jul,
   volume = 326,
    pages = {414-427},
      doi = {10.1002/asna.200510366},
   adsurl = {http://adsabs.harvard.edu/abs/2005AN....326..414B}
}

@ARTICLE{BeckEA1996,
   author = {{Beck}, R. and {Brandenburg}, A. and {Moss}, D. and {Shukurov}, A. and 
  {Sokoloff}, D.},
    title = "{Galactic Magnetism: Recent Developments and Perspectives}",
  journal = {\araa},
     year = 1996,
   volume = 34,
    pages = {155-206},
      doi = {10.1146/annurev.astro.34.1.155},
   adsurl = {http://adsabs.harvard.edu/abs/1996ARA%26A..34..155B}
}

@ARTICLE{Beck2015,
       author = {{Beck}, Rainer},
        title = "{Magnetic fields in the nearby spiral galaxy IC 342: A multi-frequency radio polarization study}",
      journal = {\aap},
         year = 2015,
        month = jun,
       volume = {578},
          eid = {A93},
        pages = {A93},
          doi = {10.1051/0004-6361/201425572},
archivePrefix = {arXiv},
       eprint = {1502.05439},
 primaryClass = {astro-ph.GA},
       adsurl = {https://ui.adsabs.harvard.edu/abs/2015A&A...578A..93B}
}

@ARTICLE{Beck2016,
   author = {{Beck}, R.},
    title = "{Magnetic fields in spiral galaxies}",
  journal = {\araa},
archivePrefix = "arXiv",
   eprint = {1509.04522},
     year = 2016,
    month = dec,
   volume = 24,
      eid = {4},
    pages = {4},
   adsurl = {http://adsabs.harvard.edu/abs/2016A%26ARv..24....4B}
}

@ARTICLE{BrandenburgS2005,
   author = {{Brandenburg}, A. and {Subramanian}, K.},
    title = "{Astrophysical magnetic fields and nonlinear dynamo theory}",
  journal = {\physrep},
     year = 2005,
    month = oct,
   volume = 417,
    pages = {1-209},
      doi = {10.1016/j.physrep.2005.06.005},
   adsurl = {http://adsabs.harvard.edu/abs/2005PhR...417....1B}
}

@ARTICLE{BrentjensB2005,
       author = {{Brentjens}, M.~A. and {de Bruyn}, A.~G.},
        title = "{Faraday rotation measure synthesis}",
      journal = {\aap},
         year = 2005,
        month = Oct,
       volume = {441},
        pages = {1217-1228},
          doi = {10.1051/0004-6361:20052990},
       adsurl = {https://ui.adsabs.harvard.edu/#abs/2005A&A...441.1217B}
}

@ARTICLE{Burkhart2012,
   author = {{Burkhart}, B. and {Lazarian}, A. and {Gaensler}, B.~M.},
    title = "{Properties of Interstellar Turbulence from Gradients of Linear Polarization Maps}",
  journal = {\apj},
archivePrefix = "arXiv",
   eprint = {1111.3544},
     year = 2012,
    month = apr,
   volume = 749,
      eid = {145},
    pages = {145},
      doi = {10.1088/0004-637X/749/2/145},
   adsurl = {http://adsabs.harvard.edu/abs/2012ApJ...749..145B}
}

@ARTICLE{Burn1966,
       author = {{Burn}, B.~J.},
        title = "{On the depolarization of discrete radio sources by Faraday dispersion}",
      journal = {\mnras},
         year = 1966,
        month = Jan,
       volume = {133},
        pages = {67},
          doi = {10.1093/mnras/133.1.67},
       adsurl = {https://ui.adsabs.harvard.edu/#abs/1966MNRAS.133...67B}
}

@ARTICLE{Chandran2000,
   author = {{Chandran}, B.~D.~G.},
    title = "{Confinement and Isotropization of Galactic Cosmic Rays by Molecular-Cloud Magnetic Mirrors When Turbulent Scattering Is Weak}",
  journal = {\apj},
     year = 2000,
    month = jan,
   volume = 529,
    pages = {513-535},
      doi = {10.1086/308232},
   adsurl = {http://adsabs.harvard.edu/abs/2000ApJ...529..513C}
}

@ARTICLE{CrutcherEA2010,
       author = {{Crutcher}, Richard M. and {Wandelt}, Benjamin and {Heiles}, Carl and
         {Falgarone}, Edith and {Troland}, Thomas H.},
        title = "{Magnetic Fields in Interstellar Clouds from Zeeman Observations: Inference of Total Field Strengths by Bayesian Analysis}",
      journal = {\apj},
         year = 2010,
        month = dec,
       volume = {725},
       number = {1},
        pages = {466-479},
          doi = {10.1088/0004-637X/725/1/466},
       adsurl = {https://ui.adsabs.harvard.edu/abs/2010ApJ...725..466C}
}

@ARTICLE{FletcherEA2011,
       author = {{Fletcher}, A. and {Beck}, R. and {Shukurov}, A. and {Berkhuijsen},
        E.~M. and {Horellou}, C.},
        title = "{Magnetic fields and spiral arms in the galaxy M51}",
      journal = {\mnras},
         year = 2011,
        month = Apr,
       volume = {412},
        pages = {2396-2416},
          doi = {10.1111/j.1365-2966.2010.18065.x},
       adsurl = {https://ui.adsabs.harvard.edu/#abs/2011MNRAS.412.2396F}
}

@ARTICLE{GaenslerEA2005,
   author = {{Gaensler}, B.~M. and {Haverkorn}, M. and {Staveley-Smith}, L. and 
  {Dickey}, J.~M. and {McClure-Griffiths}, N.~M. and {Dickel}, J.~R. and 
  {Wolleben}, M.},
    title = "{The Magnetic Field of the Large Magellanic Cloud Revealed Through Faraday Rotation}",
  journal = {Science},
   eprint = {astro-ph/0503226},
     year = 2005,
    month = mar,
   volume = 307,
    pages = {1610-1612},
      doi = {10.1126/science.1108832},
   adsurl = {http://adsabs.harvard.edu/abs/2005Sci...307.1610G}
}

@ARTICLE{GaenslarEA2011,
   author = {{Gaensler}, B.~M. and {Haverkorn}, M. and {Burkhart}, B. and 
  {Newton-McGee}, K.~J. and {Ekers}, R.~D. and {Lazarian}, A. and 
  {McClure-Griffiths}, N.~M. and {Robishaw}, T. and {Dickey}, J.~M. and 
  {Green}, A.~J.},
    title = "{Low-Mach-number turbulence in interstellar gas revealed by radio polarization gradients}",
  journal = {\nat},
archivePrefix = "arXiv",
   eprint = {1110.2896},
 primaryClass = "astro-ph.GA",
     year = 2011,
    month = oct,
   volume = 478,
    pages = {214-217},
      doi = {10.1038/nature10446},
   adsurl = {http://adsabs.harvard.edu/abs/2011Natur.478..214G}
}

@ARTICLE{Garrington1988,
       author = {{Garrington}, S.~T. and {Leahy}, J.~P. and {Conway}, R.~G. and {Laing},
        R.~A.},
        title = "{A systematic asymmetry in the polarization properties of double radio
        sources with one jet}",
      journal = {\nat},
         year = 1988,
        month = Jan,
       volume = {331},
        pages = {147-149},
          doi = {10.1038/331147a0},
       adsurl = {https://ui.adsabs.harvard.edu/#abs/1988Natur.331..147G}
}

@ARTICLE{HaugenEA2004,
   author = {{Haugen}, N.~E. and {Brandenburg}, A. and {Dobler}, W.},
    title = "{Simulations of nonhelical hydromagnetic turbulence}",
  journal = {\pre},
   eprint = {astro-ph/0307059},
     year = 2004,
    month = jul,
   volume = 70,
   number = 1,
      eid = {016308},
    pages = {016308},
      doi = {10.1103/PhysRevE.70.016308},
   adsurl = {http://adsabs.harvard.edu/abs/2004PhRvE..70a6308H}
}

@ARTICLE{Han2017,
       author = {{Han}, J.~L.},
        title = "{Observing Interstellar and Intergalactic Magnetic Fields}",
      journal = {\araa},
         year = 2017,
        month = aug,
       volume = {55},
       number = {1},
        pages = {111-157},
          doi = {10.1146/annurev-astro-091916-055221},
       adsurl = {https://ui.adsabs.harvard.edu/abs/2017ARA&A..55..111H}
}

@ARTICLE{HaverkornEA2008,
   author = {{Haverkorn}, M. and {Brown}, J.~C. and {Gaensler}, B.~M. and 
  {McClure-Griffiths}, N.~M.},
    title = "{The Outer Scale of Turbulence in the Magnetoionized Galactic Interstellar Medium}",
  journal = {\apj},
archivePrefix = "arXiv",
   eprint = {0802.2740},
     year = 2008,
    month = jun,
   volume = 680,
      eid = {362-370},
    pages = {362-370},
      doi = {10.1086/587165},
   adsurl = {http://adsabs.harvard.edu/abs/2008ApJ...680..362H}
}

@INPROCEEDINGS{Haverkorn2015,
   author = {{Haverkorn}, M.},
    title = "{Magnetic Fields in the Milky Way}",
booktitle = {Magnetic Fields in Diffuse Media},
     year = 2015,
   series = {Astrophysics and Space Science Library},
   volume = 407,
archivePrefix = "arXiv",
   eprint = {1406.0283},
   editor = {{Lazarian}, A. and {de Gouveia Dal Pino}, E.~M. and {Melioli}, C.
  },
    pages = {483},
      doi = {10.1007/978-3-662-44625-6_17},
   adsurl = {http://adsabs.harvard.edu/abs/2015ASSL..407..483H}
}

@ARTICLE{Kazantsev1968,
   author = {{Kazantsev}, A.~P.},
    title = "{Enhancement of a Magnetic Field by a Conducting Fluid}",
  journal = {Soviet Journal of Experimental and Theoretical Physics},
     year = 1968,
    month = may,
   volume = 26,
    pages = {1031},
   adsurl = {http://adsabs.harvard.edu/abs/1968JETP...26.1031K}
}

@ARTICLE{Laing1988,
       author = {{Laing}, R.~A.},
        title = "{The sidedness of jets and depolarization in powerful extragalactic radio
        sources}",
      journal = {\nat},
         year = 1988,
        month = Jan,
       volume = {331},
        pages = {149-151},
          doi = {10.1038/331149a0},
       adsurl = {https://ui.adsabs.harvard.edu/#abs/1988Natur.331..149L}
}

@ARTICLE{LivingstonEA2021,
       author = {{Livingston}, J.~D. and {McClure-Griffiths}, N.~M. and {Gaensler}, B.~M. and {Seta}, A. and {Alger}, M.~J.},
        title = "{Heightened Faraday complexity in the inner 1 kpc of the galactic centre}",
      journal = {\mnras},
         year = 2021,
        month = apr,
       volume = {502},
       number = {3},
        pages = {3814-3828},
          doi = {10.1093/mnras/stab253},
archivePrefix = {arXiv},
       eprint = {2102.01139},
 primaryClass = {astro-ph.GA},
       adsurl = {https://ui.adsabs.harvard.edu/abs/2021MNRAS.502.3814L}
}

@ARTICLE{LivingstonEA2022,
       author = {{Livingston}, J.~D. and {McClure-Griffiths}, N.~M. and {Mao}, S.~A. and {Ma}, Y.~K. and {Gaensler}, B.~M. and {Heald}, G. and {Seta}, A.},
        title = "{A radio polarization study of magnetic fields in the Small Magellanic Cloud}",
      journal = {\mnras},
         year = 2022,
        month = feb,
       volume = {510},
       number = {1},
        pages = {260-275},
          doi = {10.1093/mnras/stab3375},
archivePrefix = {arXiv},
       eprint = {2112.04044},
 primaryClass = {astro-ph.GA},
       adsurl = {https://ui.adsabs.harvard.edu/abs/2022MNRAS.510..260L}
}

@ARTICLE{MinterS1996,
   author = {{Minter}, A.~H. and {Spangler}, S.~R.},
    title = "{Observation of Turbulent Fluctuations in the Interstellar Plasma Density and Magnetic Field on Spatial Scales of 0.01 to 100 Parsecs}",
  journal = {\apj},
     year = 1996,
    month = feb,
   volume = 458,
    pages = {194},
      doi = {10.1086/176803},
   adsurl = {http://adsabs.harvard.edu/abs/1996ApJ...458..194M}
}

@ARTICLE{MaoEA2008,
       author = {{Mao}, S.~A. and {Gaensler}, B.~M. and {Stanimirovi{\'c}}, S. and
        {Haverkorn}, M. and {McClure-Griffiths}, N.~M. and {Staveley-
        Smith}, L. and {Dickey}, J.~M.},
        title = "{A Radio and Optical Polarization Study of the Magnetic Field in the
        Small Magellanic Cloud}",
      journal = {\apj},
         year = 2008,
        month = Dec,
       volume = {688},
        pages = {1029-1049},
          doi = {10.1086/590546},
       adsurl = {https://ui.adsabs.harvard.edu/#abs/2008ApJ...688.1029M}
}

@ARTICLE{MaoEA2015,
       author = {{Mao}, S.~A. and {Zweibel}, E. and {Fletcher}, A. and {Ott}, J. and
         {Tabatabaei}, F.},
        title = "{Properties of the Magneto-ionic Medium in the Halo of M51 Revealed by Wide-band Polarimetry}",
      journal = {\apj},
         year = 2015,
        month = feb,
       volume = {800},
       number = {2},
          eid = {92},
        pages = {92},
          doi = {10.1088/0004-637X/800/2/92},
archivePrefix = {arXiv},
       eprint = {1412.8320},
 primaryClass = {astro-ph.GA},
       adsurl = {https://ui.adsabs.harvard.edu/abs/2015ApJ...800...92M}
}

@ARTICLE{MaoEA2017,
       author = {{Mao}, S.~A. and {Carilli}, C. and {Gaensler}, B.~M. and {Wucknitz}, O.
        and {Keeton}, C. and {Basu}, A. and {Beck}, R. and {Kronberg},
        P.~P. and {Zweibel}, E.},
        title = "{Detection of microgauss coherent magnetic fields in a galaxy five
        billion years ago}",
      journal = {Nature Astronomy},
         year = 2017,
        month = Sep,
       volume = {1},
        pages = {621-626},
          doi = {10.1038/s41550-017-0218-x},
       adsurl = {https://ui.adsabs.harvard.edu/#abs/2017NatAs...1..621M}
}

@ARTICLE{MossS1996,
   author = {{Moss}, D. and {Shukurov}, A.},
    title = "{Turbulence and magnetic fields in elliptical galaxies.}",
  journal = {\mnras},
     year = 1996,
    month = mar,
   volume = 279,
    pages = {229-239},
      doi = {10.1093/mnras/279.1.229},
   adsurl = {http://adsabs.harvard.edu/abs/1996MNRAS.279..229M}
}

@ARTICLE{Planck2016fil,
   author = {{Planck Collaboration\vspace{0cm}} and {Ade}, P.~A.~R. and {Aghanim}, N. and 
  {Alves}, M.~I.~R. and {Arnaud}, M. and {Arzoumanian}, D. and 
  {Aumont}, J. and {Baccigalupi}, C. and {Banday}, A.~J. and {Barreiro}, R.~B. and 
  {Bartolo}, N. and {Battaner}, E. and {Benabed}, K. and {Benoit-L{\'e}vy}, A. and 
  {Bernard}, J.-P. and {Bern{\'e}}, O. and {Bersanelli}, M. and 
  {Bielewicz}, P. and {Bonaldi}, A. and {Bonavera}, L. and {Bond}, J.~R. and 
  {Borrill}, J. and {Bouchet}, F.~R. and {Boulanger}, F. and {Bracco}, A. and 
  {Burigana}, C. and {Calabrese}, E. and {Cardoso}, J.-F. and 
  {Catalano}, A. and {Chamballu}, A. and {Chiang}, H.~C. and {Christensen}, P.~R. and 
  {Clements}, D.~L. and {Colombi}, S. and {Colombo}, L.~P.~L. and 
  {Combet}, C. and {Couchot}, F. and {Crill}, B.~P. and {Curto}, A. and 
  {Cuttaia}, F. and {Danese}, L. and {Davies}, R.~D. and {Davis}, R.~J. and 
  {de Bernardis}, P. and {de Rosa}, A. and {de Zotti}, G. and 
  {Delabrouille}, J. and {Dickinson}, C. and {Diego}, J.~M. and 
  {Donzelli}, S. and {Dor{\'e}}, O. and {Douspis}, M. and {Ducout}, A. and 
  {Dupac}, X. and {Elsner}, F. and {En{\ss}lin}, T.~A. and {Eriksen}, H.~K. and 
  {Falgarone}, E. and {Ferri{\`e}re}, K. and {Finelli}, F. and 
  {Forni}, O. and {Frailis}, M. and {Fraisse}, A.~A. and {Franceschi}, E. and 
  {Frejsel}, A. and {Galeotta}, S. and {Galli}, S. and {Ganga}, K. and 
  {Ghosh}, T. and {Giard}, M. and {Giraud-H{\'e}raud}, Y. and 
  {Gjerl{\o}w}, E. and {Gonz{\'a}lez-Nuevo}, J. and {G{\'o}rski}, K.~M. and 
  {Gregorio}, A. and {Gruppuso}, A. and {Guillet}, V. and {Hansen}, F.~K. and 
  {Hanson}, D. and {Harrison}, D.~L. and {Hern{\'a}ndez-Monteagudo}, C. and 
  {Herranz}, D. and {Hildebrandt}, S.~R. and {Hivon}, E. and {Hobson}, M. and 
  {Holmes}, W.~A. and {Huffenberger}, K.~M. and {Hurier}, G. and 
  {Jaffe}, A.~H. and {Jaffe}, T.~R. and {Jones}, W.~C. and {Juvela}, M. and 
  {Keskitalo}, R. and {Kisner}, T.~S. and {Knoche}, J. and {Kunz}, M. and 
  {Kurki-Suonio}, H. and {Lagache}, G. and {Lamarre}, J.-M. and 
  {Lasenby}, A. and {Lawrence}, C.~R. and {Leonardi}, R. and {Levrier}, F. and 
  {Liguori}, M. and {Lilje}, P.~B. and {Linden-V{\o}rnle}, M. and 
  {L{\'o}pez-Caniego}, M. and {Lubin}, P.~M. and {Mac{\'{\i}}as-P{\'e}rez}, J.~F. and 
  {Maffei}, B. and {Mandolesi}, N. and {Mangilli}, A. and {Maris}, M. and 
  {Martin}, P.~G. and {Mart{\'{\i}}nez-Gonz{\'a}lez}, E. and {Masi}, S. and 
  {Matarrese}, S. and {Mazzotta}, P. and {Melchiorri}, A. and 
  {Mendes}, L. and {Mennella}, A. and {Migliaccio}, M. and {Mitra}, S. and 
  {Miville-Desch{\^e}nes}, M.-A. and {Moneti}, A. and {Montier}, L. and 
  {Morgante}, G. and {Mortlock}, D. and {Munshi}, D. and {Murphy}, J.~A. and 
  {Naselsky}, P. and {Nati}, F. and {Natoli}, P. and {N{\o}rgaard-Nielsen}, H.~U. and 
  {Noviello}, F. and {Novikov}, D. and {Novikov}, I. and {Oppermann}, N. and 
  {Pagano}, L. and {Pajot}, F. and {Paladini}, R. and {Paoletti}, D. and 
  {Pasian}, F. and {Perrotta}, F. and {Pettorino}, V. and {Piacentini}, F. and 
  {Piat}, M. and {Pierpaoli}, E. and {Pietrobon}, D. and {Plaszczynski}, S. and 
  {Pointecouteau}, E. and {Polenta}, G. and {Pratt}, G.~W. and 
  {Puget}, J.-L. and {Rachen}, J.~P. and {Rebolo}, R. and {Reinecke}, M. and 
  {Remazeilles}, M. and {Renault}, C. and {Renzi}, A. and {Ricciardi}, S. and 
  {Ristorcelli}, I. and {Rocha}, G. and {Rosset}, C. and {Rossetti}, M. and 
  {Roudier}, G. and {Rubi{\~n}o-Mart{\'{\i}}n}, J.~A. and {Rusholme}, B. and 
  {Sandri}, M. and {Savelainen}, M. and {Savini}, G. and {Scott}, D. and 
  {Soler}, J.~D. and {Stolyarov}, V. and {Sutton}, D. and {Suur-Uski}, A.-S. and 
  {Sygnet}, J.-F. and {Tauber}, J.~A. and {Terenzi}, L. and {Toffolatti}, L. and 
  {Tomasi}, M. and {Tristram}, M. and {Tucci}, M. and {Valenziano}, L. and 
  {Valiviita}, J. and {Van Tent}, B. and {Vielva}, P. and {Villa}, F. and 
  {Wade}, L.~A. and {Wandelt}, B.~D. and {Yvon}, D. and {Zacchei}, A. and 
  {Zonca}, A.},
    title = "{Planck intermediate results. XXXIII. Signature of the magnetic field geometry of interstellar filaments in dust polarization maps}",
  journal = {\aap},
archivePrefix = "arXiv",
   eprint = {1411.2271},
     year = 2016,
    month = feb,
   volume = 586,
      eid = {A136},
    pages = {A136},
      doi = {10.1051/0004-6361/201425305},
   adsurl = {http://adsabs.harvard.edu/abs/2016A%26A...586A.136P}
}

@ARTICLE{KierdorfEA2020,
       author = {{Kierdorf}, M. and {Mao}, S.~A. and {Beck}, R. and {Basu}, A. and {Fletcher}, A. and {Horellou}, C. and {Tabatabaei}, F. and {Ott}, J. and {Haverkorn}, M.},
        title = "{The magnetized disk-halo transition region of M 51}",
      journal = {\aap},
         year = 2020,
        month = oct,
       volume = {642},
          eid = {A118},
        pages = {A118},
          doi = {10.1051/0004-6361/202037847},
archivePrefix = {arXiv},
       eprint = {2007.00702},
 primaryClass = {astro-ph.GA},
       adsurl = {https://ui.adsabs.harvard.edu/abs/2020A&A...642A.118K}
}

@ARTICLE{Rincon2019,
       author = {{Rincon}, Fran{\c{c}}ois},
        title = "{Dynamo theories}",
      journal = {Journal of Plasma Physics},
         year = "2019",
        month = "Aug",
       volume = {85},
       number = {4},
          eid = {205850401},
        pages = {205850401},
          doi = {10.1017/S0022377819000539},
archivePrefix = {arXiv},
       eprint = {1903.07829},
 primaryClass = {physics.plasm-ph},
       adsurl = {https://ui.adsabs.harvard.edu/abs/2019JPlPh..85d2001R}
}

@PROCEEDINGS{RuzmaikinEA1988,
    title = "{Magnetic fields of galaxies}",
booktitle = {Astrophysics and Space Science Library},
     year = 1988,
   series = {Astrophysics and Space Science Library},
   volume = 133,
   editor = {{Ruzmaikin}, A.~A. and {Sokoloff}, D.~D. and {Shukurov}, A.~M.
  },
      doi = {10.1007/978-94-009-2835-0},
   adsurl = {http://adsabs.harvard.edu/abs/1988ASSL..133.....R}
}

@BOOK{RL79,
       author = {{Rybicki}, George B. and {Lightman}, Alan P.},
        title = "{Radiative processes in astrophysics}",
    booktitle = {A Wiley-Interscience Publication},
         year = 1979,
       adsurl = {https://ui.adsabs.harvard.edu/#abs/1979rpa..book.....R}
}

@ARTICLE{SchekochihinEA02,
       author = {{Schekochihin}, A.~A. and {Cowley}, S.~C. and {Hammett}, G.~W. and
        {Maron}, J.~L. and {McWilliams}, J.~C.},
        title = "{A model of nonlinear evolution and saturation of the turbulent MHD
        dynamo}",
      journal = {New Journal of Physics},
         year = 2002,
        month = Oct,
       volume = {4},
        pages = {84},
          doi = {10.1088/1367-2630/4/1/384},
       adsurl = {https://ui.adsabs.harvard.edu/#abs/2002NJPh....4...84S}
}

@ARTICLE{SchekochihinEA2004,
   author = {{Schekochihin}, A.~A. and {Cowley}, S.~C. and {Taylor}, S.~F. and 
  {Maron}, J.~L. and {McWilliams}, J.~C.},
    title = "{Simulations of the Small-Scale Turbulent Dynamo}",
  journal = {\apj},
   eprint = {astro-ph/0312046},
     year = 2004,
    month = sep,
   volume = 612,
    pages = {276-307},
      doi = {10.1086/422547},
   adsurl = {http://adsabs.harvard.edu/abs/2004ApJ...612..276S}
}

@ARTICLE{Subramanian1998,
       author = {{Subramanian}, Kandaswamy},
        title = "{Can the turbulent galactic dynamo generate large-scale magnetic fields?}",
      journal = {\mnras},
         year = 1998,
        month = mar,
       volume = {294},
        pages = {718-728},
          doi = {10.1046/j.1365-8711.1998.01284.x10.1111/j.1365-8711.1998.01284.x},
archivePrefix = {arXiv},
       eprint = {astro-ph/9707280},
 primaryClass = {astro-ph},
       adsurl = {https://ui.adsabs.harvard.edu/abs/1998MNRAS.294..718S}
}

@ARTICLE{Subramanian2016,
       author = {{Subramanian}, Kandaswamy},
        title = "{The origin, evolution and signatures of primordial magnetic fields}",
      journal = {Reports on Progress in Physics},
         year = 2016,
        month = jul,
       volume = {79},
       number = {7},
          eid = {076901},
        pages = {076901},
          doi = {10.1088/0034-4885/79/7/076901},
archivePrefix = {arXiv},
       eprint = {1504.02311},
 primaryClass = {astro-ph.CO},
       adsurl = {https://ui.adsabs.harvard.edu/abs/2016RPPh...79g6901S}
}

@ARTICLE{SetaEA2018,
   author = {{Seta}, A. and {Shukurov}, A. and {Wood}, T.~S. and {Bushby}, P.~J. and 
  {Snodin}, A.~P.},
    title = "{Relative distribution of cosmic rays and magnetic fields}",
  journal = {\mnras},
archivePrefix = "arXiv",
   eprint = {1708.07499},
     year = 2018,
    month = feb,
   volume = 473,
    pages = {4544-4557},
      doi = {10.1093/mnras/stx2606},
   adsurl = {http://adsabs.harvard.edu/abs/2018MNRAS.473.4544S}
}

@ARTICLE{SetaB2019,
       author = {{Seta}, Amit and {Beck}, Rainer},
        title = "{Revisiting the Equipartition Assumption in Star-Forming Galaxies}",
      journal = {Galaxies},
         year = "2019",
        month = "Apr",
       volume = {7},
       number = {2},
        pages = {45},
          doi = {10.3390/galaxies7020045},
archivePrefix = {arXiv},
       eprint = {1903.11856},
 primaryClass = {astro-ph.GA},
       adsurl = {https://ui.adsabs.harvard.edu/abs/2019Galax...7...45S}
}

@article{SetaEA2020,
  title = {Saturation mechanism of the fluctuation dynamo at ${\mathrm{Pr}}_{M} \ensuremath{\ge} 1$},
  author = {Seta, Amit and Bushby, Paul J. and Shukurov, Anvar and Wood, Toby S.},
  journal = {Phys. Rev. Fluids},
  volume = {5},
  issue = {4},
  pages = {043702},
  numpages = {23},
  year = {2020},
  month = {Apr},
  publisher = {American Physical Society},
  doi = {10.1103/PhysRevFluids.5.043702},
  url = {https://link.aps.org/doi/10.1103/PhysRevFluids.5.043702}
}

@ARTICLE{SetaF2020,
       author = {{Seta}, Amit and {Federrath}, Christoph},
        title = "{Seed magnetic fields in turbulent small-scale dynamos}",
      journal = {\mnras},
         year = 2020,
        month = sep,
       volume = {499},
       number = {2},
        pages = {2076-2086},
          doi = {10.1093/mnras/staa2978},
archivePrefix = {arXiv},
       eprint = {2009.12024},
 primaryClass = {astro-ph.GA},
       adsurl = {https://ui.adsabs.harvard.edu/abs/2020MNRAS.499.2076S}
}

@ARTICLE{SetaEA2021,
       author = {{Seta}, Amit and {Rodrigues}, Luiz Felippe S. and {Federrath}, Christoph and {Hales}, Christopher A.},
        title = "{Magnetic Fields in Elliptical Galaxies: An Observational Probe of the Fluctuation Dynamo Action}",
      journal = {\apj},
         year = 2021,
        month = jan,
       volume = {907},
       number = {1},
          eid = {2},
        pages = {2},
          doi = {10.3847/1538-4357/abd2bb},
archivePrefix = {arXiv},
       eprint = {2012.02329},
 primaryClass = {astro-ph.GA},
       adsurl = {https://ui.adsabs.harvard.edu/abs/2021ApJ...907....2S}
}

@ARTICLE{SetaF2022,
       author = {{Seta}, Amit and {Federrath}, Christoph},
        title = "{Turbulent dynamo in the two-phase interstellar medium}",
      journal = {\mnras},
         year = 2022,
        month = jul,
       volume = {514},
       number = {1},
        pages = {957-976},
          doi = {10.1093/mnras/stac1400},
archivePrefix = {arXiv},
       eprint = {2202.08324},
 primaryClass = {astro-ph.GA},
       adsurl = {https://ui.adsabs.harvard.edu/abs/2022MNRAS.514..957S}
}

@ARTICLE{ShahS2021,
       author = {{Shah}, Hilay and {Seta}, Amit},
        title = "{Magnetic fields in elliptical galaxies: using the Laing-Garrington effect in radio galaxies and polarized emission from background radio sources}",
      journal = {\mnras},
         year = 2021,
        month = nov,
       volume = {508},
       number = {1},
        pages = {1371-1388},
          doi = {10.1093/mnras/stab2500},
archivePrefix = {arXiv},
       eprint = {2108.12793},
 primaryClass = {astro-ph.GA},
       adsurl = {https://ui.adsabs.harvard.edu/abs/2021MNRAS.508.1371S}
}

@ARTICLE{SobeyEA2019,
       author = {{Sobey}, C. and {Bilous}, A.~V. and {Grie{\ss}meier}, J. -M. and
         {Hessels}, J.~W.~T. and {Karastergiou}, A. and {Keane}, E.~F. and
         {Kondratiev}, V.~I. and {Kramer}, M. and {Michilli}, D. and
         {Noutsos}, A. and {Pilia}, M. and {Polzin}, E.~J. and
         {Stappers}, B.~W. and {Tan}, C.~M. and {van Leeuwen}, J. and
         {Verbiest}, J.~P.~W. and {Weltevrede}, P. and {Heald}, G. and
         {Alves}, M.~I.~R. and {Carretti}, E. and {En{\ss}lin}, T. and
         {Haverkorn}, M. and {Iacobelli}, M. and {Reich}, W. and {Van Eck}, C.},
        title = "{Low-frequency Faraday rotation measures towards pulsars using LOFAR: probing the 3D Galactic halo magnetic field}",
      journal = {\mnras},
         year = 2019,
        month = apr,
       volume = {484},
       number = {3},
        pages = {3646-3664},
          doi = {10.1093/mnras/stz214},
archivePrefix = {arXiv},
       eprint = {1901.07738},
 primaryClass = {astro-ph.GA},
       adsurl = {https://ui.adsabs.harvard.edu/abs/2019MNRAS.484.3646S}
}

@ARTICLE{SokoloffEA1998,
   author = {{Sokoloff}, D.~D. and {Bykov}, A.~A. and {Shukurov}, A. and 
  {Berkhuijsen}, E.~M. and {Beck}, R. and {Poezd}, A.~D.},
    title = "{Depolarization and Faraday effects in galaxies}",
  journal = {\mnras},
     year = 1998,
    month = aug,
   volume = 299,
    pages = {189-206},
      doi = {10.1046/j.1365-8711.1998.01782.x},
   adsurl = {http://adsabs.harvard.edu/abs/1998MNRAS.299..189S}
}

@ARTICLE{StilEA2011,
       author = {{Stil}, J.~M. and {Taylor}, A.~R. and {Sunstrum}, C.},
        title = "{Structure in the Rotation Measure Sky}",
      journal = {\apj},
         year = 2011,
        month = jan,
       volume = {726},
       number = {1},
          eid = {4},
        pages = {4},
          doi = {10.1088/0004-637X/726/1/4},
archivePrefix = {arXiv},
       eprint = {1010.5299},
 primaryClass = {astro-ph.GA},
       adsurl = {https://ui.adsabs.harvard.edu/abs/2011ApJ...726....4S}
}

@ARTICLE{WilkinBS2007,
   author = {{Wilkin}, S.~L. and {Barenghi}, C.~F. and {Shukurov}, A.},
    title = "{Magnetic Structures Produced by the Small-Scale Dynamo}",
  journal = {\prl},
     year = 2007,
    month = sep,
   volume = 99,
   number = 13,
      eid = {134501},
    pages = {134501},
      doi = {10.1103/PhysRevLett.99.134501},
   adsurl = {http://adsabs.harvard.edu/abs/2007PhRvL..99m4501W}
}

@ARTICLE{ZaroubiEA2015,
   author = {{Zaroubi}, S. and {Jeli{\'c}}, V. and {de Bruyn}, A.~G. and 
  {Boulanger}, F. and {Bracco}, A. and {Kooistra}, R. and {Alves}, M.~I.~R. and 
  {Brentjens}, M.~A. and {Ferri{\`e}re}, K. and {Ghosh}, T. and 
  {Koopmans}, L.~V.~E. and {Levrier}, F. and {Miville-Desch{\^e}nes}, M.-A. and 
  {Montier}, L. and {Pandey}, V.~N. and {Soler}, J.~D.},
    title = "{Galactic interstellar filaments as probed by LOFAR and Planck}",
  journal = {\mnras},
archivePrefix = "arXiv",
   eprint = {1508.06652},
     year = 2015,
    month = nov,
   volume = 454,
    pages = {L46-L50},
      doi = {10.1093/mnrasl/slv123},
   adsurl = {http://adsabs.harvard.edu/abs/2015MNRAS.454L..46Z}
}

@ARTICLE{ZeldovichEA1987,
  author={{Zel'dovich}, {\relax Ya}.~B. and {Molchanov}, S.~A. and {Ruzmaikin}, A.~A. and {Sokoloff}, D.~D.},
  title="{Intermittency in random media}",
  journal="{Soviet Physics Uspekhi}",
  volume=30,
  number=5,
  pages=353,
  year=1987
}

@ARTICLE{ElmegreenS2004,
       author = {{Elmegreen}, Bruce G. and {Scalo}, John},
        title = "{Interstellar Turbulence I: Observations and Processes}",
      journal = {\araa},
         year = 2004,
        month = sep,
       volume = {42},
       number = {1},
        pages = {211-273},
          doi = {10.1146/annurev.astro.41.011802.094859},
archivePrefix = {arXiv},
       eprint = {astro-ph/0404451},
 primaryClass = {astro-ph},
       adsurl = {https://ui.adsabs.harvard.edu/abs/2004ARA&A..42..211E}
}

@ARTICLE{SetaF2021dyn,
       author = {{Seta}, Amit and {Federrath}, Christoph},
        title = "{Saturation mechanism of the fluctuation dynamo in supersonic turbulent plasmas}",
      journal = {Physical Review Fluids},
         year = 2021,
        month = oct,
       volume = {6},
       number = {10},
          eid = {103701},
        pages = {103701},
          doi = {10.1103/PhysRevFluids.6.103701},
archivePrefix = {arXiv},
       eprint = {2109.11698},
 primaryClass = {astro-ph.GA},
       adsurl = {https://ui.adsabs.harvard.edu/abs/2021PhRvF...6j3701S}
}

@ARTICLE{Jaffe2010,
       author = {{Jaffe}, T.~R. and {Leahy}, J.~P. and {Banday}, A.~J. and {Leach}, S.~M. and {Lowe}, S.~R. and {Wilkinson}, A.},
        title = "{Modelling the Galactic magnetic field on the plane in two dimensions}",
      journal = {\mnras},
         year = 2010,
        month = jan,
       volume = {401},
       number = {2},
        pages = {1013-1028},
          doi = {10.1111/j.1365-2966.2009.15745.x},
archivePrefix = {arXiv},
       eprint = {0907.3994},
 primaryClass = {astro-ph.GA},
       adsurl = {https://ui.adsabs.harvard.edu/abs/2010MNRAS.401.1013J}
}

@ARTICLE{Jansson2012b,
       author = {{Jansson}, Ronnie and {Farrar}, Glennys R.},
        title = "{The Galactic Magnetic Field}",
      journal = {\apjl},
         year = 2012,
        month = dec,
       volume = {761},
       number = {1},
          eid = {L11},
        pages = {L11},
          doi = {10.1088/2041-8205/761/1/L11},
archivePrefix = {arXiv},
       eprint = {1210.7820},
 primaryClass = {astro-ph.GA},
       adsurl = {https://ui.adsabs.harvard.edu/abs/2012ApJ...761L..11J}
}

@ARTICLE{VanEck2017,
       author = {{Van Eck}, C.~L. and {Haverkorn}, M. and {Alves}, M.~I.~R. and {Beck}, R. and {de Bruyn}, A.~G. and {En{\ss}lin}, T. and {Farnes}, J.~S. and {Ferri{\`e}re}, K. and {Heald}, G. and {Horellou}, C. and {Horneffer}, A. and {Iacobelli}, M. and {Jeli{\'c}}, V. and {Mart{\'\i}-Vidal}, I. and {Mulcahy}, D.~D. and {Reich}, W. and {R{\"o}ttgering}, H.~J.~A. and {Scaife}, A.~M.~M. and {Schnitzeler}, D.~H.~F.~M. and {Sobey}, C. and {Sridhar}, S.~S.},
        title = "{Faraday tomography of the local interstellar medium with LOFAR: Galactic foregrounds towards IC 342}",
      journal = {\aap},
         year = 2017,
        month = jan,
       volume = {597},
          eid = {A98},
        pages = {A98},
          doi = {10.1051/0004-6361/201629707},
archivePrefix = {arXiv},
       eprint = {1612.00710},
 primaryClass = {astro-ph.GA},
       adsurl = {https://ui.adsabs.harvard.edu/abs/2017A&A...597A..98V}
}

@ARTICLE{Sobey2021,
       author = {{Sobey}, C. and {Johnston}, S. and {Dai}, S. and {Kerr}, M. and {Manchester}, R.~N. and {Oswald}, L.~S. and {Parthasarathy}, A. and {Shannon}, R.~M. and {Weltevrede}, P.},
        title = "{A polarization census of bright pulsars using the ultrawideband receiver on the Parkes radio telescope}",
      journal = {\mnras},
         year = 2021,
        month = jun,
       volume = {504},
       number = {1},
        pages = {228-247},
          doi = {10.1093/mnras/stab861},
archivePrefix = {arXiv},
       eprint = {2103.13838},
 primaryClass = {astro-ph.HE},
       adsurl = {https://ui.adsabs.harvard.edu/abs/2021MNRAS.504..228S}
}

@ARTICLE{Pandhi2024,
       author = {{Pandhi}, Ayush and {Pleunis}, Ziggy and {Mckinven}, Ryan and {Gaensler}, B.~M. and {Su}, Jianing and {Ng}, Cherry and {Bhardwaj}, Mohit and {Brar}, Charanjot and {Cassanelli}, Tomas and {Cook}, Amanda and {Curtin}, Alice P. and {Kaspi}, Victoria M. and {Lazda}, Mattias and {Leung}, Calvin and {Li}, Dongzi and {Masui}, Kiyoshi W. and {Michilli}, Daniele and {Nimmo}, Kenzie and {Pearlman}, Aaron B. and {Petroff}, Emily and {Rafiei-Ravandi}, Masoud and {Sand}, Ketan R. and {Scholz}, Paul and {Shin}, Kaitlyn and {Smith}, Kendrick and {Stairs}, Ingrid},
        title = "{Polarization Properties of 128 Nonrepeating Fast Radio Bursts from the First CHIME/FRB Baseband Catalog}",
      journal = {\apj},
         year = 2024,
        month = jun,
       volume = {968},
       number = {2},
          eid = {50},
        pages = {50},
          doi = {10.3847/1538-4357/ad40aa},
archivePrefix = {arXiv},
       eprint = {2401.17378},
 primaryClass = {astro-ph.HE},
       adsurl = {https://ui.adsabs.harvard.edu/abs/2024ApJ...968...50P}
}

@ARTICLE{Gaensler2004,
       author = {{Gaensler}, B.~M. and {Beck}, R. and {Feretti}, L.},
        title = "{The origin and evolution of cosmic magnetism}",
      journal = {\nar},
         year = 2004,
        month = dec,
       volume = {48},
       number = {11-12},
        pages = {1003-1012},
          doi = {10.1016/j.newar.2004.09.003},
archivePrefix = {arXiv},
       eprint = {astro-ph/0409100},
 primaryClass = {astro-ph},
       adsurl = {https://ui.adsabs.harvard.edu/abs/2004NewAR..48.1003G}
}

@ARTICLE{Haverkorn2004,
       author = {{Haverkorn}, M. and {Gaensler}, B.~M. and {McClure-Griffiths}, N.~M. and {Dickey}, John M. and {Green}, A.~J.},
        title = "{Magnetic Fields and Ionized Gas in the Inner Galaxy: An Outer Scale for Turbulence and the Possible Role of H II Regions}",
      journal = {\apj},
         year = 2004,
        month = jul,
       volume = {609},
       number = {2},
        pages = {776-784},
          doi = {10.1086/421341},
archivePrefix = {arXiv},
       eprint = {astro-ph/0403655},
 primaryClass = {astro-ph},
       adsurl = {https://ui.adsabs.harvard.edu/abs/2004ApJ...609..776H}
}

@ARTICLE{Seta2024,
       author = {{Seta}, Amit and {Federrath}, Christoph},
        title = "{Structure functions with higher-order stencils as a probe to separate small- and large-scale magnetic fields}",
      journal = {\mnras},
         year = 2024,
        month = sep,
       volume = {533},
       number = {2},
        pages = {1875-1886},
          doi = {10.1093/mnras/stae1935},
archivePrefix = {arXiv},
       eprint = {2408.04156},
 primaryClass = {astro-ph.GA},
       adsurl = {https://ui.adsabs.harvard.edu/abs/2024MNRAS.533.1875S}
}

@ARTICLE{MMGPS,
       author = {{Padmanabh}, P.~V. and {Barr}, E.~D. and {Sridhar}, S.~S. and {Rugel}, M.~R. and {Damas-Segovia}, A. and {Jacob}, A.~M. and {Balakrishnan}, V. and {Berezina}, M. and {Bernadich}, M.~C. and {Brunthaler}, A. and {Champion}, D.~J. and {Freire}, P.~C.~C. and {Khan}, S. and {Kl{\"o}ckner}, H. -R. and {Kramer}, M. and {Ma}, Y.~K. and {Mao}, S.~A. and {Men}, Y.~P. and {Menten}, K.~M. and {Sengupta}, S. and {Venkatraman Krishnan}, V. and {Wucknitz}, O. and {Wyrowski}, F. and {Bezuidenhout}, M.~C. and {Buchner}, S. and {Burgay}, M. and {Chen}, W. and {Clark}, C.~J. and {K{\"u}nkel}, L. and {Nieder}, L. and {Stappers}, B. and {Legodi}, L.~S. and {Nyamai}, M.~M.},
        title = "{The MPIfR-MeerKAT Galactic Plane Survey - I. System set-up and early results}",
      journal = {\mnras},
         year = 2023,
        month = sep,
       volume = {524},
       number = {1},
        pages = {1291-1315},
          doi = {10.1093/mnras/stad1900},
archivePrefix = {arXiv},
       eprint = {2303.09231},
 primaryClass = {astro-ph.HE},
       adsurl = {https://ui.adsabs.harvard.edu/abs/2023MNRAS.524.1291P}
}

@ARTICLE{GaenslerEA2025,
       author = {{Gaensler}, B.~M. and {Heald}, G.~H. and {McClure-Griffiths}, N.~M. and {Anderson}, C.~S. and {Van Eck}, C.~L. and {West}, J.~L. and {Thomson}, A.~J.~M. and {Leahy}, J.~P. and {Rudnick}, L. and {Ma}, Y.~K. and {Akahori}, Takuya and {G{\"u}rkan}, G. and {Landecker}, T.~L. and {Mao}, S.~A. and {O'Sullivan}, S.~P. and {Raja}, W. and {Sun}, X. and {Vernstrom}, T. and {Baidoo}, Lerato and {Carretti}, Ettore and {Taylor}, A.~R. and {Willis}, A.~G. and {Osinga}, Erik and {Livingston}, J.~D. and {Alexander}, E.~L. and {Alonso-L{\'o}pez}, David and {Amaral}, A.~D. and {An}, T. and {Bracco}, Andrea and {Bradbury}, S. and {Br{\"u}ggen}, Marcus and {Eswaraiah}, Chakali and {En{\ss}lin}, Torsten and {Galvin}, T.~J. and {Haverkorn}, Marijke and {Hopkins}, A.~M. and {Hutschenreuter}, Sebastian and {Ideguchi}, Shinsuke and {Jaswanth}, S. and {Jung}, S. Lyla and {Kaczmarek}, J.~F. and {Kothes}, Roland and {Lazarevi{\'c}}, Sanja and {Leahy}, Denis and {Loi}, Francesca and {Marvil}, Joshua R. and {Norris}, Ray and {Pandhi}, Ayush and {Price}, Jason M. and {Riseley}, C.~J. and {Ryder}, P. and {Seta}, Amit and {Shaw}, Vasundhara and {Shen}, A.~X. and {Sobey}, C. and {Stil}, J. and {Stuardi}, Chiara and {Upasana}, Gupta and {Vanderwoude}, Shannon and {Velovi{\'c}}, Velibor},
        title = "{The Polarisation Sky Survey of the Universe's Magnetism (POSSUM): Science goals and survey description}",
      journal = {\pasa},
         year = 2025,
        month = jun,
       volume = {42},
          eid = {e091},
        pages = {e091},
          doi = {10.1017/pasa.2025.10031},
archivePrefix = {arXiv},
       eprint = {2505.08272},
 primaryClass = {astro-ph.GA},
       adsurl = {https://ui.adsabs.harvard.edu/abs/2025PASA...42...91G}
}

@ARTICLE{MaEA2019a,
       author = {{Ma}, Yik Ki and {Mao}, S.~A. and {Stil}, Jeroen and {Basu}, Aritra and {West}, Jennifer and {Heiles}, Carl and {Hill}, Alex S. and {Betti}, S.~K.},
        title = "{A broad-band spectro-polarimetric view of the NVSS rotation measure catalogue - I. Breaking the n{\ensuremath{\pi}}-ambiguity}",
      journal = {\mnras},
         year = 2019,
        month = aug,
       volume = {487},
       number = {3},
        pages = {3432-3453},
          doi = {10.1093/mnras/stz1325},
archivePrefix = {arXiv},
       eprint = {1905.04313},
 primaryClass = {astro-ph.GA},
       adsurl = {https://ui.adsabs.harvard.edu/abs/2019MNRAS.487.3432M}
}

@ARTICLE{BaidooEA2023,
       author = {{Baidoo}, Lerato and {Perley}, Richard A. and {Eilek}, Jean and {Smirnov}, Oleg and {Vacca}, Valentina and {En{\ss}lin}, Torsten},
        title = "{A Wideband Polarization Observation of Hydra A with the Jansky Very Large Array}",
      journal = {\apj},
         year = 2023,
        month = sep,
       volume = {955},
       number = {1},
          eid = {16},
        pages = {16},
          doi = {10.3847/1538-4357/acebc5},
archivePrefix = {arXiv},
       eprint = {2308.05805},
 primaryClass = {astro-ph.GA},
       adsurl = {https://ui.adsabs.harvard.edu/abs/2023ApJ...955...16B}
}

@ARTICLE{FarnsworthEA2011,
       author = {{Farnsworth}, Damon and {Rudnick}, Lawrence and {Brown}, Shea},
        title = "{Integrated Polarization of Sources at {\ensuremath{\lambda}} \raisebox{-0.5ex}\textasciitilde 1 m and New Rotation Measure Ambiguities}",
      journal = {\aj},
         year = 2011,
        month = jun,
       volume = {141},
       number = {6},
          eid = {191},
        pages = {191},
          doi = {10.1088/0004-6256/141/6/191},
archivePrefix = {arXiv},
       eprint = {1103.4149},
 primaryClass = {astro-ph.CO},
       adsurl = {https://ui.adsabs.harvard.edu/abs/2011AJ....141..191F}
}

@ARTICLE{OSullivanEA2012,
       author = {{O'Sullivan}, S.~P. and {Brown}, S. and {Robishaw}, T. and {Schnitzeler}, D.~H.~F.~M. and {McClure-Griffiths}, N.~M. and {Feain}, I.~J. and {Taylor}, A.~R. and {Gaensler}, B.~M. and {Landecker}, T.~L. and {Harvey-Smith}, L. and {Carretti}, E.},
        title = "{Complex Faraday depth structure of active galactic nuclei as revealed by broad-band radio polarimetry}",
      journal = {\mnras},
         year = 2012,
        month = apr,
       volume = {421},
       number = {4},
        pages = {3300-3315},
          doi = {10.1111/j.1365-2966.2012.20554.x},
archivePrefix = {arXiv},
       eprint = {1201.3161},
 primaryClass = {astro-ph.CO},
       adsurl = {https://ui.adsabs.harvard.edu/abs/2012MNRAS.421.3300O}
}

@ARTICLE{RanchodEA2024,
       author = {{Ranchod}, S. and {Mao}, S.~A. and {Deane}, R. and {Sridhar}, S.~S. and {Damas-Segovia}, A. and {Livingston}, J.~D. and {Ma}, Y.~K.},
        title = "{The Galactic latitude dependency of Faraday complexity in the S-PASS/ATCA RM catalogue}",
      journal = {\aap},
         year = 2024,
        month = jun,
       volume = {686},
          eid = {A104},
        pages = {A104},
          doi = {10.1051/0004-6361/202348993},
archivePrefix = {arXiv},
       eprint = {2403.13500},
 primaryClass = {astro-ph.GA},
       adsurl = {https://ui.adsabs.harvard.edu/abs/2024A&A...686A.104R}
}

@ARTICLE{MaEA2025,
       author = {{Ma}, Yik Ki and {Seta}, Amit and {McClure-Griffiths}, N.~M. and {Van Eck}, C.~L. and {Mao}, S.~A. and {Ordog}, A. and {Brown}, J.~C. and {Kovacs}, T.~O. and {Akahori}, Takuya and {Kurahara}, K. and {Oberhelman}, L. and {Anderson}, C.~S.},
        title = "{A new window into the sub-parsec scale magnetic field in the Milky Way? Unveiling small-scale magneto-ionic structures with Faraday complexity}",
      journal = {\mnras},
         year = 2025,
        month = jul,
       volume = {541},
       number = {1},
        pages = {306-336},
          doi = {10.1093/mnras/staf1000},
archivePrefix = {arXiv},
       eprint = {2506.18968},
 primaryClass = {astro-ph.GA},
       adsurl = {https://ui.adsabs.harvard.edu/abs/2025MNRAS.541..306M}
}

@ARTICLE{BasuEA2017,
       author = {{Basu}, Aritra and {Mao}, S.~A. and {Kepley}, Amanda A. and {Robishaw}, Timothy and {Zweibel}, Ellen G. and {Gallagher}, III, John. S.},
        title = "{Detection of an {\ensuremath{\sim}}20 kpc coherent magnetic field in the outskirt of merging spirals: the Antennae galaxies}",
      journal = {\mnras},
         year = 2017,
        month = jan,
       volume = {464},
       number = {1},
        pages = {1003-1017},
          doi = {10.1093/mnras/stw2369},
archivePrefix = {arXiv},
       eprint = {1609.04266},
 primaryClass = {astro-ph.GA},
       adsurl = {https://ui.adsabs.harvard.edu/abs/2017MNRAS.464.1003B}
}

@ARTICLE{MorganEA2018,
       author = {{Morgan}, J.~S. and {Macquart}, J. -P. and {Ekers}, R. and {Chhetri}, R. and {Tokumaru}, M. and {Manoharan}, P.~K. and {Tremblay}, S. and {Bisi}, M.~M. and {Jackson}, B.~V.},
        title = "{Interplanetary Scintillation with the Murchison Widefield Array I: a sub-arcsecond survey over 900 deg$^{2}$ at 79 and 158 MHz}",
      journal = {\mnras},
         year = 2018,
        month = jan,
       volume = {473},
       number = {3},
        pages = {2965-2983},
          doi = {10.1093/mnras/stx2284},
archivePrefix = {arXiv},
       eprint = {1709.00312},
 primaryClass = {astro-ph.IM},
       adsurl = {https://ui.adsabs.harvard.edu/abs/2018MNRAS.473.2965M}
}

@ARTICLE{OSullivanEA2023,
       author = {{O'Sullivan}, S.~P. and {Shimwell}, T.~W. and {Hardcastle}, M.~J. and {Tasse}, C. and {Heald}, G. and {Carretti}, E. and {Br{\"u}ggen}, M. and {Vacca}, V. and {Sobey}, C. and {Van Eck}, C.~L. and {Horellou}, C. and {Beck}, R. and {Bilicki}, M. and {Bourke}, S. and {Botteon}, A. and {Croston}, J.~H. and {Drabent}, A. and {Duncan}, K. and {Heesen}, V. and {Ideguchi}, S. and {Kirwan}, M. and {Lawlor}, L. and {Mingo}, B. and {Nikiel-Wroczy{\'n}ski}, B. and {Piotrowska}, J. and {Scaife}, A.~M.~M. and {van Weeren}, R.~J.},
        title = "{The Faraday Rotation Measure Grid of the LOFAR Two-metre Sky Survey: Data Release 2}",
      journal = {\mnras},
         year = 2023,
        month = mar,
       volume = {519},
       number = {4},
        pages = {5723-5742},
          doi = {10.1093/mnras/stac3820},
archivePrefix = {arXiv},
       eprint = {2301.07697},
 primaryClass = {astro-ph.CO},
       adsurl = {https://ui.adsabs.harvard.edu/abs/2023MNRAS.519.5723O}
}

@ARTICLE{PanopoulouEA2025,
       author = {{Panopoulou}, G.~V. and {Zucker}, C. and {Clemens}, D. and {Pelgrims}, V. and {Soler}, J.~D. and {Clark}, S.~E. and {Alves}, J. and {Goodman}, A. and {Becker Tjus}, J.},
        title = "{The magnetic field of the Radcliffe wave: Starlight polarization at the nearest approach to the Sun}",
      journal = {\aap},
         year = 2025,
        month = feb,
       volume = {694},
          eid = {A97},
        pages = {A97},
          doi = {10.1051/0004-6361/202450991},
archivePrefix = {arXiv},
       eprint = {2406.03765},
 primaryClass = {astro-ph.GA},
       adsurl = {https://ui.adsabs.harvard.edu/abs/2025A&A...694A..97P}
}

@ARTICLE{AngaritaEA2024,
       author = {{Angarita}, Y. and {Versteeg}, M.~J.~F. and {Haverkorn}, M. and {Marchal}, A. and {Rodrigues}, C.~V. and {Magalh{\~a}es}, A.~M. and {Santos-Lima}, R. and {Kawabata}, Koji S.},
        title = "{Interstellar Polarization Survey. IV. Characterizing the Magnetic Field Strength and Turbulent Dispersion Using Optical Starlight Polarization in the Diffuse Interstellar Medium}",
      journal = {\aj},
         year = 2024,
        month = jul,
       volume = {168},
       number = {1},
          eid = {47},
        pages = {47},
          doi = {10.3847/1538-3881/ad4b14},
archivePrefix = {arXiv},
       eprint = {2405.07347},
 primaryClass = {astro-ph.GA},
       adsurl = {https://ui.adsabs.harvard.edu/abs/2024AJ....168...47A}
}

@ARTICLE{Schnitzeler2010,
       author = {{Schnitzeler}, D.~H.~F.~M.},
        title = "{The latitude dependence of the rotation measures of NVSS sources}",
      journal = {\mnras},
         year = 2010,
        month = nov,
       volume = {409},
       number = {1},
        pages = {L99-L103},
          doi = {10.1111/j.1745-3933.2010.00957.x},
archivePrefix = {arXiv},
       eprint = {1011.0737},
 primaryClass = {astro-ph.GA},
       adsurl = {https://ui.adsabs.harvard.edu/abs/2010MNRAS.409L..99S}
}

@ARTICLE{RudnickOwen2014,
       author = {{Rudnick}, L. and {Owen}, F.~N.},
        title = "{The Distribution of Polarized Radio Sources >15 {\ensuremath{\mu}}Jy in GOODS-N}",
      journal = {\apj},
         year = 2014,
        month = apr,
       volume = {785},
       number = {1},
          eid = {45},
        pages = {45},
          doi = {10.1088/0004-637X/785/1/45},
archivePrefix = {arXiv},
       eprint = {1402.3637},
 primaryClass = {astro-ph.GA},
       adsurl = {https://ui.adsabs.harvard.edu/abs/2014ApJ...785...45R}
}

@ARTICLE{PoraykoEA2019,
       author = {{Porayko}, N.~K. and {Noutsos}, A. and {Tiburzi}, C. and {Verbiest}, J.~P.~W. and {Horneffer}, A. and {K{\"u}nsem{\"o}ller}, J. and {Os{\l}owski}, S. and {Kramer}, M. and {Schnitzeler}, D.~H.~F.~M. and {Anderson}, J.~M. and {Br{\"u}ggen}, M. and {Grie{\ss}meier}, J. -M. and {Hoeft}, M. and {Schwarz}, D.~J. and {Serylak}, M. and {Wucknitz}, O.},
        title = "{Testing the accuracy of the ionospheric Faraday rotation corrections through LOFAR observations of bright northern pulsars}",
      journal = {\mnras},
         year = 2019,
        month = mar,
       volume = {483},
       number = {3},
        pages = {4100-4113},
          doi = {10.1093/mnras/sty3324},
archivePrefix = {arXiv},
       eprint = {1812.01463},
 primaryClass = {astro-ph.IM},
       adsurl = {https://ui.adsabs.harvard.edu/abs/2019MNRAS.483.4100P}
}

@ARTICLE{WangEA2011,
       author = {{Wang}, Chen and {Han}, J.~L. and {Lai}, Dong},
        title = "{The Faraday rotation in the pulsar magnetosphere}",
      journal = {\mnras},
         year = 2011,
        month = oct,
       volume = {417},
       number = {2},
        pages = {1183-1191},
          doi = {10.1111/j.1365-2966.2011.19333.x},
archivePrefix = {arXiv},
       eprint = {1105.2602},
 primaryClass = {astro-ph.SR},
       adsurl = {https://ui.adsabs.harvard.edu/abs/2011MNRAS.417.1183W}
}

@ARTICLE{Laing1980,
       author = {{Laing}, R.~A.},
        title = "{A model for the magnetic-field structure in extended radio sources.}",
      journal = {\mnras},
         year = 1980,
        month = nov,
       volume = {193},
        pages = {439-449},
          doi = {10.1093/mnras/193.3.439},
       adsurl = {https://ui.adsabs.harvard.edu/abs/1980MNRAS.193..439L}
}

@article{Takayanagi1973,
	Adsurl = {http://adsabs.harvard.edu/abs/1973PASJ...25..327T},
	Author = {{Takayanagi}, K.},
	Journal = {\pasj},
	Pages = {327},
	Title = {{Molecule Formation in Dense Interstellar Clouds}},
	Volume = 25,
	Year = 1973}
\end{document}